\documentclass[useAMS,usenatbib]{mn2e}

\usepackage{psfig,epsfig,supertabular,wrapfig,placeins}
\usepackage{graphicx,amsmath,amssymb,subfigure,multirow,units}
\usepackage{marvosym}
\usepackage{graphics}
\usepackage{natbib}
\usepackage{amssymb}
\usepackage{amsmath}
\usepackage{euscript}
\usepackage{setspace}
\usepackage{fancyhdr}
\usepackage{afterpage,rotating,url}
\usepackage{times}

\newcommand{\msun}{\mbox{\,$\rm M_{\odot}$\,}}        
\newcommand{\Lsub}{\mbox{\,$ L_{250}$\,}} 
\newcommand{\Msed}{\mbox{\,$ M_{\rm sed}$\,}}
\newcommand{\Miso}{\mbox{\,$M_{\rm iso}$\,}}
\newcommand{\Md}{\mbox{\,$M_{\rm d}$\,}}
\newcommand{\asec}{\ensuremath{^{\prime\prime}}}

\def\mic{$\mu $m\,}

\title[Evolution of dust mass]{Herschel\thanks{Herschel is an ESA
    space observatory with science instruments provided by
    European-led Principal Investigator consortia and with important
    participation from NASA}-ATLAS: Rapid evolution of dust in
  galaxies over the last 5 billion years}

\author[L. Dunne et al.]
{L. Dunne$^{1}$\thanks{E-mail:loretta.dunne@nottingham.ac.uk}, 
~H. L. Gomez$^2$, 
~E.~da Cunha$^{3,4}$, 
~S.~Charlot$^5$,
~S.~Dye$^2$, 
~S.~Eales$^2$, 
~S.J.~Maddox$^{1}$
\newauthor
K. Rowlands$^1$,
~D.J.B.~Smith$^1$,
~R.~Auld$^2$,
~M.~Baes$^6$, 
~D.G.~Bonfield$^{7}$, 
~N.~Bourne$^1$ 
\newauthor
S.~Buttiglione$^8$,  
~A.~Cava$^9$, 
~D.~L.~Clements$^{10}$, 
~K.~E.~K.~Coppin$^{11,12}$, 
~A.~Cooray$^{13}$, 
~A.~Dariush$^2$
\newauthor
G.~de Zotti$^{8,14}$, 
~S.~Driver$^{15}$,   
~J.~Fritz$^{6}$, 
~J. Geach$^{11,12}$, 
~R.~Hopwood$^{13}$, 
~E.~Ibar$^{16}$, 
~R.J.~Ivison$^{16,17}$ 
\newauthor
M.J.~Jarvis$^{7}$, 
~L.~Kelvin$^{15}$, 
~E.~Pascale$^2$,
~M.~Pohlen$^2$,
~C.~Popescu$^{18}$  
~E.E.~Rigby$^1$,
~A.~Robotham$^{15}$ 
\newauthor
G.~Rodighiero$^{19}$, 
~A.E.~Sansom$^{18}$, 
~S.~Serjeant$^{20}$, 
~P.~Temi$^{21}$, 
~M.~Thompson$^{7}$, 
~R.~Tuffs$^{22}$ 
\newauthor
P.~van der Werf$^{16,23}$, 
C.~Vlahakis$^{24}$ \\
  \\ $^{1}$School of Physics \&\ Astronomy, Nottingham University, University Park Campus, Nottingham, NG7 2RD, UK \\ 
  $^2$School of Physics \&\ Astronomy, Cardiff University, Queen
  Buildings, The Parade, Cardiff, CF24 3AA, UK \\
  $^3$Max Planck Institute for Astronomy, Konigstuhl 17, 69117, Heidelberg, Germany\\
  $^4$Department of Physics, University of Crete, PO Box 2208, 71003 Heraklion, Greece\\
  $^5$Institut d'Astrophysique de Paris, CNRS, Universit\'e Pierre \& Marie Curie, UMR 7095, 98bis bd Arago, 75014 Paris, France\\
  $^6$Sterrenkundig Observatorium, Universiteit Gent, Krijgslaan 281
  S9, B-9000 Gent, Belgium\\
  $^{7}$Centre for Astrophysics, Science \&\ Technology Research
  Institute, University of Hertfordshire, Hatfield, Herts, AL10 9AB,
  UK \\ 
  $^8$INAF-Osservatorio Astronomico di Padova, Vicolo Osservatorio 5, I-35122, Padova, Italy\\
  $^9$Departmento de Astrof\'{\i}sica, Facultad de CC. F\'{\i}sicas, Universidad Complutense de Madrid, E-28040 Madrid, Spain\\ 
  $^{10}$Department of Physics and Astronomy, University of California, Irvine, CA 92697, USA\\
  $^{11}$Department of Physics, McGill University, Ernest Rutherford Building, 3600 Rue University, Montreal, Quebec, H3A 2T8, Canada\\
  $^{12}$Institute for Computational Cosmology, Durham University, South Road, Durham, DH1 3LE, UK\\
  $^{13}$Physics Department, Imperial College, Prince Consort Road, London, SW7 2AZ\\
  $^{14}$SISSA, Via Bonomea 265, I-34136 Trieste, Italy\\
  $^{15}$SUPA, School of Physics and Astronomy, University of St. Andrews, North Haugh, St. Andrews, KY16 9SS, UK\\
  $^{16}$Uk Astronomy Technology Centre, Royal Observatory, Edinburgh,
  EH9 3HJ, UK\\
  $^{17}$SUPA, Institute for Astronomy, University of Edinbugh, Royal Observatory, Blackford Hill, Edinbugh EH9 3HJ, UK \\
  $^{18}$Jeremiah Horrocks Institute, University of Central Lancashire, Preston PR1 2HE, UK\\
  $^{19}$University of Padova, Department of Astronomy, Vicolo Osservatorio 3, I-35122, Padova, Italy\\
  $^{20}$Astrophysics Branch, NASA Ames Research Center, Mail Stop 2456,
  Moffett Field, CA 94035, USA\\
  $^{21}$Dept. of Physics and Astronomy, The Open University, Milton Keynes, MK7 6AA\\
  $^{22}$Max Planck Institut fuer Kernphysik, Saupfercheckweg 1, D-69117 Heidelberg, Germany\\
  $^{23}$Leiden Observatory, Leiden University, P.O. Box 9513, NL -2300 RA Leiden \\
  $^{24}$Departamento de Astronomia, Universidad de Chile, Casilla 36-D, Santiago, Chile
}
\begin{document}

\date{\vspace*{-5em}\today}

\pagerange{\pageref{firstpage}--\pageref{lastpage}} \pubyear{2002}

\maketitle

\label{firstpage}

\begin{abstract}
We present the first direct and unbiased measurement of the evolution
of the dust mass function of galaxies over the past 5 billion years of
cosmic history using data from the Science Demonstration Phase of the
{\em Herschel}-ATLAS. The sample consists of galaxies selected at
250\mic which have reliable counterparts from SDSS at $z<0.5$, and
contains 1867 sources. Dust masses are calculated using both a single
temperature grey-body model for the spectral energy distribution and
also using a model with multiple temperature components. The dust
temperature for either model shows no trend with redshift. Splitting
the sample into bins of redshift reveals a strong evolution in the
dust properties of the most massive galaxies. At $z=0.4-0.5$, massive
galaxies had dust masses about five times larger than in the local
Universe. At the same time, the dust-to-stellar mass ratio was about
3--4 times larger, and the optical depth derived from fitting the
UV--sub-mm data with an energy balance model was also higher. This
increase in the dust content of massive galaxies at high redshift is
difficult to explain using standard dust evolution models and requires
a rapid gas consumption timescale together with either a more
top-heavy IMF, efficient mantle growth, less dust destruction or
combinations of all three. This evolution in dust mass is likely to be
associated with a change in overall ISM mass, and points to an
enhanced supply of fuel for star formation at earlier cosmic epochs.
\end{abstract}


\section{Introduction}
The evolution of the dust content of galaxies is an important and
poorly understood topic. Dust is responsible for obscuring the UV and
optical light from galaxies and thus introduces biases into our
measures of galaxy properties based on their stellar light (Driver et
al. 2007). The energy absorbed by dust is re-emitted at longer
infra-red and sub-millimetre (sub-mm) wavelengths, providing a means
of recovering the stolen starlight. Dust emission is often used as an
indicator of the current star formation rate in galaxies - although
this calibration makes the assumption that young, massive stars are
the main source of heating for the dust and that the majority of the
UV photons from the young stars are absorbed and re-radiated by dust
(Kennicutt et al. 1998, 2009; Calzetti et al. 2007). Many surveys of
dust emission from 24--850\mic\ (Saunders et al. 1990; Blain et
al. 1999; Le Floc'h et al. 2005; Gruppioni et al. 2010; Dye et
al. 2010; Eales et al. 2010) have noted the very strong evolution
present in these bands and this is usually ascribed to a decrease in
the star formation rate density over the past 8 billion years of
cosmic history ($z\sim1$: Madau et al. 1996, Hopkins 2004). The
interpretation of this evolution is complicated by the fact
that the dust luminosity of a galaxy is a function of both the dust
content and the temperature of the dust. It is pertinent to now ask
the question ``{\em What drives the evolution in the FIR luminosity
  density?\/}'', is it an increase in dust heating (due to enhanced
star formation activity) or an increase the dust content of galaxies
(due to their higher gas content in the past) -- or both?

Dust is thought to be produced by both low-intermediate mass AGB stars
(Gehrz 1989; Ferrarotti \& Gail 2006; Sargent et al. 2010) and by
massive stars when they explode as supernovae at the end of their
short lives (Rho et al. 2008; Dunne et al. 2009; Barlow et
al. 2010). Thus, the dust mass in a galaxy should be related to its
current and past star formation history. Dust is also destroyed
through astration and via supernovae shocks (Jones et al. 1994), and
may also reform through accretion in both the dense and diffuse ISM
(Zhukovska et al. 2008; Inoue 2003; Tielens 1998). The life cycle
of dust is thus a complicated process which many have attempted to
model (Morgan \& Edmunds 2003; Dwek et al. 1998; Calura et al. 2008,
Gomez et al. 2010; Gall, Anderson \& Hjorth 2011) and yet the basic
statistic describing the dust content of galaxies - the dust mass
function (DMF) - is not well determined.

The first attempts to measure the dust mass function were made by
Dunne et al. (2000; hereafter D00) and Dunne \& Eales (2001; hereafter
DE01) as part of the SLUGS survey using a sample of {\em IRAS} bright
galaxies observed with SCUBA at 450 and 850\mic. Vlahakis, Dunne \&
Eales (2005; hereafter VDE05) improved on this by adding an optically
selected sample with sub-mm observations.  These combined studies,
however, comprised less than 200 objects - none of which were selected
on the basis of their dust mass. These studies were also at very low-z
and did not allow for a determination of evolution. A high-z dust mass
function was estimated by Dunne, Eales \& Edmunds (2003; hereafter
DEE03) using data from deep sub-mm surveys. This showed considerable
evolution with galaxies at the high mass end requiring an order of
magnitude more dust at $z\sim 2.5$ compared to today (for pure
luminosity evolution), though with generous caveats due to the
difficulties in making this measurement. Finally, Eales et al. (2009)
used BLAST data from 250--500\mic\ and also concluded that there was
strong evolution in the dust mass function between $z=0-1$ but were
also limited by small number statistics and confusion in the BLAST
data due to their large beam size.

In this paper, we present the first direct measurement of the space
density of galaxies as a function of dust mass out to $z=0.5$. Our
sample is an order of magnitude larger than previous studies, and is
the first which is near `dust mass' selected. We then use this sample
to study the evolution of dust mass in galaxies over the past $\sim 5$
billion years of cosmic history in conjunction with the elementary
dust evolution model of Edmunds (2001).

The new sample which allows us to study the dust mass function in this
way comes from the {\em Herschel}-Astrophysical Terahertz Large Area
Survey (H-ATLAS; Eales et al., 2010), which is the largest open-time
key project currently being carried out with the {\em Herschel} Space
Observatory (Pilbratt et al., 2010). H-ATLAS will survey in excess of
550 deg$^2$ in five bands centered on 100, 160, 250, 350 and
500$\mu$m, using the PACS (Poglitsch et al., 2010) and SPIRE
instruments (Griffin et al., 2010). The observations consist of two
scans in parallel mode reaching 5$\sigma$ point source sensitivities
of 132, 126, 32, 36 and 45 mJy in the 100, 160, 250, 350 \&\ 500$\mu$m
bands respectively, with beam sizes of approximately 9\asec, 13\asec,
18\asec, 25\asec\ and 35\asec. The SPIRE and PACS map-making are
described in the papers by Pascale et al. (2011) and Ibar et
al. (2010), while the catalogues are described in Rigby et
al. (2011). One of the primary aims of the {\em Herschel}-ATLAS is to
obtain the first unbiased survey of the local Universe at sub-mm
wavelengths, and as a result was designed to overlap with existing
large optical and infrared surveys. These Science Demonstration Phase
(SDP) observations are centered on the 9$^h$ field of the Galaxy And
Mass Assembly (GAMA; Driver et al. 2011) survey. The SDP field covers
14.4 sq. deg and comprises approximately one thirtieth of the eventual
full H-ATLAS sky coverage.

In section \ref{Sample} we describe the sample that we have chosen to
use for this analysis and the completeness corrections required. In
section \ref{masses} we describe how we have derived luminosities and
dust masses from the {\em Herschel} data, while in section \ref{DMF},
we present the dust mass function and evaluate its evolution. Section
\ref{models} compares the DMF to models of dust evolution in order to
explain the origin of the strong evolution. Throughout we use a
cosmology with $\Omega_m = 0.27,\,\Omega_{\Lambda} =0.73$ and $H_o =
71\, \rm{km\,s^{-1}\,Mpc^{-1}}$.

\section{Sample definitions}
\label{Sample}

The sub-mm catalogue used in this work is based on the $>5\sigma$ at
250\mic catalogue from Rigby et al. (2011), which contains 6610
sources. The 250\mic fluxes of sources selected in this way have been
shown to be unaffected by flux boosting, see Rigby et al. (2011) for a
thorough description. Sources from this catalogue are matched to
optical counterparts from SDSS DR7 (Abazajian et al. 2009) down to a
limiting magnitude of $r$-modelmag =22.4 using a Likelihood Ratio (LR)
technique (e.g. Sutherland \& Saunders 1992). The method is described
in detail in Smith et al. (2011). Briefly, each optical galaxy within
10\asec of a 250\mic source is assigned a reliability, $R$, which is
the probability that it is truly associated with the 250\mic
emission. This method accounts for the possibility that true IDs are
below the optical flux limit, the positional uncertainties of both
samples, and deals with sharing the likelihoods when there are
multiple counterparts. For our study we have used a reliability cut of
$R\geq 0.8$ as this ensures a low contamination rate ($<5$ percent)
which leaves 2423 250\mic sources with reliable counterparts. The LR
method tells us that $\sim 3800$ counterparts should be present in the SDSS
catalogue, however we can only unambiguously associate around 64
percent of these. Our sample is thus low in contamination but
incomplete (we will deal specifically with the incompleteness of the
ID process in the next section). A further cut was made to this sample
to remove any stars or unresolved objects, this was done using a
star-galaxy separation technique based on optical/IR colour and size,
similar to that used by Baldry et al. (2010). Only six objects in the
final reliable ID catalogue have `stellar or QSO IDs' and so required
removal. We also removed the five sources which were identified as
being lensed by Negrello et al. (2010).

We then used the GAMA database (Driver et al. 2011) to obtain
spectroscopic redshifts for as many of the sources as possible (GAMA
target selection is based on SDSS so no further matching is
required). These are supplemented by public redshifts from SDSS DR7
(Abazajian et al., 2009), 2SLAQ-LRG (Cannon et al., 2006), 2SLAQ-QSO
(Croom et al., 2009) and 6dFGRS (Jones et al., 2009). Where
spectroscopic redshifts were not available we used photometric
redshifts which were produced for H-ATLAS using SDSS and UKIDSS-LAS
(Lawrence et al. 2007) data and the ANNz method (Collister \& Lahav
2004). This method is described fully in Smith et al. (2011).

Section~\ref{sec:comp} shows that we can quantify the statistical
completeness of the IDs out to $z=0.5$ and we choose this as the
redshift limit of the current study. The total number of sources in
the final sample is 1867 with 1095 spectroscopic redshifts. With this
sample, the number of expected false IDs (summing $1-R$, see Smith et
al. 2011) is 60 (or 3.2 percent).


\begin{figure}
\centering
\includegraphics[width=0.45\textwidth,angle=-90]{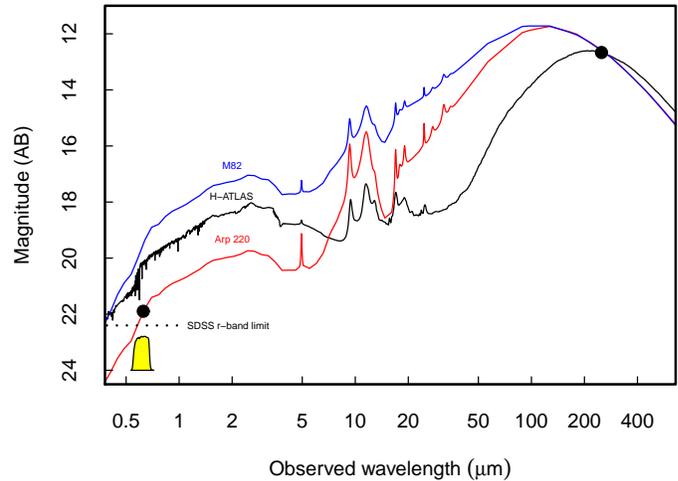}
\caption{Templates for three galaxies showing the range of optical
  fluxes expected for galaxies which are at the SPIRE flux limit of
  $S_{250}=32$ mJy at $z=0.5$; the limit of our study. The templates
  are for M82 (a typical starburst), a {\em Herschel\/}-ATLAS template
  derived from our survey data by Smith et al. in prep and Arp 220, a
  highly obscured local ULIRG. The SDSS limit of $r=22.4$ is
  shown as a horizontal dotted line and even a galaxy as obscured as
  Arp 220 is still visible as an ID to our optical limit at
  $z=0.5$. The yellow shape represent the SDSS-$r$ band filter which
  was used to compute the optical flux}
\label{fig:templates}
\end{figure}

\begin{figure*}
\centering
\includegraphics[width=0.5\textwidth]{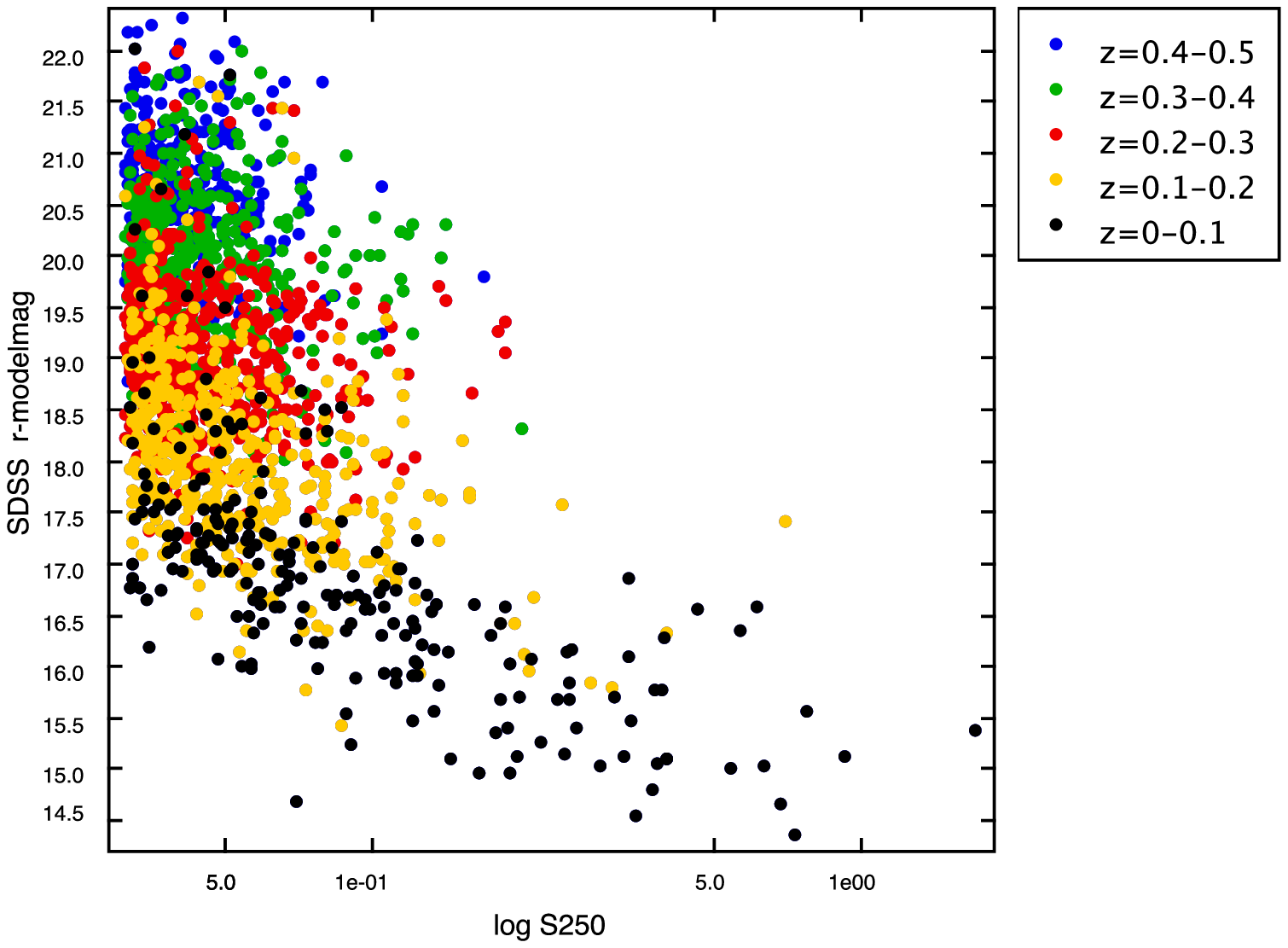}
\includegraphics[width=0.49\textwidth]{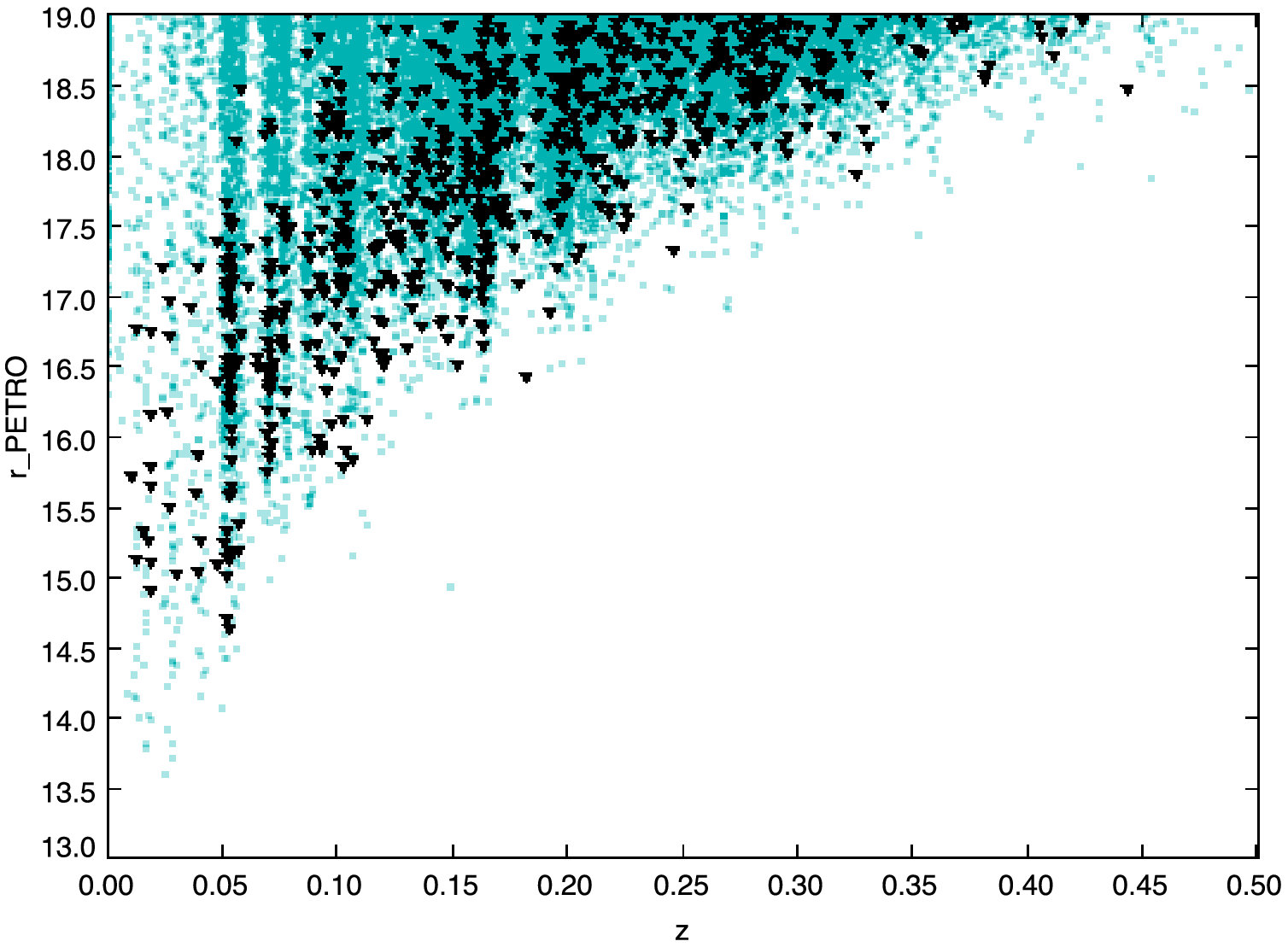}
\caption{{\bf Left:} SDSS $r$-modelmag as a function of 250\mic
  flux. There is no strong correlation apart from at the brightest
  fluxes. Only 4 galaxies lie within 0.4 mag of the flux limit used
  for IDs ($r<22.4$) at $z<0.5$ and so we consider that optical
  incompleteness is not a serious problem for this sample. {\bf
    Right:} $r$-mag versus redshift for all sources in GAMA-9 (pale
  blue squares) and SPIRE IDs with $R\geq 0.8$ (black triangles). {\em
    Herschel} sources tend to be larger mass optical galaxies and so
  the SDSS flux-limit does not affect our ability to optically identify H-ATLAS
  sources until $z\sim 0.5$. Note that the right panel uses the
  brighter limit of $r<19$ appropriate for the GAMA redshift survey. }
\label{fig:rcomps}
\end{figure*}

\subsection{Completeness corrections}
\label{sec:comp}

There are three sources of incompleteness in this current sample.
\begin{enumerate}
\item{{\em Sub-mm Catalogue Incompleteness ($C_s$):\/} This is due to
  the 250\mic\ flux limit of the survey and the efficacy of the source
  extraction process. The catalogue number density completeness has
  been estimated through simulations and presented by Rigby et
  al. (2011). Apart from the very small range of flux near to the
  limit, at $32-34$ mJy the catalogue is $>80$ percent
  complete. Correction factors are applied to each source in turn based on
  its flux following Tables 1 and 2 in Rigby et al. (2011). The
  largest correction is in the flux range 32--32.7 mJy and is a factor
  2.17, this applies to 124 sources out of a total of 1867 at
  $z<0.5$.}

\item{{\em ID Incompleteness ($C_z$):\/} The LR method measures in an
    empirical way a quantity $Q_o$, which is the fraction of SPIRE
    sources with counterparts above the flux limit in the optical
    survey. However, it is not possible to unambiguously identify all
    these counterparts with $>80$ percent confidence due to positional
    uncertainties, close secondaries and the random probability of
    finding a background source within that search radius. Smith et
    al. (2011) have estimated a completeness for reliable IDs as a
    function of redshift. This allows us to make a statistical
    number density correction in redshift slices for the sources which
    should have a counterpart above the SDSS limit in that redshift
    slice, but which do not have $R\geq 0.8$. This correction is
    applied to each source and is listed in
    Table~\ref{tab:z_comp}. The ID incompleteness is a function of
    redshift (not unexpectedly) with corrections of a factor $\sim 2$
    needing to be applied in the highest redshift bins.}

\begin{table}
\caption{The percentage completeness of our reliable ID catalogue as a
  function of redshift, as taken from Smith et
  al. (2011). The correction factor used in the luminosity function
  is denoted by $C_z$.}  \centering
  \begin{tabular}{|c|c|c|c|}
    \hline
    $z$ & Completeness (\%) & $C_z$ \\
    \hline
    0.0 -- 0.1 & 93.2 &1.07\\
    0.1 -- 0.2 & 83.2 &1.20\\
    0.2 -- 0.3 & 74.2 &1.35\\
    0.3 -- 0.4 & 55.6 &1.80\\
    0.4 -- 0.5 & 53.1 &1.88\\
    \hline
    \label{tab:z_comp}
  \end{tabular}
\end{table}

\item{{\em Optical catalogue incompleteness ($C_r$):\/} This
  correction is required because the SDSS catalogue from which we made
  the identifications is itself incomplete as we approach the optical
  flux limit of $r=22.4$. We ascertained the completeness using the
  background source catalogue used in the ID analysis of Smith et
  al. (2011), containing all sources which passed the star-galaxy
  separation at $r$-modelmag $<22.4$ in the primary SDSS DR7 catalogue
  in a region of $\sim 35$ degrees centered on the SDP field. We
  fitted a linear slope to the logarithmic number counts in the range
  $r=19-21.5$ and extrapolated this to fainter magnitudes. We then
  used the difference between observed and expected number counts to
  estimate completeness. The results are presented in
  Table~\ref{tab:r_comp} and show that completeness is above 80
  percent to $r=21.8$, falling to 50 percent by $r=22.2$. By
  restricting our analysis to $z<0.5$ we keep 97 percent of the
  sources below $r\sim 22$ and so in the range of acceptable
  completeness. It is possible, in principle, for there to be some
  form of optical incompleteness in the sample which is not corrected
  for with the above prescription, e.g. a population of objects which
  begin to appear at high redshifts in the H-ATLAS sample but which
  are not well represented in SDSS. Such a population could
  conceivably consist of very obscured star-bursts. To test our
  susceptibility to this, we estimate the SDSS $r$ magnitude of a
  highly obscured galaxy with an SED like that of Arp 220 ($A_v = 15$)
  at our 250\mic flux limit at the redshift limit of $z=0.5$ and find
  that it would still be detected in our
  sample. Figure~\ref{fig:templates} shows three different SED
  templates normalised to $S_{250}=32$ mJy at $z=0.5$: M82, an H-ATLAS
  based template appropriate for sources at $z=0.5$ from Smith et
  al. in prep, and Arp 220. All templates less obscured than Arp 220 are
  easily visible at our optical flux limit. We will therefore proceed
  on the assumption that no such new populations exist below the
  optical limit in our highest $z$ bins.}

Figure~\ref{fig:rcomps}a plots $r$-mag as a function of 250\mic
flux. A galaxy with $S_{250}$ below $\sim 100$mJy can have a
wide range of optical magnitude ($r$-mag = 16.5-22.0), and while
optical magnitude is a strong function of redshift this is not the
case for the sub-mm flux. Figure~\ref{fig:rcomps}b shows $r$-mag as a
function of redshift for all galaxies in the GAMA 9hr (Driver et
al. 2011) spectroscopic sample (cyan), as well as the reliable SPIRE
IDs (black). This shows a lack of {\em Herschel} sources at the
fainter magnitudes at low redshifts (i.e. the lowest absolute
magnitudes or stellar masses).\footnote{The limit of GAMA is $r\sim
  19$ which is brighter than the SDSS limit used for H-ATLAS IDs
  ($r\sim 22.4$).}  It appears that H-ATLAS is less sensitive to low
stellar mass galaxies than the SDSS (due to them having lower dust
masses) and so only at high-z does the $r$-band limit preclude the
identification of {\em Herschel} sources.

\end{enumerate}

\begin{table}
  \caption{The percentage completeness as a function of $r$ magnitude
    for the catalogue used to make the identifications to
     H-ATLAS sources.  The correction factor used in the luminosity function is denoted by $C_r$.}  \centering
  \begin{tabular}{|c|c|c|}
    \hline
    $r$ mag & Completeness (\%) & $C_r$ \\
    \hline
    21.6   & 91.1 & 1.10\\
    21.7   & 87.6 & 1.14\\
    21.8   & 82.8 & 1.21\\
    21.9   & 77.7 & 1.29\\
    22.0   & 70.5 & 1.42\\
    22.1  & 61.6 & 1.62\\
    22.2  & 52.5 & 1.90\\
    22.3  & 42.8 & 2.33\\
    22.4  & 17.0 & 5.88\\
    \hline
    \label{tab:r_comp}
  \end{tabular}
\end{table}

\section{Dust mass and Luminosity}
\label{masses}

The {\em Herschel} fluxes are translated into monochromatic
rest-frame 250\mic\ luminosities following

\begin{equation}
\Lsub = 4 \pi D^2\, (1+z)\, S_{250}\, K
\end{equation}
where \Lsub is in $\rm{W Hz^{-1}}$, $D$ is the co-moving distance,
$S_{250}$ is the observed flux density at 250\mic and $K$ is the
K-correction which is given by:

\begin{equation}
K = \left(\frac{\nu_{\rm{obs}}}{\nu_{\rm obs(1+z)}}\right)^{3+\beta} \frac{e^{(h\nu_{\rm obs(1+z)}/kT_{\rm{iso}})}-1}{e^{(h\nu_{\rm{obs}}/kT_{\rm iso})}-1}
\end{equation}

\noindent where $\nu_{\rm{obs}}$ is the observed frequency at 250\mic,
$\nu_{\rm{obs(1+z)}}$ is the rest-frame frequency and $\rm{T_{iso}}$
and $\beta$ are the temperature and emissivity index describing the
global SED shape.

In order to derive the values required for the K-correction, a simple
grey body SED of the form $S \propto \nu^{\beta}\,B(\nu, T)$ was
fitted to the PACS and SPIRE fluxes as described in Dye et al. (2010),
with a fixed dust emissivity index of $\beta=1.5$ and a temperature
range of 10--50 K. Where insufficient data points are available for
the fit (300/1867), the median temperature of 26 K from the galaxies
which could be fitted was used. With only SPIRE data on the
Rayleigh-Jeans side of the SED (as is the case for most sources), only
the combination of $\beta$ and $\rm{T_{iso}}$ is well constrained,
with the two parameters being inversely correlated by the fit; good
fits are obtained with $\beta = 1.5-2.0$. These simple grey body fits
can be performed for the majority of sources and are accurate at
representing the flux between rest-frame 250--166\mic (relevant for
our redshift range) and so are suitable for applying the K-correction.

A dust mass can also be calculated from the observed 250\mic flux
density and the grey body temperature as:

\begin{equation}
M_{\rm iso} = \frac{S_{250}\, D^2\, (1+z)\, K}{\kappa_{250}\,B(\nu_{250}, T_{\rm iso})}
\label{eqn:md}
\end{equation}

where $\kappa_{250}$ is the dust mass absorption coefficient which we
take to be equal to $0.89 \,\rm{m^2 \, kg^{-1}}$ at 250\mic
(equivalent to scaling $\kappa_{850} = 0.077 \,\rm{m^2 \, kg^{-1}}$,
as used by D00, James et al. 2002, da Cunha, Charlot \& Elbaz 2008
with a $\beta=2$). It also lies within the range of values found for
the diffuse ISM in the Milky Way and other nearby galaxies (Boulanger
et al. 1996; Sodroski et al. 1997; Bianchi et al. 1999; Planck
Collaboration 2011a). The dust mass via
Eq.~\ref{eqn:md} scales as $M_d \propto T^{-2.4}$ at $z\sim 0$ for
temperatures around 20\,K; changing the temperature from 20--30 K
results in a reduction in mass by a factor 2.6. At $z=0.5$ this
dependence is steeper since the peak of the dust emission is shifted
to longer wavelengths so the observed frame is even further from the
Rayleigh-Jeans regime. Changing from $\beta=1.5$ to 2.0 reduces the
temperatures by $\sim 3$K and increases the dust masses by
$\sim 30-50$ percent.

\begin{figure*}
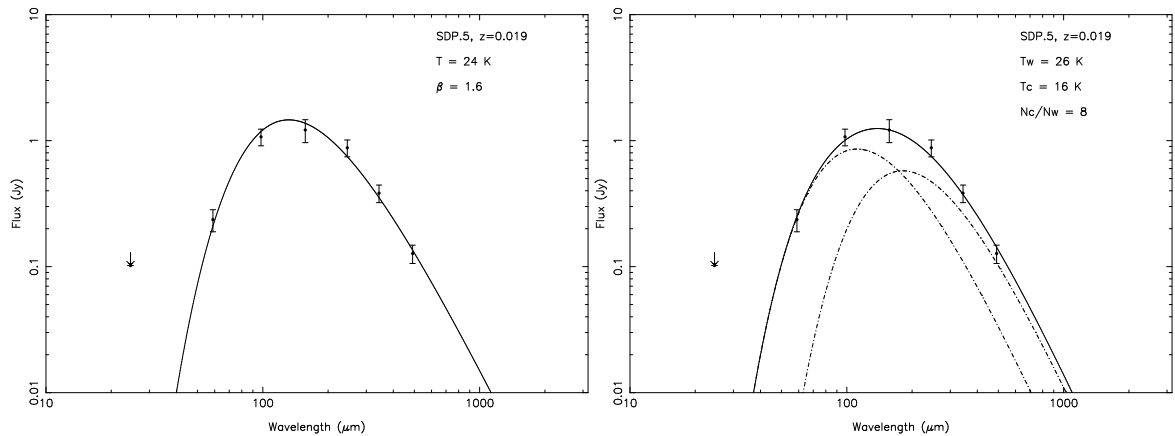

\centering
\subfigure{\includegraphics[width=0.32\textwidth,angle=-90]{sdp5_iso.ps}}
\subfigure{\includegraphics[width=0.32\textwidth,angle=-90]{sdp5.ps}}\\
\caption{\label{fig:SEDsimple} {\bf Top row:} Isothermal and
  2-component SEDs for an H-ATLAS sources with a well sampled
  SED. Redshifts and fitted parameters are shown in each panel. For
  the isothermal fits $T$ and $\beta$ were free to vary while for the
  2-component fits $\beta$ was fixed to be 2. The parameter $N_c/N_w$
  is the ratio of cold/warm mass. }

\end{figure*}

This isothermal dust mass estimate can be biased low as it is now well
established that dust exists at a range of temperatures in
galaxies. Only dust in close proximity to sources of heating
(e.g. star forming regions) will be warm enough to emit at $\lambda
\leq 100$\mic\, but this small fraction of dust (by mass) can strongly
influence the temperature of the isothermal fits. The bulk of the ISM
(and therefore the dust) resides in the diffuse phase which is heated
by the interstellar radiation field to a cooler temperature typically
in the range of 15-20 K (Helou et al. 1986; DE01 and references
within; Popescu et al. 2002; VDE05; Draine et al. 2007; Willmer et
al. 2009; Bendo et al. 2010; Boselli et al. 2010; Kramer et al. 2010;
Bernard et al. 2010; Planck Collaboration 2011b). For more accurate
dust mass estimates we require the mass-weighted temperature of the
dust emitting at 250\mic\, which requires fitting a model with
multiple (at least two) temperature components. This is not to say
that the FIR fluxes for most of the H-ATLAS galaxies are not fitted
adequately by the single temperature model; an isothermal model and a
more realistic multi-temperature model are often degenerate in their
ability to describe the SED shape with a limited number of data
points. To illustrate this, we show in Fig~\ref{fig:SEDsimple} an
example of isothermal and 2-component SED fits to a H-ATLAS source
with a well sampled SED. Although the 2-component fit is formally
better, there is nothing to choose between them as descriptions of the
fluxes of the H-ATLAS source between 60--500\mic. DE01 studied this
issue for a sample of SLUGS galaxies with 450\mic detections, and
concluded that the best overall description of that sample was a
two-temperature model with $\beta=2$ and a cold component temperature
of $\sim 20$K.

To deal with the cold dust component, we now introduce a more
sophisticated SED model which includes dust in several physically
motivated components, following the prescription of Charlot \& Fall
(2000). The results of this fitting are presented and described in
detail in Smith et al. in prep, and outlined here in brief. This
simple, but empirically motivated, SED model fits broadband photometry
from the UV--sub-mm to estimate a wide variety of parameters (da Cunha
et al. 2008 - hereafter DCE08; da Cunha et al. 2010a). The method uses
libraries of optical and infrared models (25,000 optical and 50,000
infrared) and fits those optical-IR combinations which satisfy an
energy balance criteria to the data. The optical libraries have
stochastic star formation histories and the stellar outputs are
computed using the latest version of the Bruzual \& Charlot (2003)
population synthesis code (Charlot \& Bruzual in prep) libraries and a
Chabrier (2003) Galactic-disc Initial Mass Function (IMF). 

The dust mass in this model is computed from the sum of the masses in
various temperature components contributing to the SED, including cool
dust in the diffuse ISM, warm dust in birth clouds, hot dust
(transiently heated small grains emitting in the mid-IR) and PAHs. In
the fits to H-ATLAS sources (and SINGS galaxies; DCE08) around 90
percent of the dust mass is in the cold diffuse ISM component and this
is also the best constrained component due to the better sampling of
the FIR and sub-mm part of the SED with {\em Herschel}. Many other
studies also find that the cold dust component dominates the overall
mass, and so it is the most important one to constrain when measuring
the dust mass function (e.g. DE01; VDE05; Draine et al. 2007; Willmer
2009; Liu et al. 2010). The priors used by DCE08 for the temperatures
of the grains in equilibrium are 30--60 K for the warm component and
15--25 K for the cold component. These values agree well with
temperatures measured for local galaxies (Braine et al. 1997; Alton et
al. 1998; Hippelein et al. 2003; Popescu et al. 2002; Meijerink et
al. 2005, DE01, V05, Stevens et al. 2005, Stickel et al. 2007; Draine
et al. 2007; Willmer et al. 2009; Planck Collaboration 2011b) and also
with temperatures measured from stacking optically selected galaxies
with the same stellar mass and redshift range as our sample into {\em
  Herschel}-ATLAS maps (Bourne et al. in prep). The value of $\kappa$
used in the DCE08 model is (by design) comparable with that used in
the isothermal fits here.

The prior space of the parameters is sampled by fitting to several
million optical--FIR model combinations and returns a probability
density function (PDF) for the dust mass and other parameters
(e.g. dust temperature, stellar mass, dust luminosity, optical depth
and star formation rate) from which the median and 68 percent
confidence percentiles are taken as the estimate of the quantity and
its error.

\begin{figure}
\centering
\includegraphics[width=8.6cm]{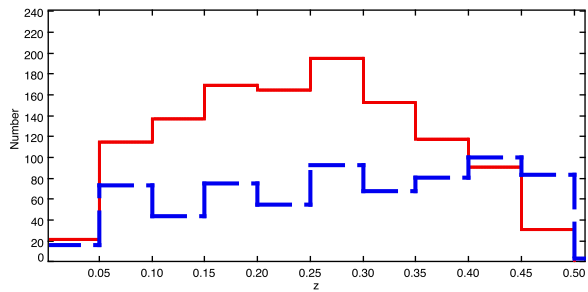}
\caption{The distribution of sources with DCE08 SED fits as a function
  of redshift (red), those without fits are shown in blue (dashed).}
\label{fig:SEDz}
\end{figure}

This model was fitted to the 60 percent of the galaxies in our sample
for which useful optical and NIR data were available from GAMA. We
fitted only to galaxies which have matched aperture photometry in
$r$-defined apertures as this best represents the total flux of the
galaxy in each band as described in Hill et al. (2011); Driver et
al. (2011). The distribution of sources with and without these SED fits as a
function of redshift is shown in Figure~\ref{fig:SEDz}. Those without
fits dominate only in the highest redshift bin, from $z=0.4-0.5$.

\begin{figure*}
\centering
\subfigure{\includegraphics[width=0.49\textwidth]{SP3_sed.ps}}
\subfigure{\includegraphics[width=0.23\textwidth]{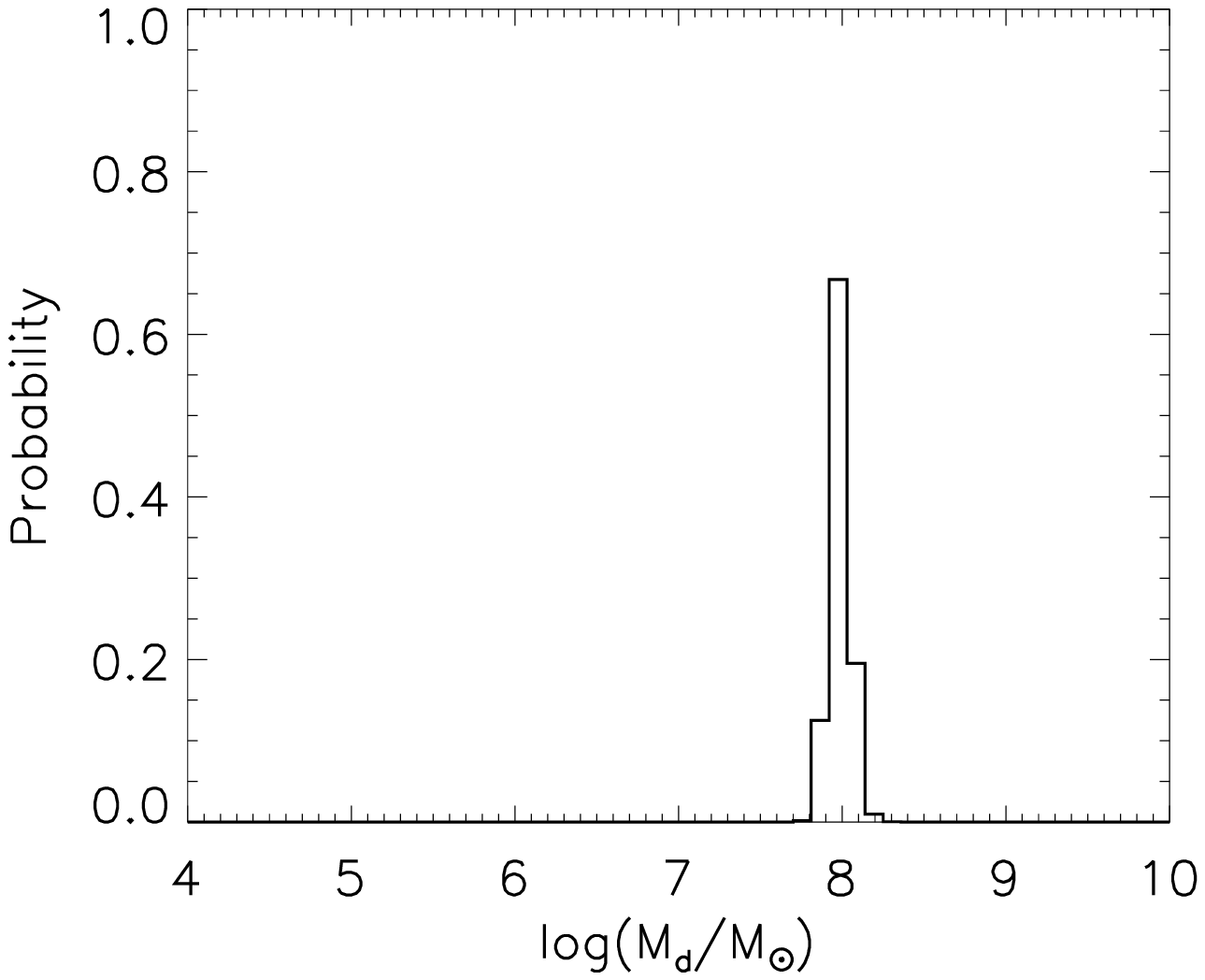}}
\subfigure{\includegraphics[width=0.23\textwidth]{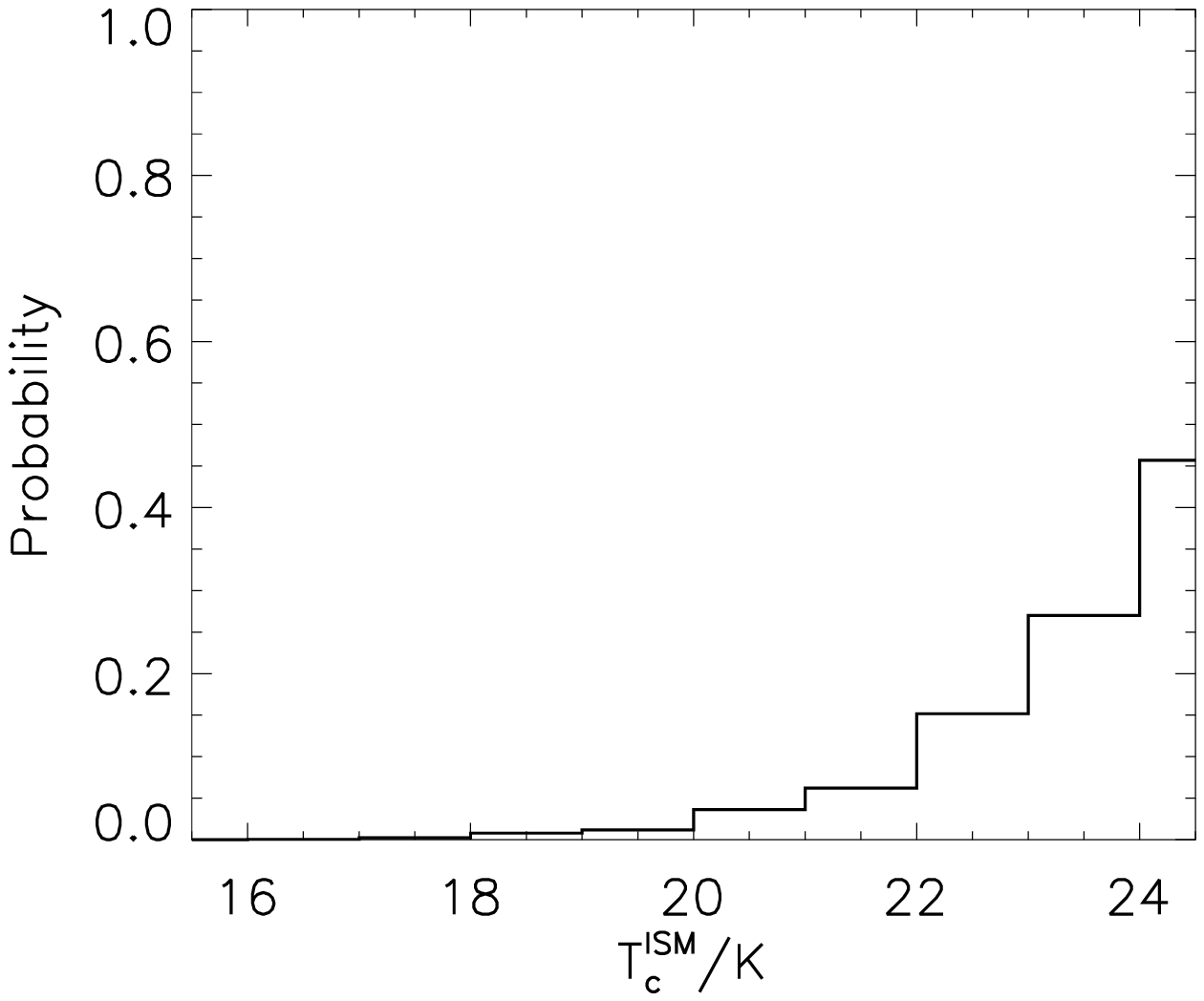}}
\subfigure{\includegraphics[width=0.49\textwidth]{SP16_sed.ps}}
\subfigure{\includegraphics[width=0.23\textwidth]{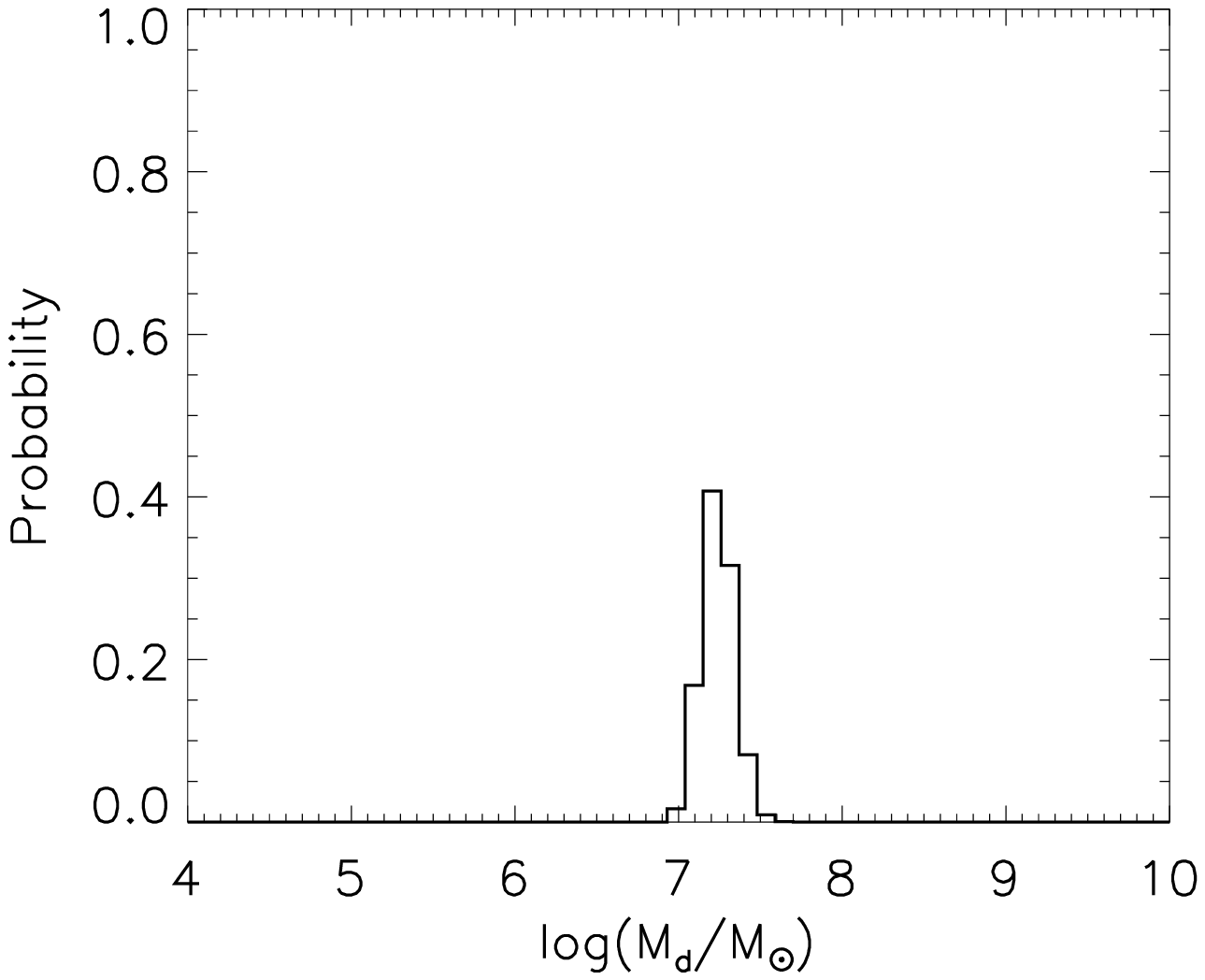}}
\subfigure{\includegraphics[width=0.23\textwidth]{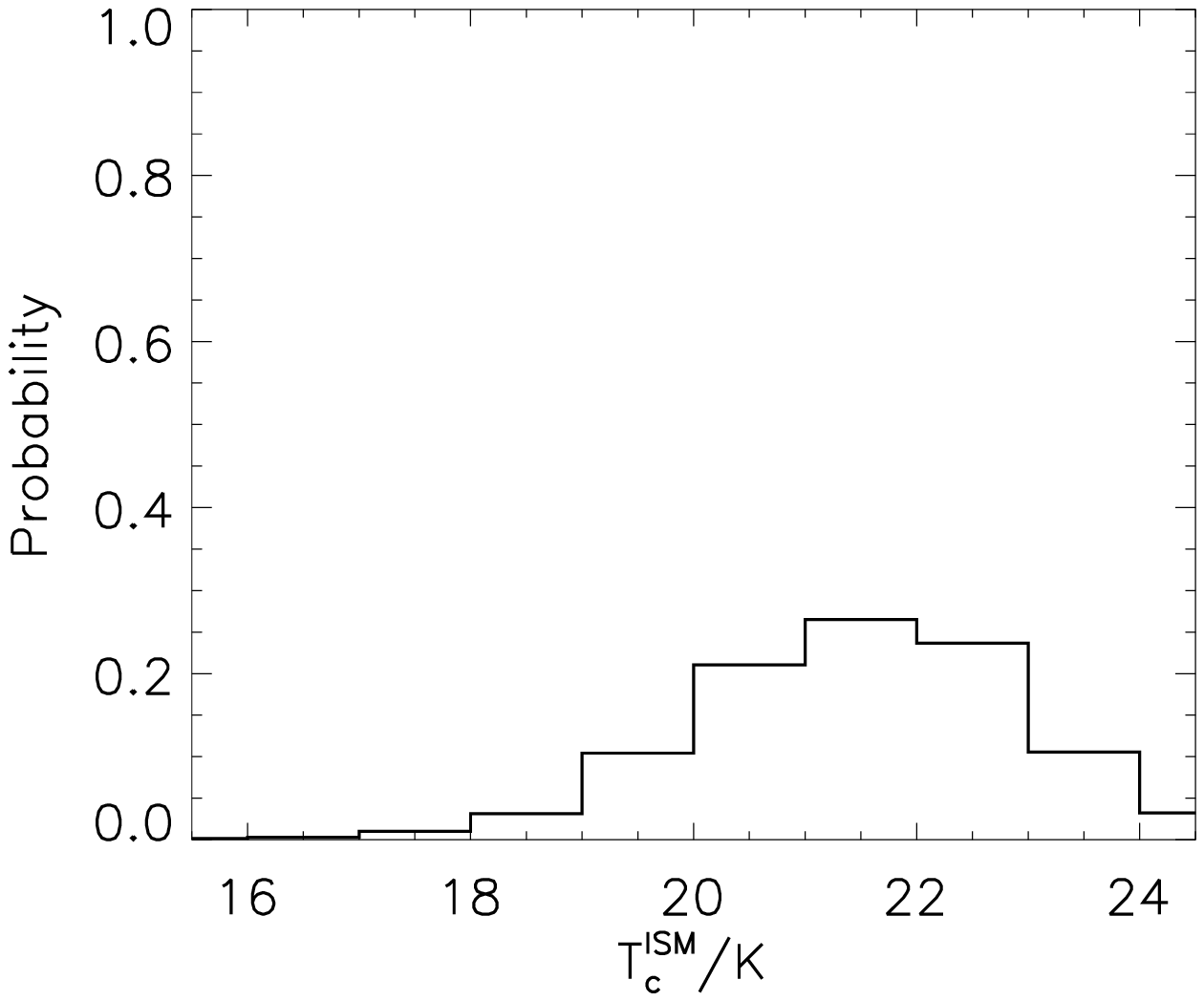}}
\subfigure{\includegraphics[width=0.49\textwidth]{SP457_sed.ps}}
\subfigure{\includegraphics[width=0.23\textwidth]{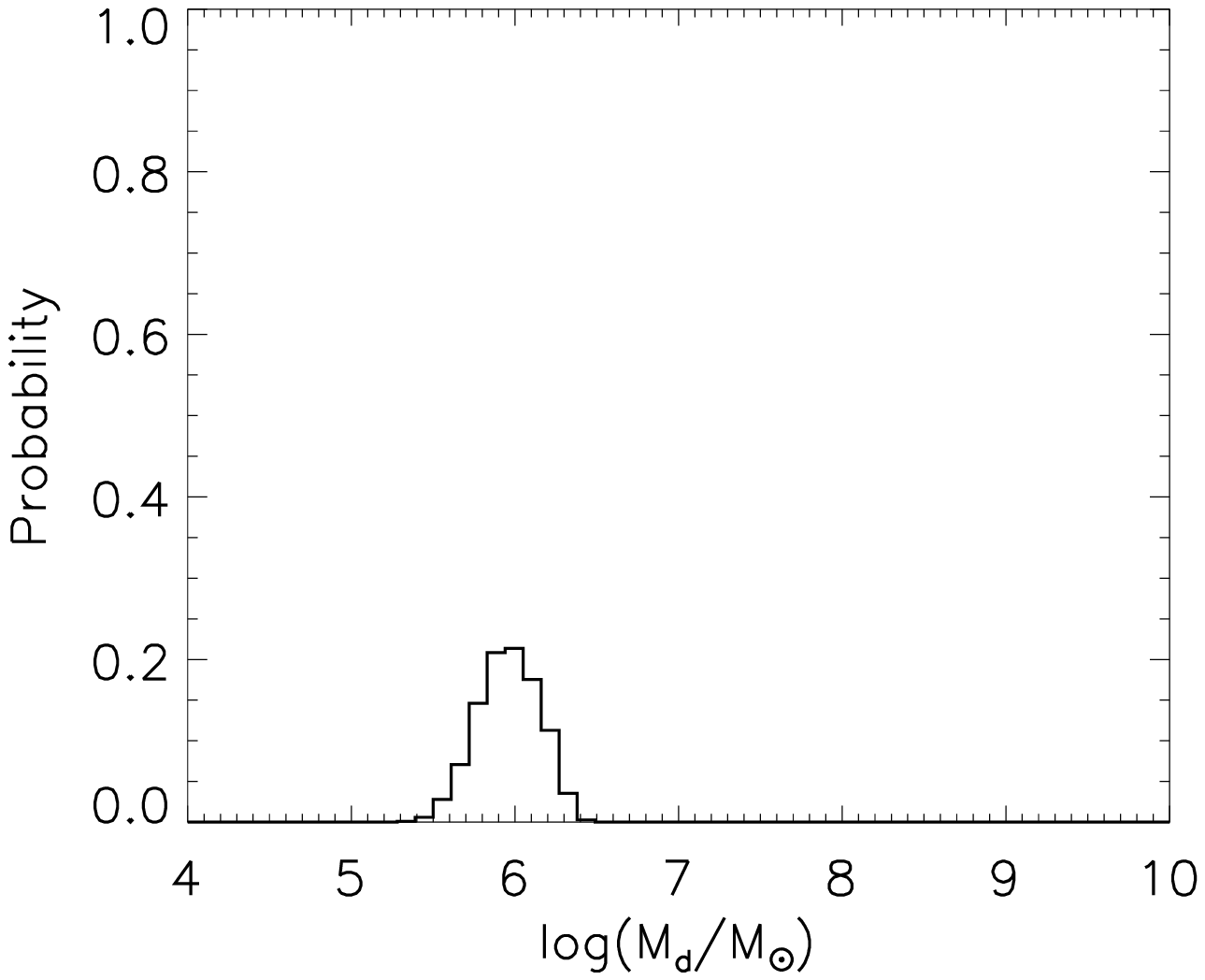}}
\subfigure{\includegraphics[width=0.23\textwidth]{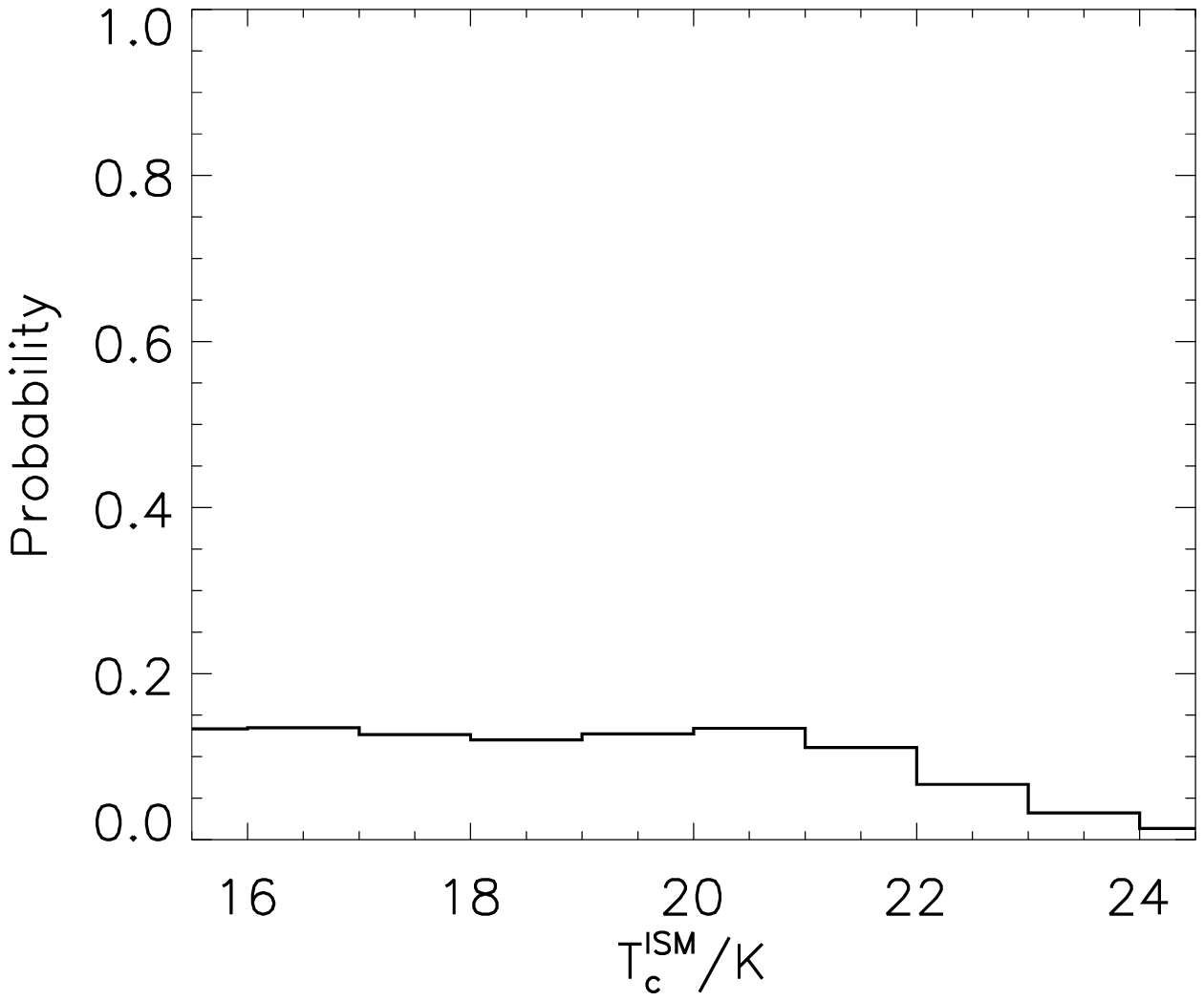}}
\subfigure{\includegraphics[width=0.49\textwidth]{SP5796_sed.ps}}
\subfigure{\includegraphics[width=0.23\textwidth]{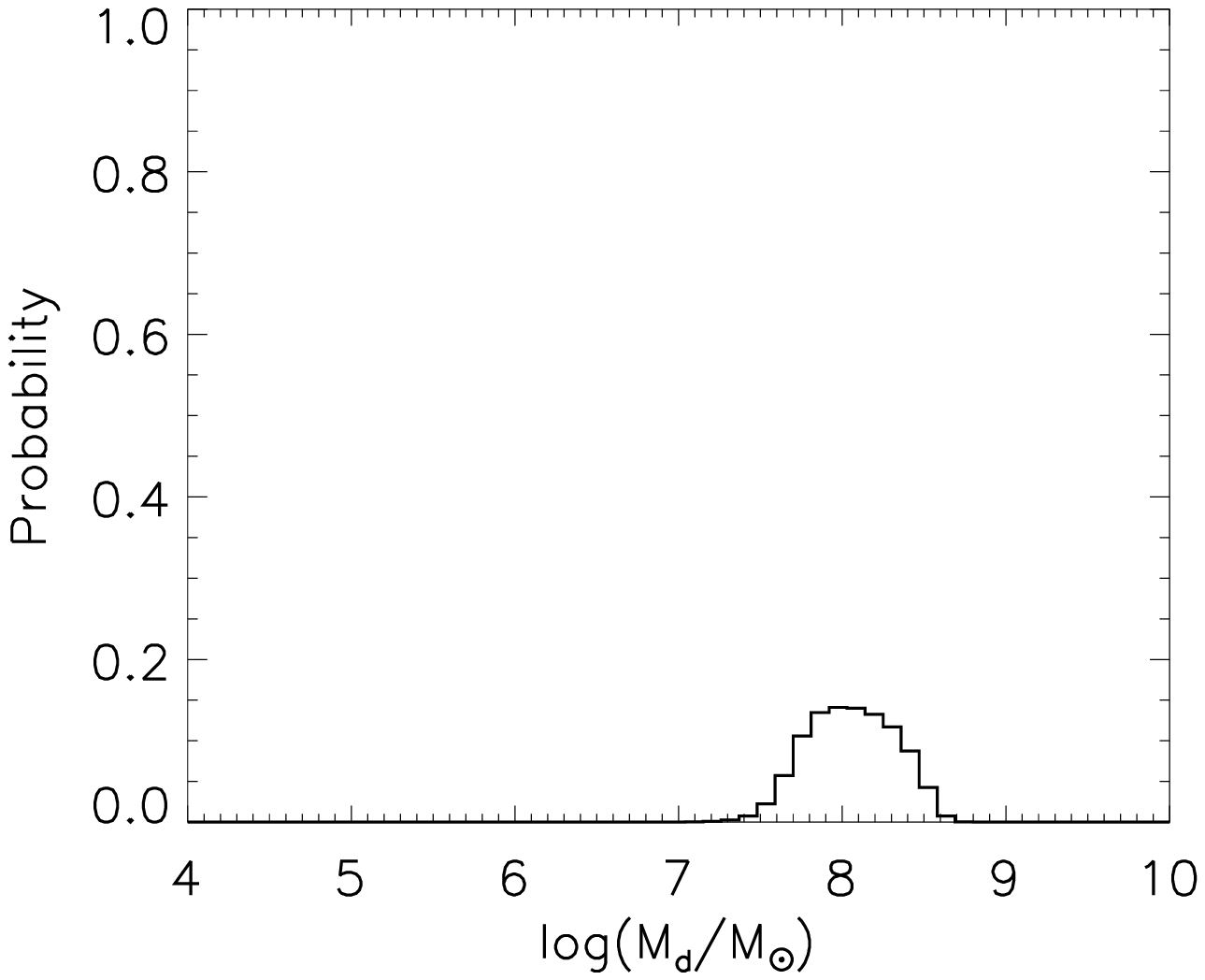}}
\subfigure{\includegraphics[width=0.23\textwidth]{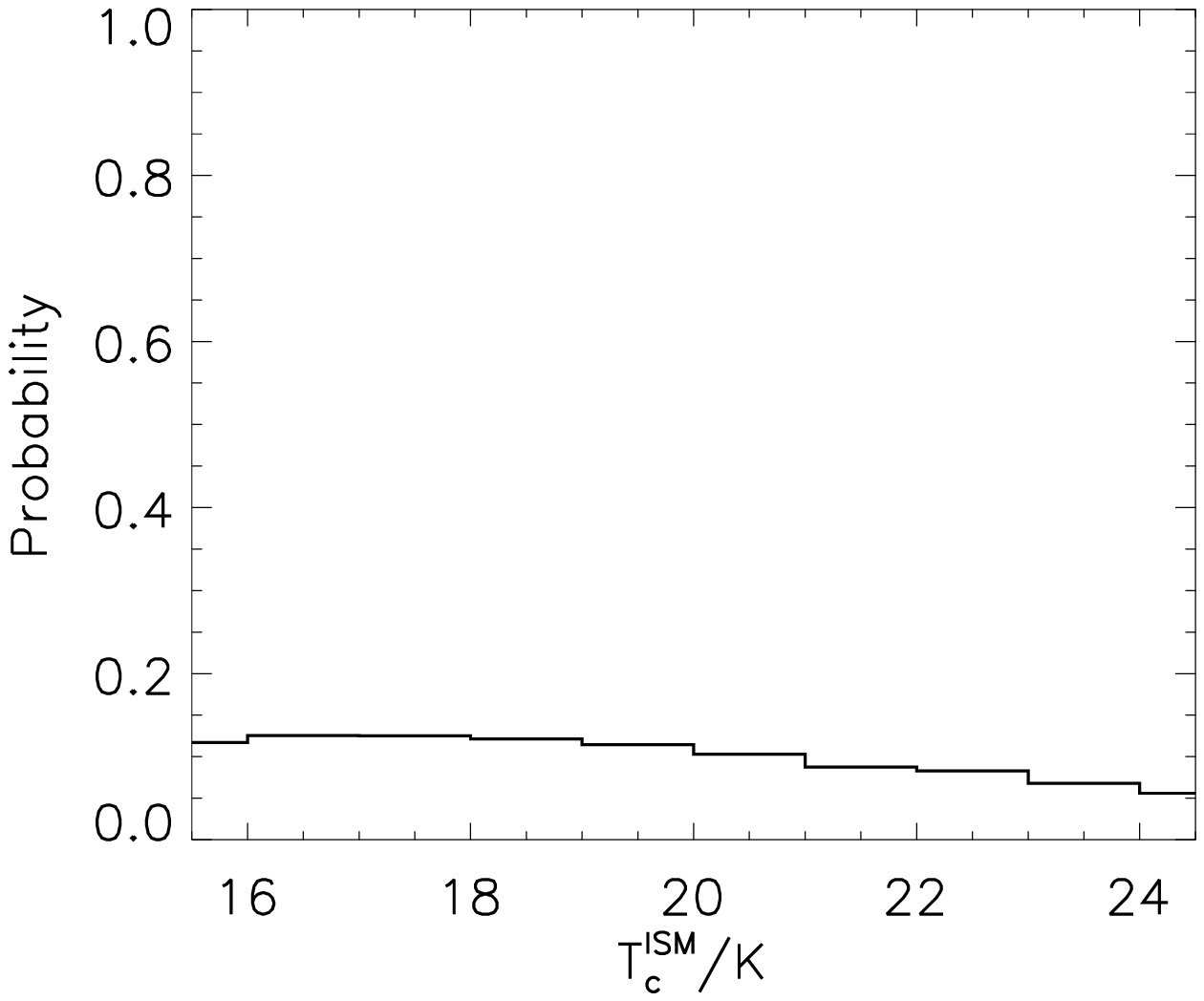}}
\caption{SED fits and probability distribution functions for dust mass
  and diffuse ISM dust temperature for a range of H-ATLAS sources. The
  black curve on the SED plot is the total attenuated starlight and
  re-radiated dust emission. Blue curve is the unattenuated
  starlight. Green is the attenuated starlight and red is the dust
  emission. The red squares show the observed photometry and errors or
  upper limits. The limit to the dust mass accuracy is our ability to
  determine the cold temperature, which is better constrained when
  there are more FIR data points available. The best constraints on
  the dust mass are $\sim 0.05$ dex and the worst are $\sim 0.27$
  dex.}
\label{fig:fits}
\end{figure*}

The errors on the dust mass range from $\pm 0.05-0.27$ dex and this
error budget includes all uncertainties in the fitting from flux
errors to changes in temperature and contribution of the various dust
components. Some typical SED fits and PDFs for the dust mass and cold
temperature parameters are shown in Figure~\ref{fig:fits}. The dust
mass is generally a well constrained parameter of these model fits;
the PDF is narrower when more IR wavelengths are available and so the
cold temperature is then better constrained.


\begin{figure}
\centering
\subfigure{\includegraphics[width=0.48\textwidth]{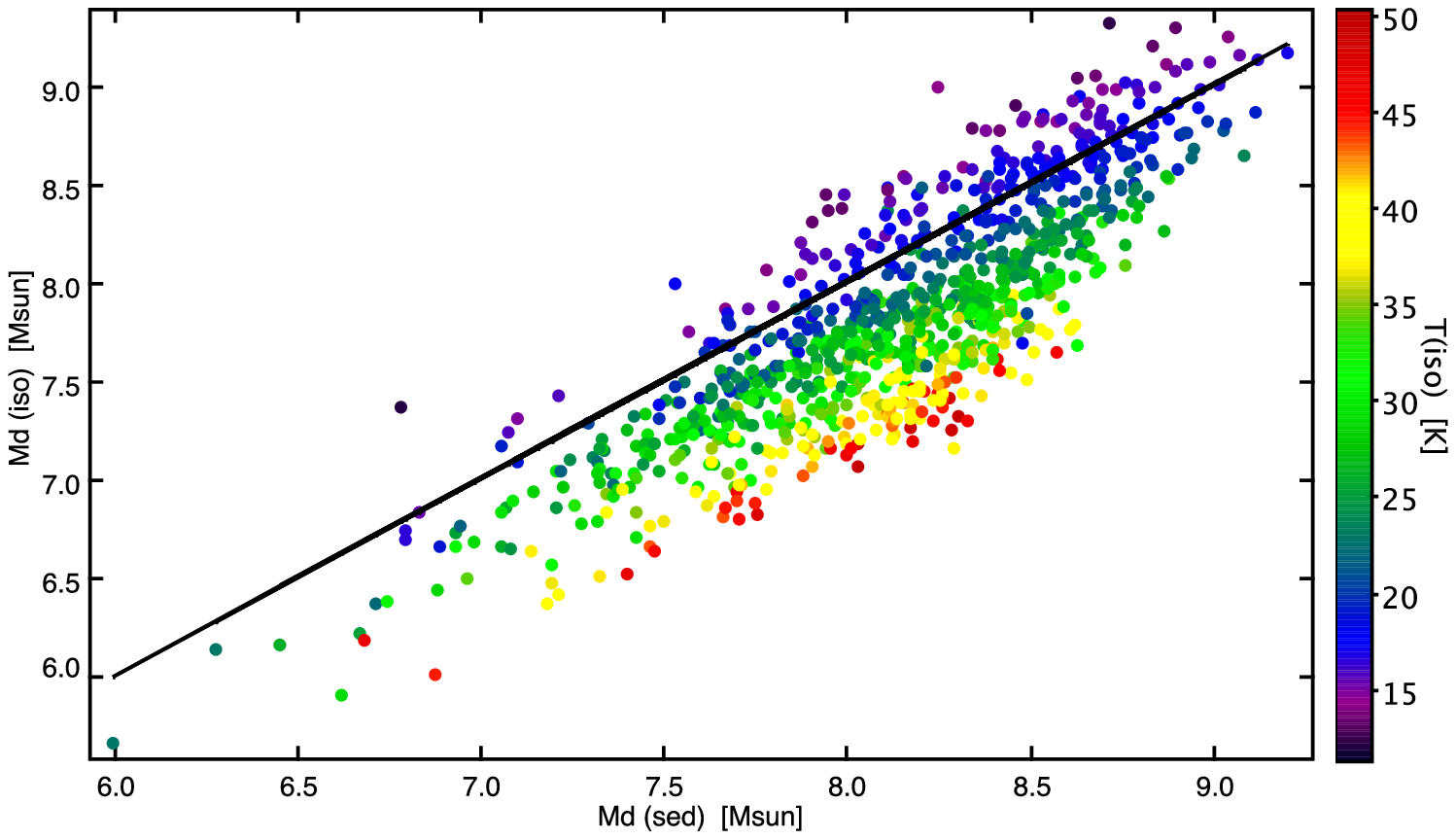}}\\
\subfigure{\includegraphics[width=0.48\textwidth]{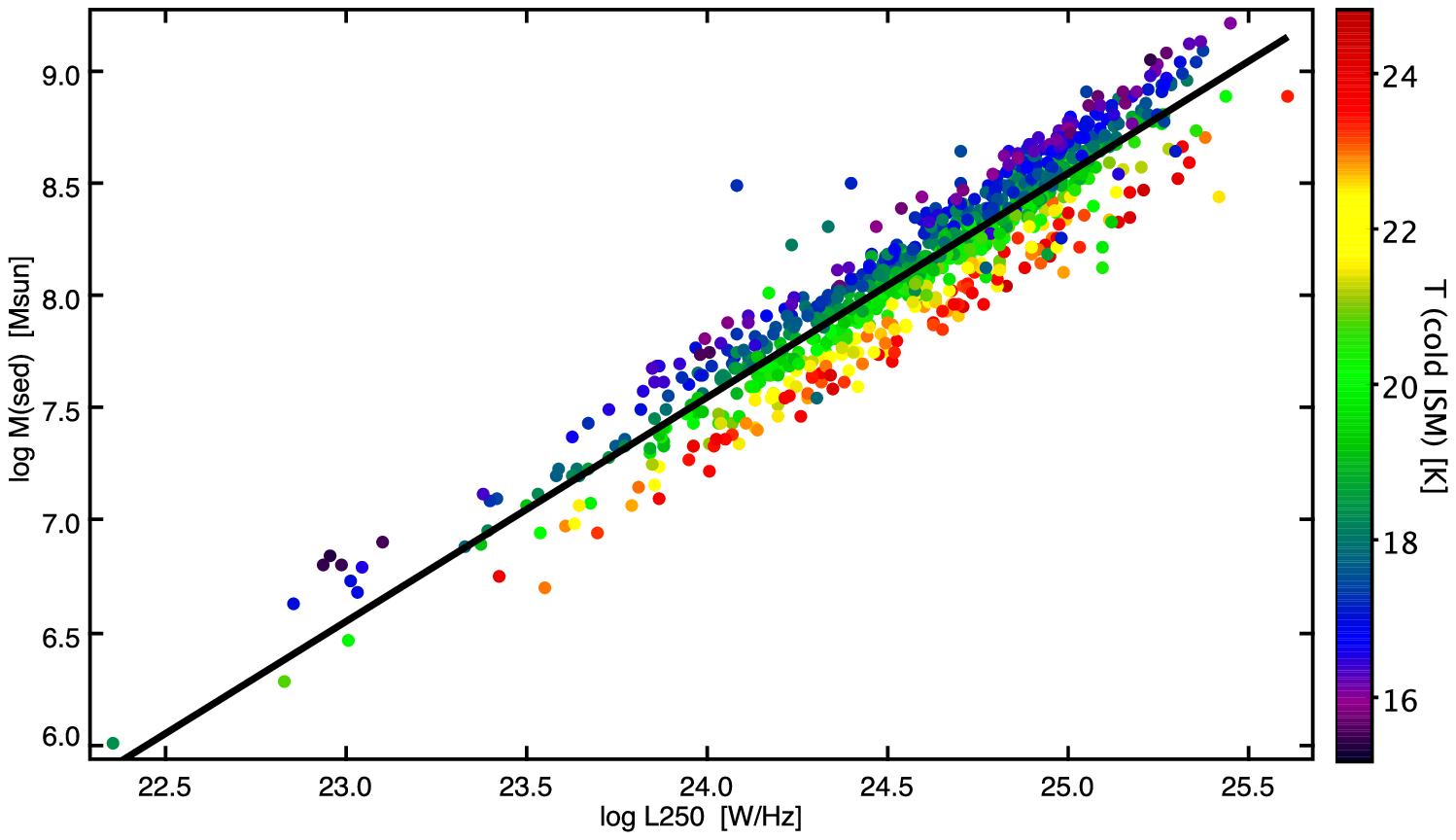}}\\
\caption{{\bf Top:} Comparison of dust masses from the DCE08 model
  (\Msed) versus the dust mass obtained by using the isothermal
  grey body fit (\Miso) using Eq.~\ref{eqn:md}. Points are
  colour-coded by the isothermal temperature. The one-to-one line is
  shown in black. There is a large difference between the two mass
  estimates which is a strong function of fitted isothermal
  temperature. {\bf Bottom:} Comparison of \Msed to \Lsub showing a reasonably
  tight and linear correlation. The best fit relationship (Eq.~\ref{eqn:lsubMd}) is
  over-plotted. The scatter in this relationship is driven by the diffuse
  ISM dust temperature, which is used to colour-code the points.}
\label{fig:mdcomp}
\end{figure}

A comparison of the isothermal dust masses (\Miso) and the full SED
based masses (\Msed) is shown in Figure~\ref{fig:mdcomp}a and there is
generally poor agreement between the two, with the scatter of a factor
2--3 related to the difference between the temperature of the
isothermal fit and that of the DCE08 model fit. This sensitivity is
because at 250\mic we are near to the peak of the black body function
for the cold temperatures appropriate to the bulk of the dust mass
(15--20\,K). At longer sub-mm wavelengths (such as 850\mic), this
temperature sensitivity is less severe, but the choice of dust
temperature used when estimating masses at rest wavelengths close to
those of {\em Herschel} is clearly important.

For sources which have insufficient UV-sub-mm data to use the DCE08
model, we need to extrapolate dust masses by comparing \Lsub with
\Msed for those sources which do have fits. The relationship is
linear, with some scatter introduced by the range in dust temperature
for the cold ISM component (Figure~\ref{fig:mdcomp}b):

\begin{equation}
\log \Msed = \log \Lsub - 16.47
\label{eqn:lsubMd}
\end{equation}

\noindent Equation~\ref{eqn:lsubMd} is used to convert \Lsub to dust
mass in cases where the full SED could not be fitted (747 sources out
of 1867). The relationship between \Msed and the cold temperature of
the diffuse ISM (which dominates the dust mass in these galaxies) is
similar to that in Eqn.~\ref{eqn:md}, since the DCE08 model fits the
sum of grey-bodies at different temperatures to the photometry. The
colder the temperature fitted, the higher the dust mass will be for a
given \Lsub. This is clearly demonstrated in Fig.~\ref{fig:mdcomp}
(bottom). In using this relationship for sources without SED fits we
are making the assumption that they will also fall on this
relationship and that there is no systematic trend in those sources
without fits (e.g. if only the highest redshift or most/least luminous
sources did not have fits). Since the galaxies without fits span a
range of redshift and luminosity and as we also find no correlation of
temperature with \Lsub for our sample (see Section~\ref{sec:temp} and
Figure~\ref{fig:lumtemp}), there should be no bias introduced in using
the relationship in Figure~\ref{fig:mdcomp}b to estimate masses for
those sources without fits.

The scatter in this figure is influenced by our choice of prior for
the temperature of the diffuse dust component (15--25 K). Allowing a
wider prior will broaden the scatter if a significant number of
sources are best fitted by hotter or colder temperatures. This issue
is explored further by Smith et al. in prep who conclude that for a
sample of galaxies with well constrained cold temperatures this prior
encompasses >80 percent of the population. Further study of the
temperatures of the populations requires a larger sample with good 5
band FIR/sub-mm photometry, which will be possible with the next Phase
of H-ATLAS data comprising 10 times the area of SDP. The difficulty in
using a wider prior is that when we lack full coverage of the
FIR/sub-mm SED (as is the case where we have only limits at PACS
wavelengths and 500\mic) there is only a weak constraint on the cold
temperature ($\sigma T_c \sim 2.5-3$ K). This can place a galaxy with
a real temperature of 15 K down at 12 K, and produces quite a bias in
the fitted dust temperature (since at 12 K the mass is very sensitive
to temperature). The model will fit arbitrarily high masses of very
cold dust since this contributes very little to the overall energy
balance. Our choice to restrict the temperature prior to the parameter
space which is preferred by observations of nearby galaxies, and by
those galaxies well sampled in H-ATLAS, potentially means we
underestimate the masses of some cooler sources, but we would prefer
to be conservative at this point.

\section{The dust mass function}
\label{DMF}

\subsection{Estimators}
\begin{figure*}
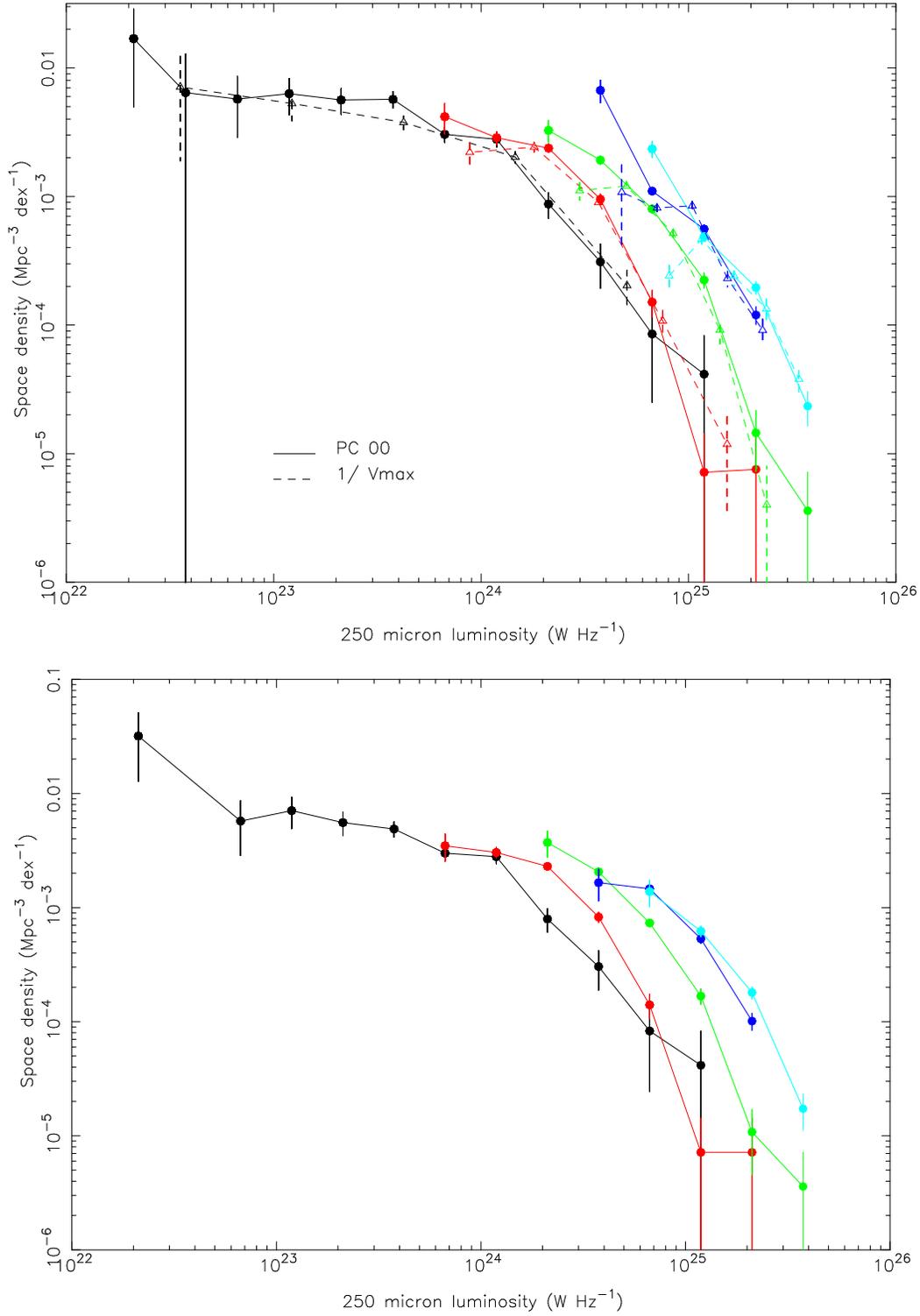

\centering
\subfigure{\includegraphics[width=0.55\textwidth, angle=-90]{250lf_va_cp.ps}}\\
\subfigure{\includegraphics[width=0.55\textwidth, angle=-90]{lf250_cpnew.ps}}
\caption{{\bf Top:} 250\mic luminosity functions calculated via the
  $\rm 1/V_{max}$ method (open triangles /dashed) and PC00 method
  (solid circles and lines) in five redshift slices of $\Delta z=0.1$
  out to $z=0.5$. Colours denote the redshifts as: black ($0<z<0.1$),
  red ($0.1<z<0.2$), green ($0.2<z<0.3$), blue ($0.3<z<0.4$) and cyan
  ($0.4<z<0.5$). The bias in $\rm 1/V_{max}$ in the lowest luminosity
  bin in each redshift slice is apparent from the turn-down in this
  bin. {\bf Bottom:} 250\mic luminosity function calculated using the
  modified PC00 estimator which includes an individual K-correction
  for each object in an $L-z$ bin. Using individual K-corrections has
  a more significant effect in the highest redshift slices as
  expected.}
\label{fig:lfcomp}
\end{figure*}

To calculate the dust mass function we use the method of Page \&
Carrera (2000; hereafter PC00) who describe a method to estimate
binned luminosity functions that is less biased than the $\rm
1/V_{max}$ method (Schmidt 1968). To begin with, we produce
measurements of the 250\mic luminosity function since this is more
directly related to the flux measurements from {\em Herschel} and
enables us to discuss the method without the added complication of
translating 250\mic luminosity into dust mass. The PC00 estimator is
given by :

\begin{equation}
\phi = \frac{\sum_{i=1}^{N} C_s\, C_z\, C_r}{\int^{L_{\rm max}}_{L_{\rm min}}\int^{z_{{\rm max}(L)}}_{z_{\rm min}} \frac{dV}{dz} \,dz \,dL}
\end{equation}

where $C_s,\,C_z\,C_r$ are the completeness corrections for each
object as described in Section~\ref{sec:comp} and the sum is over all
galaxies in a given slice of redshift and luminosity bin. $L_{\rm
  max}$ and $L_{\rm min}$ are the maximum and minimum luminosities of
the bin. $z_{\rm min}$ is the minimum redshift of the slice and
$z_{{\rm max}(L)}$ is the maximum redshift to which an object with
luminosity, $L$, can be observed given the flux limit and
K-correction, or the redshift slice maximum, whichever is the
smaller. The PC00 method has the advantage of properly calculating the
available volume for each $L-z$ bin and, in particular, it does
not overestimate the volume for objects near to the flux limit. This
prevents the artificial turn-down produced by $\rm 1/V_{max}$ in the
first luminosity bin of each redshift slice.  We compare to the $\rm
1/V_{max}$ estimator in Figure~\ref{fig:lfcomp} and confirm that the
$\rm 1/V_{max}$ estimate of the 250\mic\ luminosity function suffers
from the bias noted by PC00 due to slicing in redshift bins.

In the PC00 formalism described above, the accessible volume is not
calculated individually for each source (as for $\rm 1/V_{max}$) but
is instead calculated for each bin in the $L-z$ plane using a global
K-correction. However, we know that each object in our sample has a
different K-correction because they have different grey body SED
fits. We therefore modified the estimator such that the accessible
volume for a given $L-z$ bin is calculated for each galaxy in that bin
in turn using its grey body SED fit to generate its limiting redshift
$z_{{\rm max},i} = z(L_i,S_{\rm min},T_d)$ across the bin. These
individual contributions are then summed within the bin such that:

\begin{equation}
\phi = \sum^N_{i=1} \frac{C_s\,C_z\,C_r}{\int^{L_{\rm max}}_{L_{\rm
      min}}\int^{z_{\rm max,i}}_{z_{\rm min}} \frac{dV}{dz} \,dz \,dL}
\label{eqn:pcmod}
\end{equation}

Note that this is not the same as reverting to the $\rm 1/V_{max}$
estimator as we are still calculating the volume available for each
$L-z$ bin, however we are now being more precise about the shape
of the limiting curve for each source based on its individual
SED. This is clear from the difference in the LF calculated this way,
as shown in Figure~\ref{fig:lfcomp}(b) compared to the PC00 and
$\rm 1/V_{max}$ methods shown in Figure~\ref{fig:lfcomp}(a). This change
affected the highest redshift bins most as expected.

In this case, the error on the space density is given by 

\begin{equation}
\sigma_{\phi} = \sqrt{\sum^N_{i=1} (\phi_i)^2}
\end{equation}

where $\phi_i$ is the individual $\phi$ contribution of a galaxy to a
particular redshift and luminosity bin, and the sum is over all
galaxies in that bin. The error bars in Figure~\ref{fig:lfcomp} show
these errors.

This 250\mic luminosity function differs slightly from that presented
in Dye et al. (2010) in that the ID sample has since been updated to
include extra redshifts (1867 compared to 1688) and also to remove
stars, for which there were 130 contaminating the previous
sample\footnote{Due to using an earlier version of the LR estimate
  which combined stars and galaxies together}. While Dye et al. (2010)
did attempt to correct for incompleteness in the optical IDs of the
sub-mm sample, we are now able to extend this to correct for
incompleteness as a function of redshift, $r$-mag and sub-mm flux
which was not previously possible. The results are, however,
comparable in that strong evolution in the 250\mic LF is evident out
to $z\sim 0.4$. There is then seemingly a halt, with little evolution
between $z=0.4$ and $z=0.5$. This is still consistent with Dye et
al. (2010) within the error bars of both estimators. 

\begin{figure}
\centering
\subfigure{\includegraphics[width=0.5\textwidth]{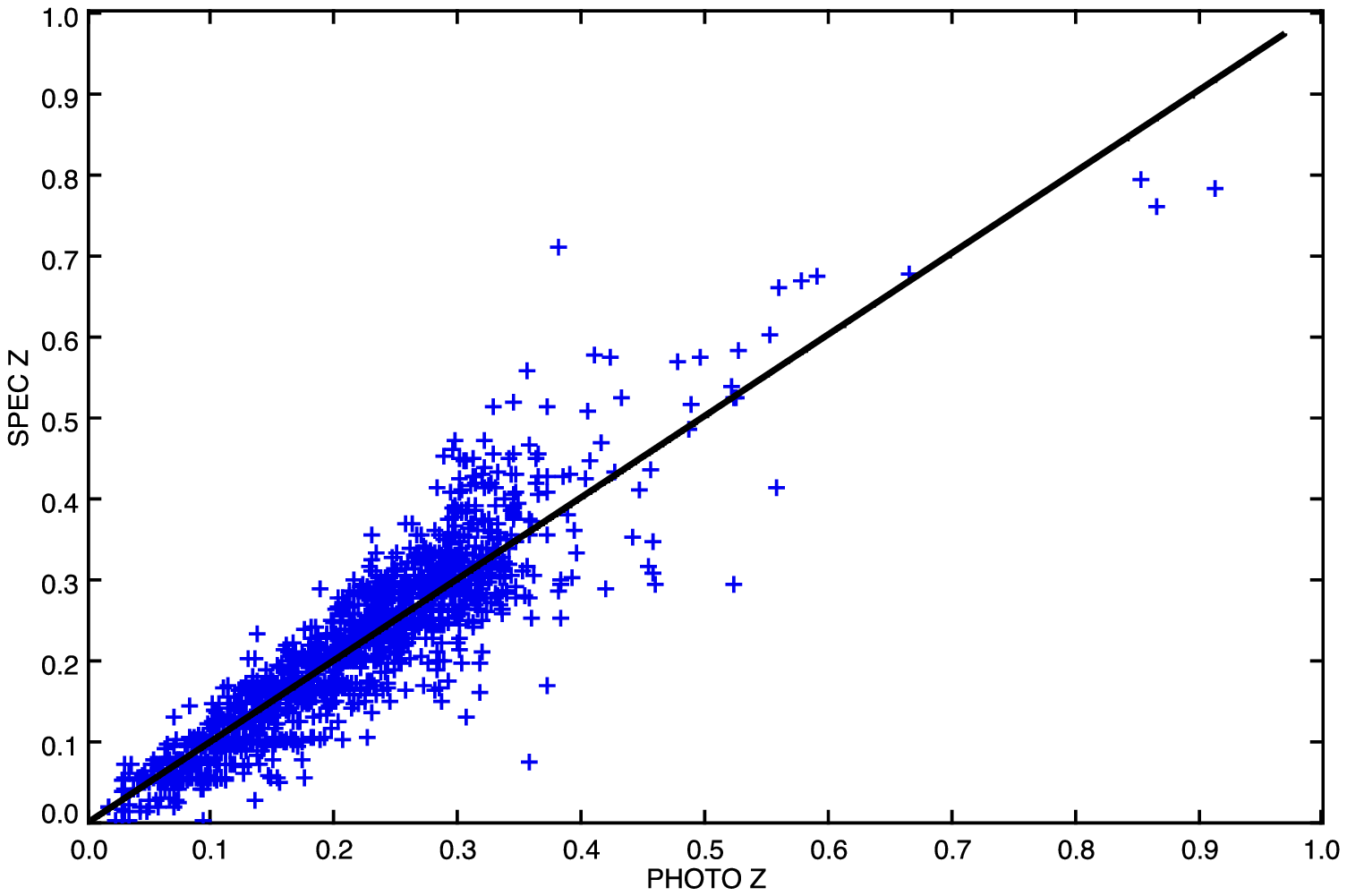}}\\
\subfigure{\includegraphics[width=0.35\textwidth, angle=-90]{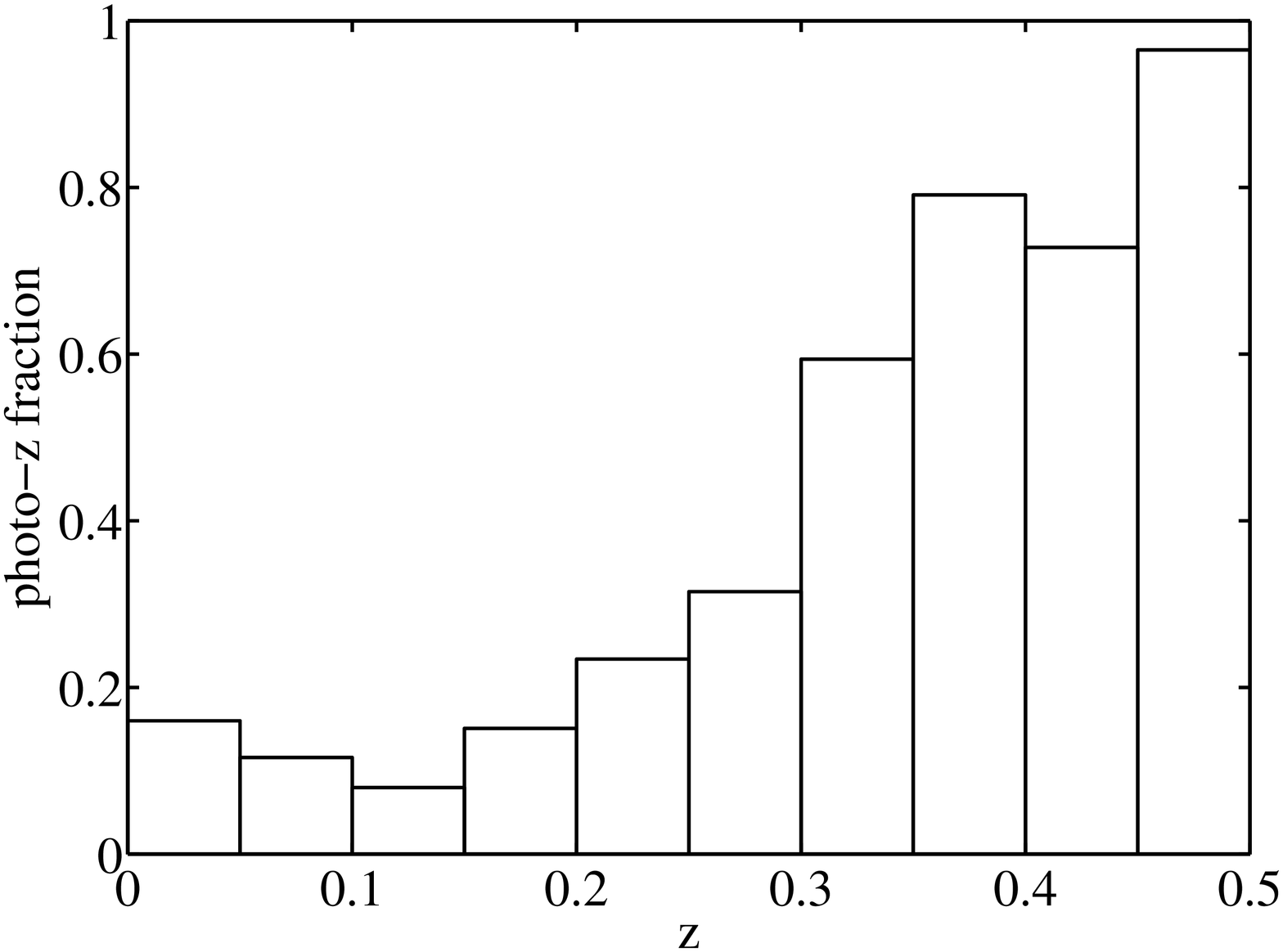}}
\caption{{\bf Top:} Comparison of spectroscopic versus photometric
  redshifts for galaxies in the H-ATLAS SDP. There is a bias in the
  photo-z at redshifts greater than $\sim 0.3$ where photo-z tend to
  underestimate the true redshift. A 1-to-1 correlation is shown by
  the solid line. {\bf Bottom:} The fraction of photo-z used in the
  luminosity function as a function of redshift. Photo-z start to
  dominate the LF at the same redshift where the photo-z bias begins.}
\label{fig:photoz}
\end{figure}

We suspect that this behaviour in the highest redshift bins is a
result of a bias in the ANNz photo-z we are
using. Figure~\ref{fig:photoz}a shows a comparison between
spectroscopic and photometric redshifts of H-ATLAS sources in the SDP
region. There is a bias above $z\sim 0.3-0.35$ where the photometric
redshifts tend to underestimate the true redshift (see Fleuren et
al. in prep). This issue is further exacerbated by the fact that this
is also the redshift at which the LF becomes dominated by photo-z
(Fig.~\ref{fig:photoz}b).

There is another potential bias in the highest-z slice due to the
optical flux limit approaching the main body of galaxies in the
sample. While we correct for the incompleteness in space density due
to the $r$-band limit, we are not able to deal with any accompanying
bias which might allow only those galaxies with lower dust-to-stellar
mass ratios into the sample at the highest redshifts (see
Fig~\ref{fig:limits} and Section~\ref{models} for more
discussion). Greater depth in optical/IR ancillary data will be
required to test the continuing evolution of the luminosity and dust
mass functions beyond $z=0.5$ and this will soon be available with
VISTA-VIKING and other deep optical imaging for the H-ATLAS regions
from VST-KIDS, INT and CFHT.

\begin{figure*}
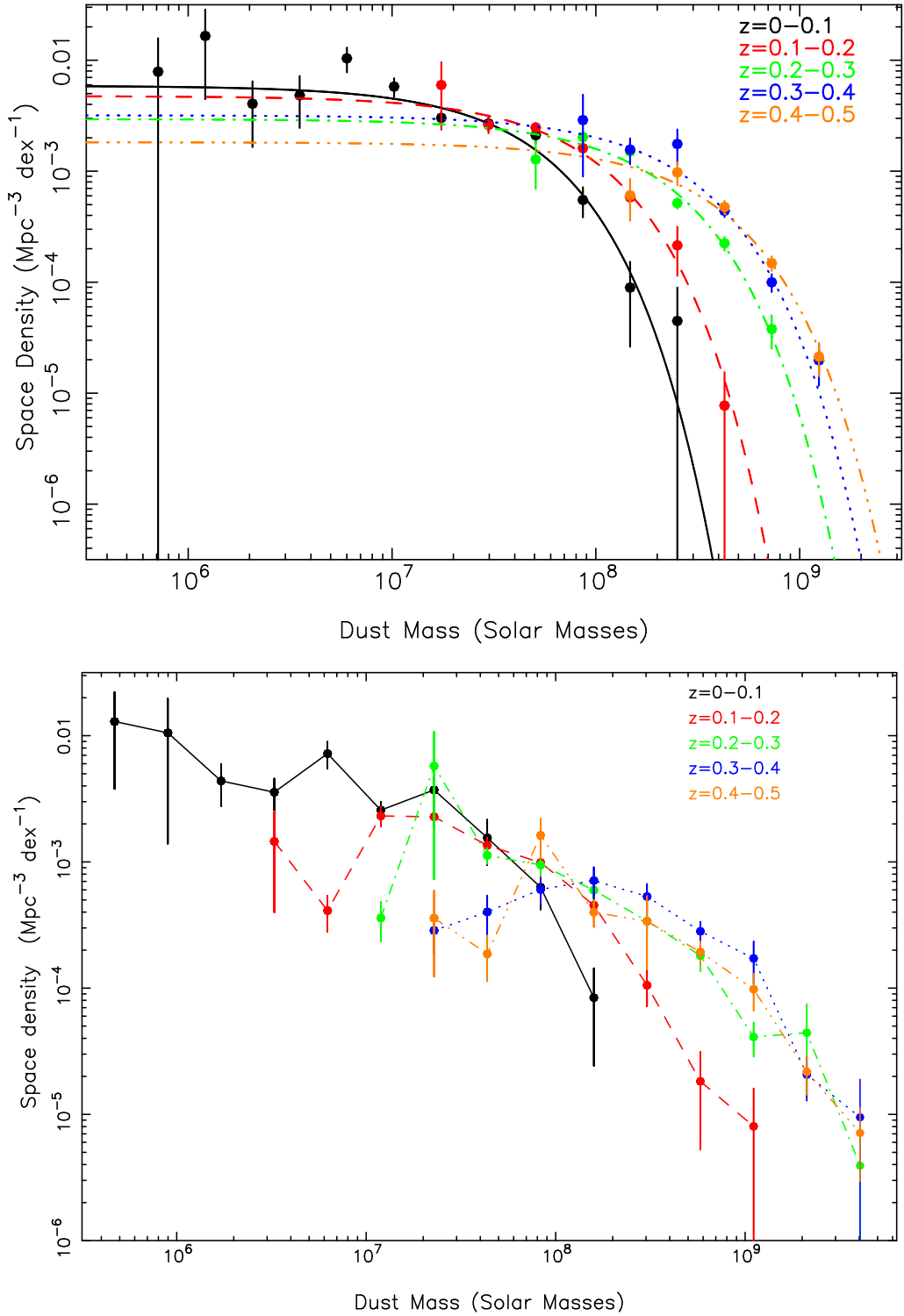

\centering
\subfigure{\includegraphics[width=0.55\textwidth, angle=-90]{dmf_lb15_z5_schec.ps}}\\
\subfigure{\includegraphics[width=0.55\textwidth, angle=-90]{dmf_lb15_z5_miso.ps}}
\caption{Dust mass functions using the modified PC00 estimator
  calculated in 5 redshift slices of $\Delta z=0.1$. {\bf Top} SED
  based dust masses. {\bf Bottom} Isothermal dust masses. The relation
  between dust mass and \Lsub has some scatter due to variation in the
  temperature of the cold ISM dust, which results in down-turns in the
  lowest mass bins in each redshift slice. The broader error on
  $M_{\rm d}$ acts to convolve the true DMF with a Gaussian of width
  approximately 0.2 dex. Schechter functions are plotted in the top
  panel with the faint-end slope fixed to that which fits best in the
  $z<0.1$ slice. Parameters for the fits are given in
  Table~\ref{tab:schec}.}
\label{fig:dmfcomp}
\end{figure*}

Having demonstrated that our modified version of the PC00 estimator
produces sensible results on the 250\mic luminosity function, we now
turn to the estimate of the dust mass function (DMF). We again use
Eqn~\ref{eqn:pcmod} however we now sum all galaxies in a bin of the
$M_d-z$ plane. We use the ratio of \Md to \Lsub to estimate the
$L_{\rm max}$ and $L_{\rm min}$ for each galaxy, which is required to
compute the individual K-correction. The results for both single
temperature masses and SED based masses are shown in
Figure~\ref{fig:dmfcomp}.

Both estimates of the dust mass function show a similar evolutionary
trend as the 250\mic LF, with the same apparent slow down at higher
redshift which we believe may be related to issues with the photo-z.
The evolution is present whichever estimate of the dust mass is used,
however, we will continue the discussion using the DMF from the DCE08
SED based dust masses (Fig.~\ref{fig:dmfcomp}a) as we believe that
this is the best possible estimate at this time.

The dust mass function also shows a down-turn in some redshift slices
at the low mass end. We do not believe that this represents a true
dearth of low mass sources at higher redshifts but rather reflects the
more complex selection function in dust mass compared to \Lsub. While
there is a strong linear relationship between our dust mass and \Lsub
(Figure~\ref{fig:mdcomp}b) there is still scatter on this relationship
due to the variation in the temperature of the cold dust in the
ISM. At fixed \Lsub warm galaxies will have smaller dust masses than
cooler ones, which leads to a sort of `Eddington' bias in the dust
masses. At the limiting \Lsub for a given redshift bin we are not as
complete as we think for low dust masses, since we can only detect
galaxies with low dust masses if their dust is warmer than
average. This in turn leads to the apparent drop in space density. In
the two highest redshift bins, the fraction of sources without SED
fits increases and so the dust masses are then directly proportional
to the 250\mic luminosity. To improve on this, we would need to use a
bi-variate dust mass/\Lsub approach for which the current data are
insufficient, however this analysis will be possible with the complete
H-ATLAS data-set.

\subsection{Dust temperatures and evolution of the LF}
\label{sec:temp}

\begin{figure*}
\centering
\subfigure{\includegraphics[width=8.6cm,height=5cm]{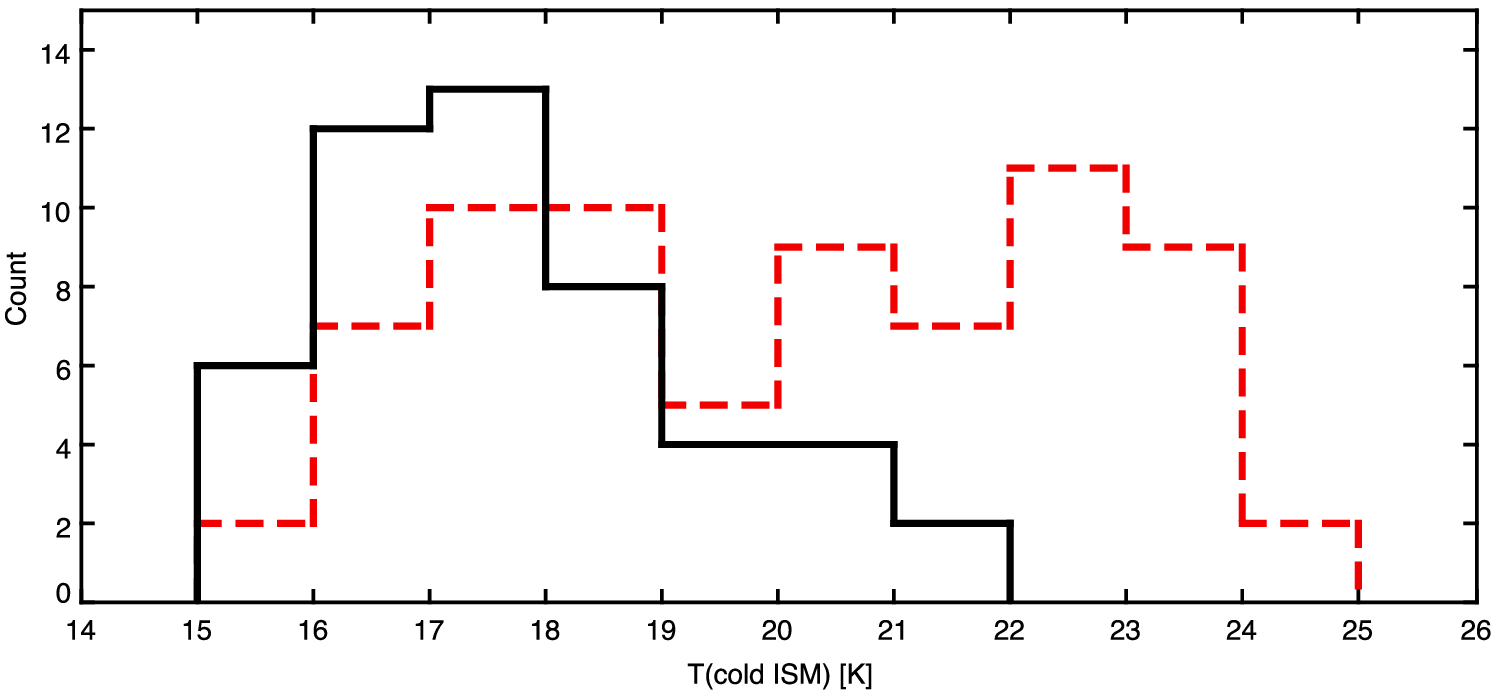}}
\subfigure{\includegraphics[width=9cm,height=6cm]{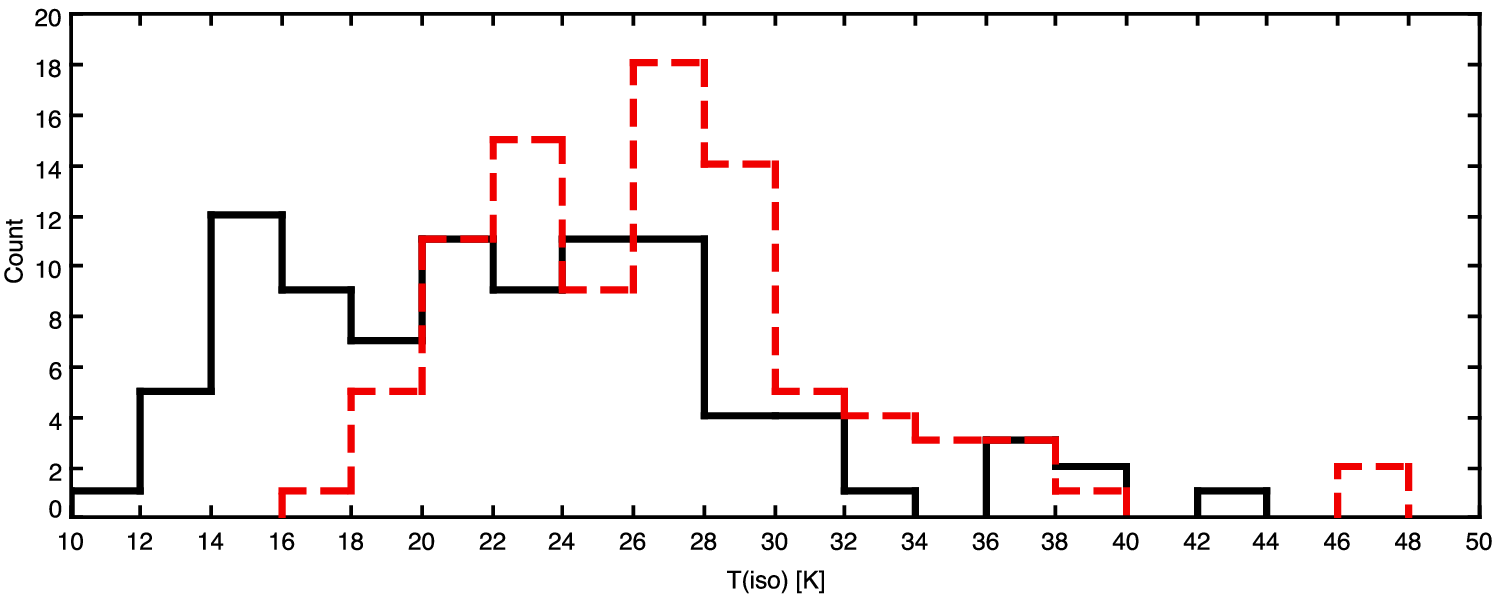}}\\
\caption{\label{fig:PACSlowz} {\bf Left:} Cold ISM temperature from
  the DCE08 fits for the lowest redshift bin which has 41 percent PACS
  detections. The sources with PACS detections are shown by the red
  dashed line while those without PACS detections but which do have a
  350\mic flux above 3$\sigma$ in addition to the 5$\sigma$ 250\mic
  point are shown in black. {\bf Right:} Same but for the isothermal
  temperature}
\end{figure*}

An important ingredient in our estimate of the dust mass is the dust
temperature. In order to interpret the increase in sub-mm luminosity
as an increase in the dust content of galaxies, we have to be wary of
potential biases in our measurements of the dust temperature. The dust
temperature is most accurately constrained when PACS data, which span
the peak of the dust SED, are available in conjunction with the longer
sub-mm data from SPIRE, which also constrains the cold
temperature. For the SDP field, the PACS data is shallow and this
results in PACS detections for only 262 galaxies. The fraction of PACS
detections as a function of redshift are 41, 21, 8, 7, 4 percent
respectively from the lowest to highest redshift bin. A comparison of
the temperatures from fits with PACS detections and without in the
lowest redshift bin (where PACS samples a representative fraction of
the population) is shown in Figure~\ref{fig:PACSlowz}. Fits without
PACS detections are included only if the 350\mic flux was greater than
3$\sigma$ in addition to the $> 5\sigma$ 250\mic flux. The left panel
shows the cold dust temperature from the DCE08 fits and it can be seen
that PACS sources have a range of cold ISM temperatures, however,
those without PACS detections tend to have mostly cooler dust in their
ISM. Smith et al. in prep show that the DCE08 fitting does tend to
underestimate the cold temperature slightly when PACS data are removed
from a fit, but this effect is of order 1--2 K and does not fully
account for the difference in these distributions. When we consider
instead the peak temperature of the SED, that given by the isothermal
fit (right panel in Fig.~\ref{fig:PACSlowz}), we see a different
trend. Now the PACS detections are found only at the higher end of the
temperature range while those without PACS detections span a wider
range of temperature. There is no bias when the PACS data are removed
from the fits (the temperatures vary randomly by $\sim \pm 3$ K when
the PACS data are removed). The sources with the coldest isothermal
temperatures ($\rm{T_{iso}} < 20 K$) are not detected by PACS, even in
the lowest redshift bin, as they either do not contain enough warm
dust, or are not massive enough, to be detected in our shallow PACS
data. We also note that where PACS does not provide a $>5 \sigma$
detection, we use the upper limit in the fitting which provides useful
constraints on dust temperature for many more H-ATLAS sources.

\begin{figure*}
\centering
\subfigure{\includegraphics[width=8.8cm]{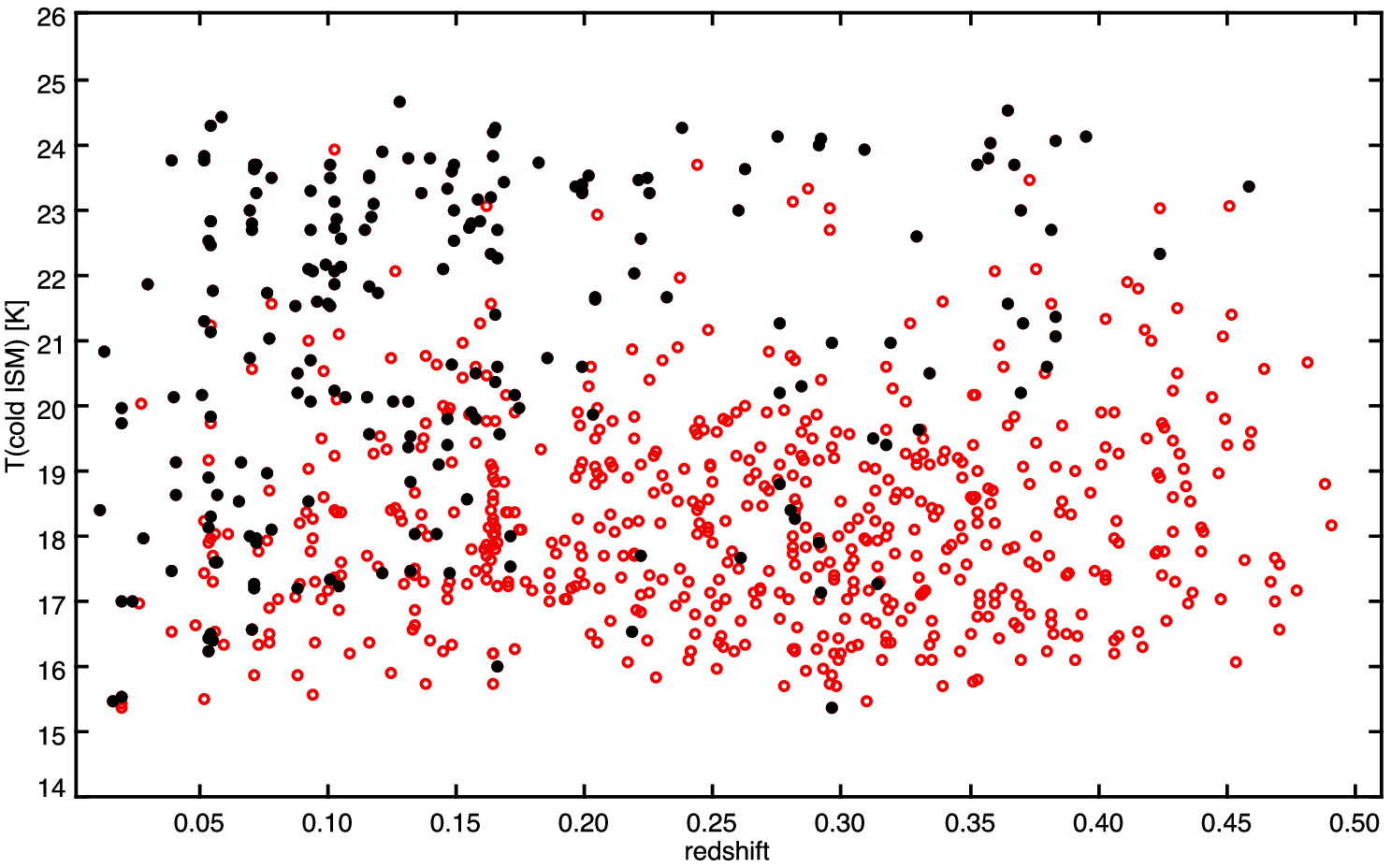}}
  \subfigure{\includegraphics[width=8.8cm]{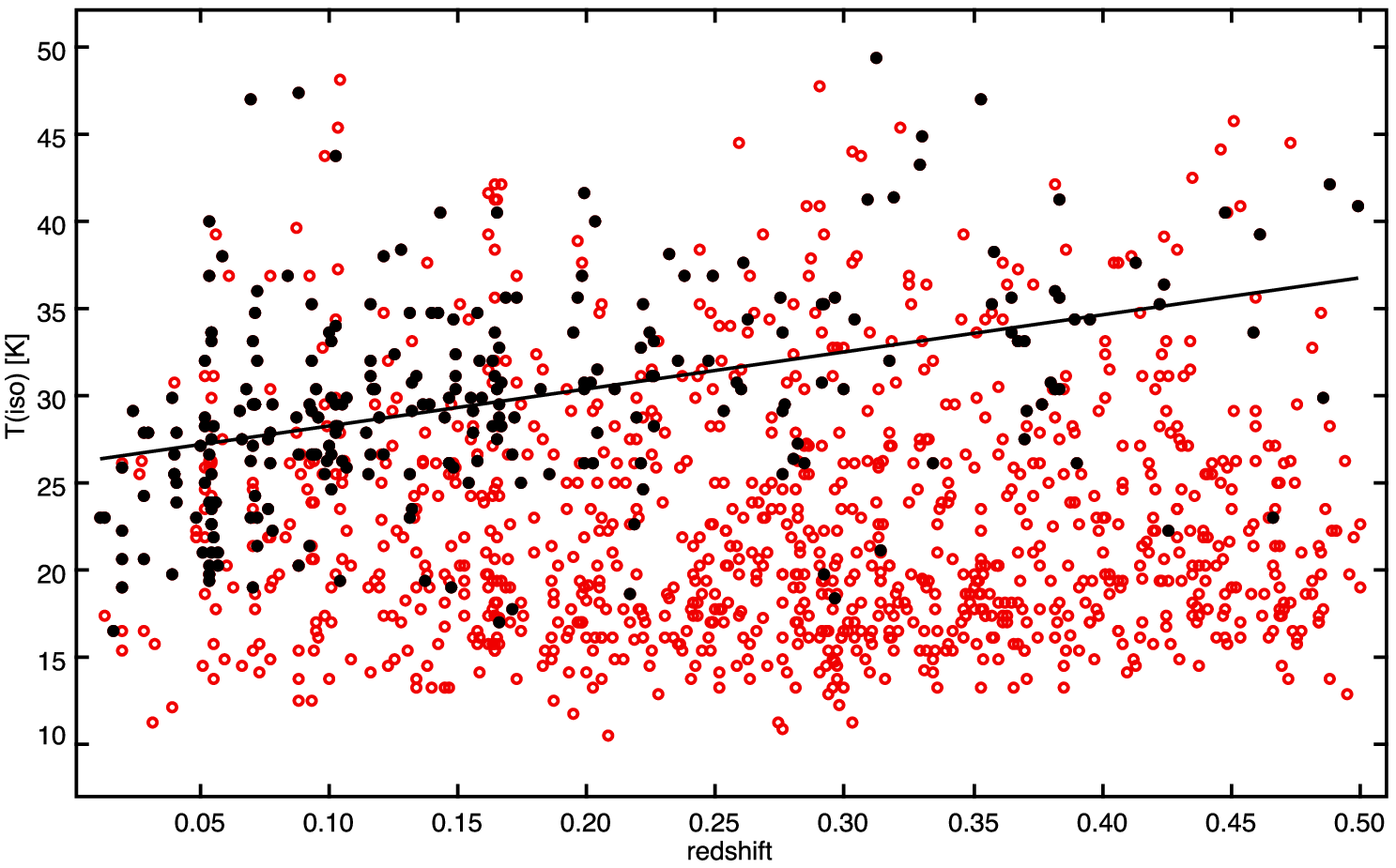}}\\ 
{\caption{\label{fig:PACSTz}
        {\bf Left:} Cold ISM temperature from the DCE08 fits versus
        redshift for sources with PACS detections (black filled) and
        which have 350\mic fluxes above 3$\sigma$ in additions to the
        5$\sigma$ 250\mic flux (red open). {\bf Right:} Same but for
        the isothermal temperature. Here there is a correlation
        between $T_{\rm{iso}}$ and redshift for the PACS detections
        ($r=0.4$).}}
\end{figure*}

The trend for PACS to detect the warmer sub-population of H-ATLAS
sources becomes more pronounced at higher redshift, as the galaxies
must be intrinsically more and more luminous to be detected by
PACS. Figure~\ref{fig:PACSTz} shows the fitted temperature versus
redshift for sources which are detected by PACS (black points) and
those not-detected by PACS but which have at least 2 good quality
sub-mm points for the fit. If we used only the PACS detected sample,
we would infer an evolution in dust temperature in this redshift range
-- but this is a selection bias due to the sensitivity of the PACS
bands to warm dust and the shallow survey limit for PACS. 

\begin{figure*}
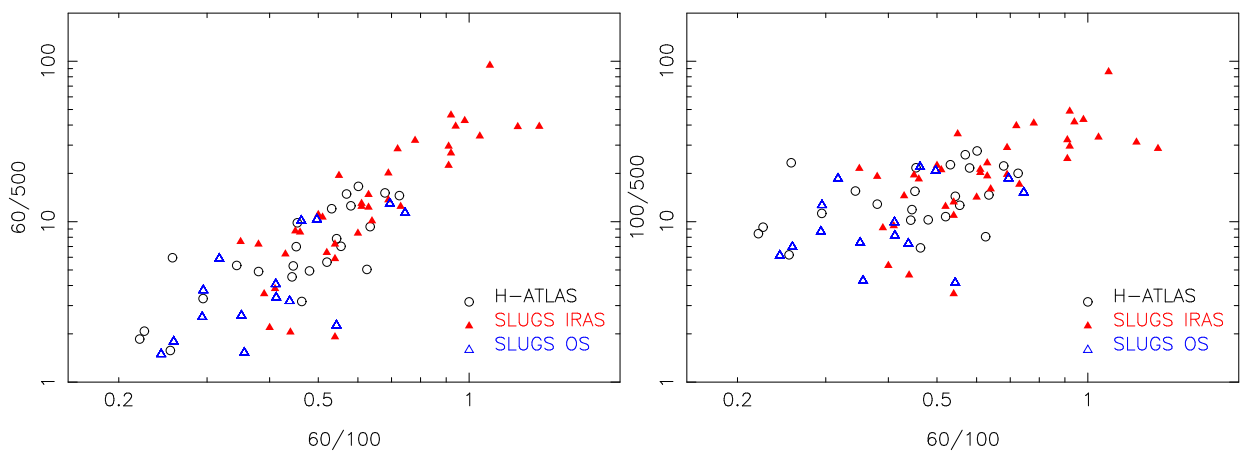

\centering
\subfigure{\includegraphics[width=0.33\textwidth,angle=-90]{colplot_hatlas2.ps}}
\subfigure{\includegraphics[width=0.33\textwidth,angle=-90]{colplot_hatlas.ps}}\\
\caption{\label{fig:cols} Colour plots for the 35 H-ATLAS galaxies with
  detections at 60, 100 and 500\mic compared to those for SLUGS
  sources detected at 450\mic from an IRAS 60\mic selected sample
  (IRAS) and an optically selected sample (OS). The SLUGS points have
  had their 450\mic fluxes adjusted downward by 37 percent to make
  them equivalent to 500\mic}
\end{figure*}

We can also look at the dust colour temperatures of the H-ATLAS
sources in comparison to other samples without the complications of
fitting models. In Figure~\ref{fig:cols} we compare the FIR/sub-mm
colours of the 35 H-ATLAS sources which have 60, 100 and 500\mic
detections at $\geq 3\sigma$ with the colours of SLUGS galaxies from
DE01 and VDE05 to see how these sub-mm selected sources compare to
those selected at 60\mic from the IRAS BGS (Soifer et al. 1989) or in
the optical. H-ATLAS fluxes at 60\mic are from the IIFSCz catalogue of
Wang \& Rowan-Robinson (2009), 100\mic fluxes are from PACS and
500\mic from SPIRE (Rigby et al. 2011). To allow a comparison between
500\mic fluxes from H-ATLAS and 450\mic fluxes from SLUGS, we reduce
the SLUGS 450\mic values by 37 percent using a standard template
suitable for SLUGS sources from DE01 (approximately $\propto \nu^3$ at
these wavelengths). All of these sources are
local. Figure~\ref{fig:cols} shows that the H-ATLAS sources are
significantly colder in their colours than the warmest end of the IRAS
sample; they overlap rather better with the optically selected SLUGS
sample. This is not surprising given our selection at 250\mic is more
sensitive to the bulk dust mass of a galaxy while that at 60\mic from
IRAS is more sensitive to warm dust (either large, warm grains in
star forming regions or small transiently heated grains). We note
that, since only a very small number (35) of H-ATLAS sources are
detected by IRAS, these few sources shown in Fig.~\ref{fig:cols} are
also likely to have `warmer' colours than the overall H-ATLAS
sample. This agrees with the findings that PACS is sensitive to only the
warmer H-ATLAS sources at higher redshifts and the SED fitting
results which show that the H-ATLAS sources contain relatively cooler
dust.

\begin{figure*}
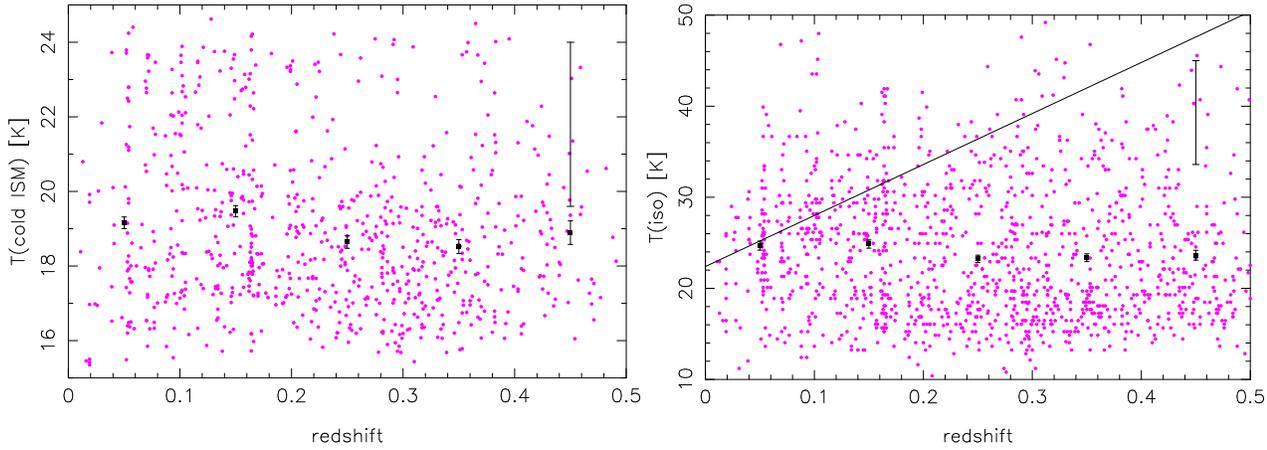

\centering
\subfigure{\includegraphics[width=0.33\textwidth, angle=-90]{tism_z_new2.ps}}
\subfigure{\includegraphics[width=0.33\textwidth, angle=-90]{tiso_z_new2.ps}}
\caption{{\bf Left:} The temperature of the cold interstellar dust
  component as a function of redshift $z$. Only sources with either a
  PACS detection or a 350\mic flux above 3$\sigma$, in addition to the
  250\mic flux, are plotted. Mean values and 1-$\sigma$ errors on the
  mean are shown as black points. The data points in magenta show the
  full distribution of the temperatures. The large error bar in the
  top right shows the average 68 percent confidence range on the
  temperature for an individual fit. {\bf Right:} The isothermal
  temperature estimated from a grey body fit versus redshift, same
  coding as before. The line plotted shows the evolution in
  temperature required in order to explain the evolution in the
  250\mic LF {\em without\/} any increase in the dust masses. Neither
  method for estimating the dust temperature shows any evolution with
  redshift.}
\label{fig:tdz}
\end{figure*}

If the evolution in the 250\mic\ LF were due simply to an increase in
the `activity' of galaxies of the same dust mass, then we should see a
corresponding increase in dust temperature with redshift and no
evolution in the DMF. To explain the amount of evolution in the
250\mic\ LF without any increase in dust mass would require an
increase in the average dust temperature of order a factor 2 over the
period $0<z<0.5$. We investigated the relationship between both the
cold ISM dust temperature from the DCE08 fits and the isothermal
grey-body temperature with redshift and found no trend for either
(Figure~\ref{fig:tdz}) at $z<0.5$, similar to the results from Amblard
et al. (2010) and inconsistent with the temperature evolution required
to explain the increase in the 250\mic luminosity density.

\begin{figure*}
\centering
\subfigure{\includegraphics[width=8.5cm]{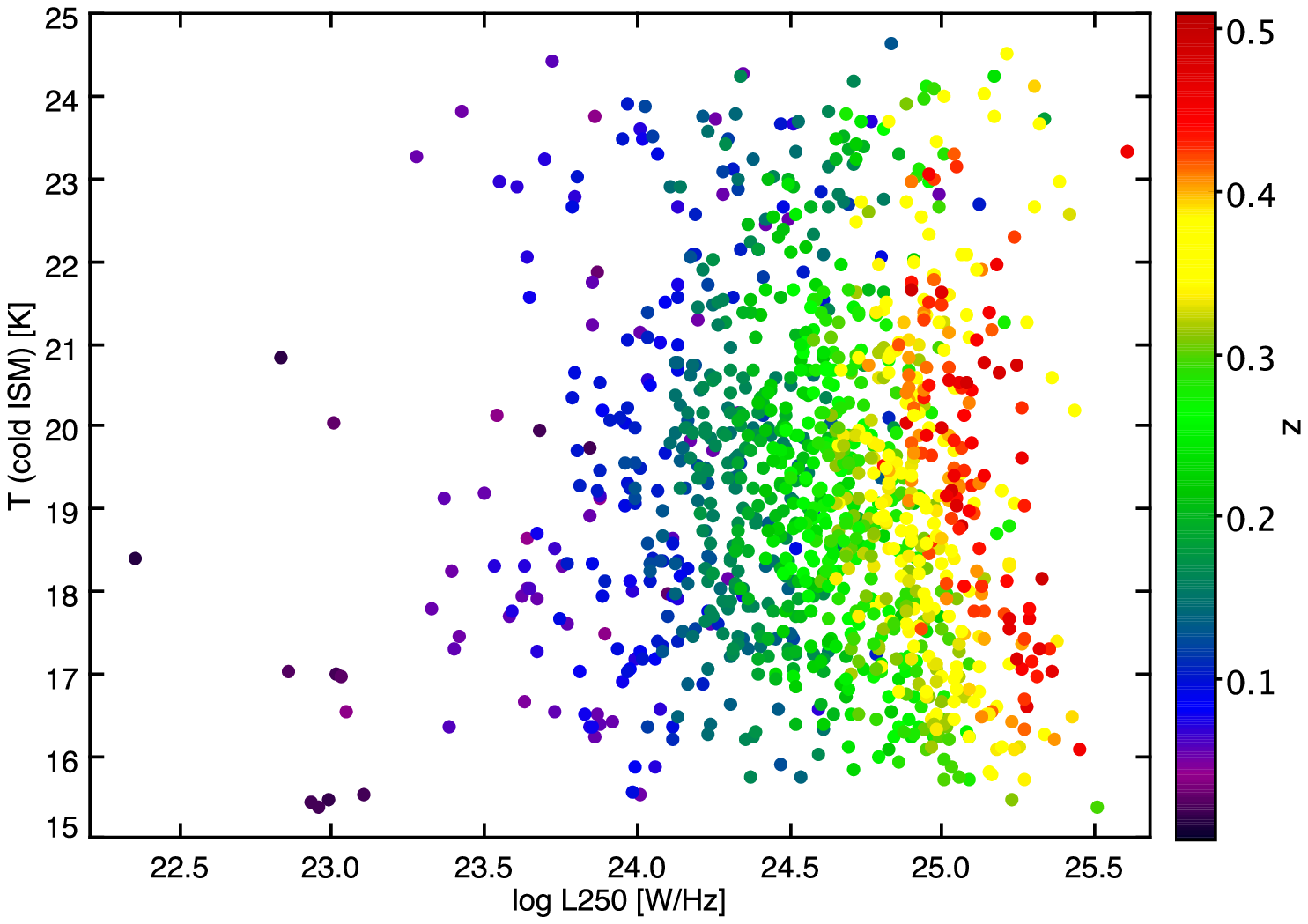}}
\subfigure{\includegraphics[width=8.5cm]{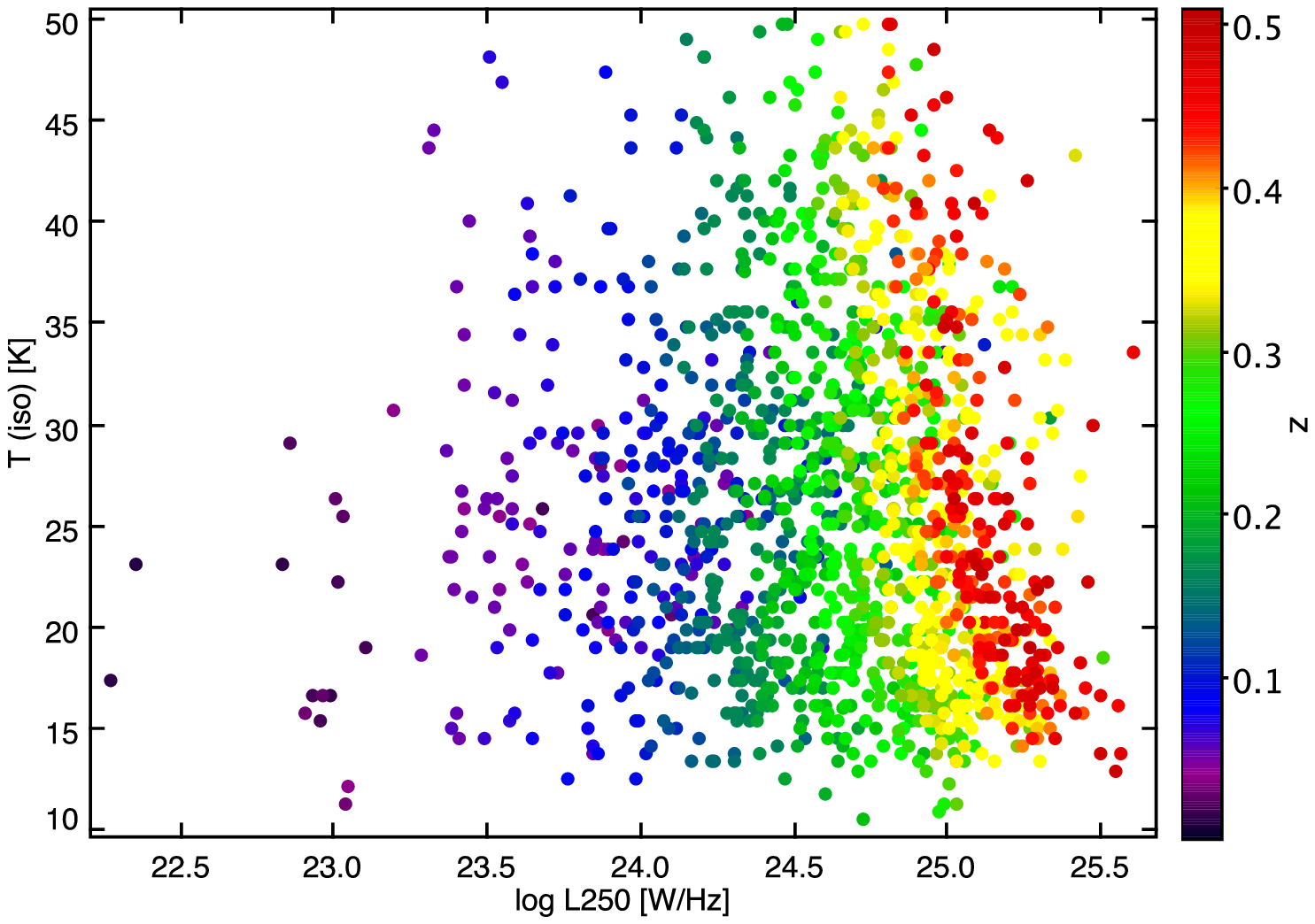}}\\
\caption{\label{fig:lumtemp} {\bf Left:} Cold dust temperature and
  \Lsub showing no correlation. The points are colour coded by
  redshift. {\bf Right:} Same as (left) but for the isothermal dust temperature}
\end{figure*}

The temperature of nearby ($z<0.1$) dusty galaxies has been shown to
be correlated with their IR luminosity (the so-called
$\rm{L_{ir}-T_d}$ relation; e.g. D00, Dale et al. 2001). A natural
explanation for this observation might be that a galaxy which has
hotter dust (for a given mass) will have a larger IR luminosity than a
similar mass galaxy with cooler dust. Recent work which extends to
more sensitive surveys and samples selected at longer wavelengths
suggests that this does not hold at higher redshifts and that galaxies
are in general cooler at a given IR luminosity than previously
believed (Coppin et al. 2008; Amblard et al. 2010; Rex et al. 2010,
Symeonidis et al. 2009, Seymour et al. 2010, Smith et al. in
prep). Symeonidis et al. (2011) suggest that this is due to more rapid
evolution of ``cold'' galaxies over the period $0.1<z<1$ than ``warm''
ones. Recent studies at other wavelengths (70--160\mic from {\em
  Spitzer\/} and PACS) seem to support this interpretation, finding
that cold galaxies are responsible for most of the increase in the IR
luminosity density over the range $0<z<0.4$ (Seymour et al. 2010;
Gruppioni et al. 2010). This is in agreement with the evolution seen
in H-ATLAS galaxies which are largely comprised of this `cold'
population. Despite our average luminosity increasing with redshift,
we see no increase in the average temperature (either isothermal or
cold ISM temperature) and indeed we also see no correlation of either
temperature with luminosity (either dust luminosity from the DCE08
model or \Lsub) for this sample (see Figure~\ref{fig:lumtemp}).

To summarise, while we are subject to uncertainties in our ability to
derive the dust masses and the exact scale of any evolution, we are
nevertheless confident that:

\begin{itemize}
\item{The evolution in the 250\mic LF out to $z=0.5$ cannot be driven
  by dust temperature increases; there must be some evolution in the
  mass of dust as well.}
\item{The H-ATLAS sources at $z\leq 0.5$ are colder than previous
  samples based on IRAS data and therefore most of the evolution at
  low redshift is driven by an increase in the luminosity or space
  density of such cooler galaxies.}
\item{H-ATLAS sources show no trend of increase in dust temperature
  with either redshift or luminosity at $z<0.5$}
\end{itemize}

\subsection{Comparison to low redshift dust mass functions}

We can compare the lowest redshift bin in the DMF ($0<z<0.1$) to
previous estimates from the SCUBA Local Universe and Galaxy Survey
(D00; VDE05), which used SCUBA to observe samples of galaxies selected
either at 60\mic from the {\em IRAS} Bright Galaxy Sample (Soifer et
al. 1989) or in the B-band from the CfA redshift survey (Huchra et
al. 1983). The {\em IRAS} SLUGS galaxies were mostly luminous
star-bursts, and in principle this should have produced an unbiased
estimate of the local dust mass function as long as there was no class
of galaxy unrepresented in the original IRAS BGS sample. However, it
was argued in D00 and VDE05 that this selection at bright 60$\mu$m
fluxes quite likely missed cold but dusty galaxies, given the small
sample size of $\sim 100$, thus may have produced a DMF which was
biased low. The optically selected SLUGS sample overcame the dust
temperature bias and did indeed show that there were very dusty
objects which were not represented as a class in the {\em IRAS} BGS
(similarly confirmed by the ISO Serendipity Survey: Stickel et
al. 2007). The directly measured DMF presented by VDE05 suffered from
small number statistics, and instead V05 followed the work of Serjeant
\& Harrison (2005) in extrapolating the {\em IRAS} PSCz (Saunders et
al. 2000) out to longer wavelengths (850\mic) using the empirical
colour-colour relations derived from the combination of {\em IRAS} and
optically selected SLUGS galaxies. This set of 850\mic\ estimates for
all {\em IRAS} PSCz sources was then converted to a dust mass assuming
a temperature of 20\,K (the average cold component temperature found
by DE01 and VDE05) and a mass opacity coefficient of $\kappa_{850} =
0.077\,\rm{m^{2} \,kg^{-1}}$. From this set of masses they then
produced an estimate of the DMF.

\begin{figure}
\centering
\includegraphics[width=0.35\textwidth, angle=-90]{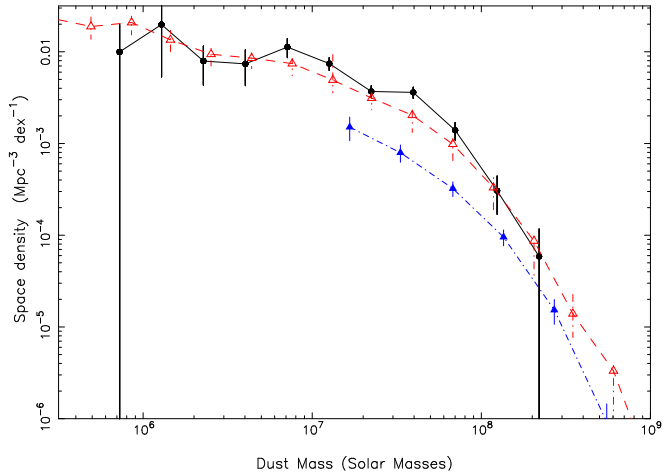}
\caption{Comparison of the local dust mass functions at $z<0.1$
    from H-ATLAS (black solid line and points) along with estimates
    from SLUGS. Blue dot-dash line and solid triangles -- directly
    measured DMF from {\em IRAS} SLUGS sample (D00, DE01). Red dashed
    line and open triangles -- extrapolated DMF from {\em IRAS} PSCz
    using sub-mm colours from the optical SLUGS sample (VDE05). The H-ATLAS
    points have been corrected for the factor 1.4 under-density in the
    GAMA-9hr field for this redshift range compared to SDSS at large.}
\label{fig:dmflowz}
\end{figure}

The DMFs are compared in Figure~\ref{fig:dmflowz}, where the black
  solid line and points are from H-ATLAS at $z<0.1$, the blue dot-dash
line and filled triangles is the SLUGS {\em IRAS} directly-measured
DMF (D00) and the red dashed line and open triangles is the DMF based
on the extrapolation of the {\em IRAS} PSCz by VDE05. In this figure, the
H-ATLAS DMF has been corrected for the known under-density of the
GAMA-9hr field relative to SDSS as required when comparing to an
all-sky measurement such as SLUGS or {\em IRAS} PSCz. This correction
is a factor of 1.4 (Driver et al. 2011). The SLUGS DMFs have been
corrected to the cosmology used in this paper, however these
corrections are small at low-z.
 
It is remarkable that despite the considerable differences in
sample size, area and selection wavelength, the SLUGS estimate from
VDE05 based on extrapolating the {\em IRAS} PSCz gives a very good
agreement to our measure. This implies that there is not a significant
population of objects in the PSCz sample, or the H-ATLAS sample which
is not represented by the combined optical and 60\mic selected SLUGS
samples (which comprised only 200 objects). Note that had VDE05 used
the {\em IRAS} data alone to measure dust masses, the results would be
extremely different. It is only that SLUGS allowed an empirical
statistical translation between {\em IRAS} colours and sub-mm flux and
from there, assumed a mass-weighted cold temperature for the bulk of
the dust that they were able to obtain such a good measure of the DMF.

The original direct measure of the DMF from the bright {\em IRAS}
SLUGS sample (blue line in Fig~\ref{fig:dmflowz}; D00) dramatically
under-estimates the dust content in the local Universe (this was also
noted by VDE05 once the optically selected sample was included). The
dust masses were derived for those objects in an identical way to the
VDE05 DMF (and very similar to our current method which has an average
measured cold temperature of between 15--19\,K), however the {\em
  IRAS} BGS simply missed objects which were dusty but did not have
enough {\em warm\/} dust to make it above the 60\mic selection. {\em
  Herschel} is able to select sources based on their total dust
content, rather than simply the small fraction of dust heated to
$>30$\,K. {\em Herschel} samples are therefore likely to contain a far
wider range of galaxies in various states of activity, so long as they
have enough material in their ISM.

\subsection{Evolution of the dust mass function}

\begin{figure}
\centering
\includegraphics[width=0.50\textwidth]{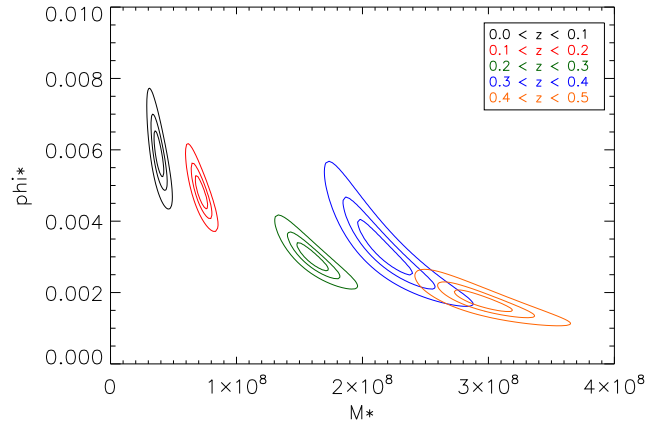}
\caption{\label{fig:chi2} $\chi^2$ confidence intervals at 68, 90, 99
  percent for $M_d^{\ast}$ and $\phi^{\ast}$ with fixed $\alpha$ for the
  five redshift bins. This shows the clear evolution of $M_d^{\ast}$
  over the interval $0<z<0.5$.}
\end{figure}

For illustration, we now fit Schechter functions (Schechter 1976) to
the dust mass functions in each redshift slice. Only in the first
redshift bin do we fit to the faint end slope $\alpha$, for other
redshift bins we keep this parameter fixed at the value which best
fits the lowest redshift bin ($\alpha=-1.01$) to avoid the
incompleteness problem mentioned above with the lowest mass bins at
high redshift. The best-fitting parameters for the slope $\alpha$,
characteristic mass $M_d^{\ast}$ and normalisation $\phi^{\ast}$ are
given in Table~\ref{tab:schec}, where the errors are calculated from
the 68 percent confidence interval from the $\chi^2$ contours. For the
lowest redshift bin, we include errors which reflect the
marginalisation over the un-plotted parameter. The $\chi^2$ contours
for $M_d^{\ast}$ and $\phi^{\ast}$ are shown in Figure~\ref{fig:chi2}.

\begin{table*}
\caption{The Schechter parameters fitted to the dust mass function}
\centering
\begin{tabular}{|c|c|c|c|c|c|c|c|c|}
\hline
Redshift & $\alpha$ & $M_d^{\ast}$ & $\phi^{\ast}$ & $\rho_d $ & $\chi^2_{\nu}$ & Cos. Var. & $\rm N_{bin}$ & $\rm z_{phot}/z_{tot}$\\
            &          & ($\times 10^7$ \msun)  & ($\times 10^{-3}\,\rm{Mpc^{-3}\,dex^{-1}}$) & ($\times 10^5$ \msun $\rm Mpc^{-3}$) &  &  \\            
\hline
$0.0-0.1$ & $-1.01^{+0.17}_{-0.14}$ & $3.83^{+0.73}_{-0.62}$ & $5.87^{+1.38}_{-1.25}$ & & 1.5 & 0.39 & 222 & 0.12  \\
\hline
$0.0-0.1$ &  $-1.01$               & $3.83^{+0.39}_{-0.43}$ & $ 5.87^{+0.59}_{-0.62}$ & $0.98\pm 0.14$ & 1.3  & 0.39 & 222 & 0.14\\
[-1ex]\\
$0.1-0.2$  & $-1.01$               & $7.23^{+0.37}_{-0.45}$ & $4.78^{+0.47}_{-0.41}$ & $1.51\pm 0.16$ & 1.0 & 0.21 & 421 & 0.14\\
[-1ex]\\
$0.2-0.3$  & $-1.01$               & $16.0^{+1.1}_{-1.2}$ & $2.97^{+0.37}_{-0.34}$ &  $2.08\pm 0.29$ & 3.0 & 0.17 &504 & 0.34\\
[-1ex]\\
$0.3-0.4$  & $-1.01$               & $21.6^{+2.0}_{-1.8}$ & $3.24^{+0.75}_{-0.74}$ & $3.06\pm 0.75$ & 0.8 & 0.17 & 416 & 0.76\\
[-1ex]\\
$0.4-0.5$  & $-1.01$               & $29.5^{+2.2}_{-2.0}$ & $1.75^{+0.31}_{-0.27}$ & $2.26\pm 0.41$ & 2.0 & 0.17 & 304 & 0.92\\
\hline
$\sim 2.5$ & $-1.08$               & 39.1             & 1.74                   & 3.11           &  \\
    \hline
    \label{tab:schec}
  \end{tabular}
\flushleft \small{The first line of the table is the fit to all three
  parameters for the lowest redshift bin with associated errors from
  the 68 percent confidence interval derived from the $\chi^2$
  contours. The following entries are where $\alpha$ is fixed to the
  best-fitting value in the lowest redshift bin. The final entry is
  the fit to the $z\sim 2.5$ DMF from DEE03 corrected to this
  cosmology and $\kappa_{250}$. Cos. Var. is the cosmic variance
  estimated using the calculator from Driver \& Robotham (2010). $\rm
  N_{bin}$ is the number of sources in that redshift bin and $\rm
  z_{phot}/z_{tot}$ is the fraction of photometric redshifts in that
  bin.}
\end{table*}


There is a strong evolution in the characteristic dust mass
$M_d^{\ast}$ with redshift, from $M^{\ast}=3.8\times10^7$ \msun at
$z<0.1$ to $M_d^{\ast}=3.0\times 10^8$ \msun at $z=0.4-0.5$. There is
seemingly a decline in $\phi^{\ast}$ over the same redshift range,
from $0.0059 - 0.0018\, \rm{Mpc^{-3}\,dex^{-1}}$ (however this could
also be due to sample incompleteness which is not corrected for
despite our best attempts). The drop in $\phi^{\ast}$ and increase in
$M_d^{\ast}$ are correlated (see Fig~\ref{fig:chi2}), and therefore we
caution against using the increase in the fitted $M_d^{\ast}$ alone as
a measure of the dust mass evolution. If we keep $\phi^{\ast}$ fixed
at $0.005 \, \rm{Mpc^{-3}\, dex^{-1}}$ (which is the average of that
for the first two redshift bins) then the $M_d^{\ast}$ of the highest
redshift bins decreases to $1.8\times 10^8$ \msun  giving an evolution
in $M_d^{\ast}$ over the range $z=0-0.5$ of a factor $\sim 5$ rather
than $\sim 8$ as is the case if the normalisation is allowed to drop.

We calculate the dust mass density in redshift slices using
Eqn.~\ref{eqn:dustdens}.

\begin{equation}
\rho_d = \Gamma(2+\alpha)\,M_d^{\ast}\,\phi^{\ast}
\label{eqn:dustdens}
\end{equation}

\begin{figure}
\centering
\includegraphics[width=0.35\textwidth,angle=-90]{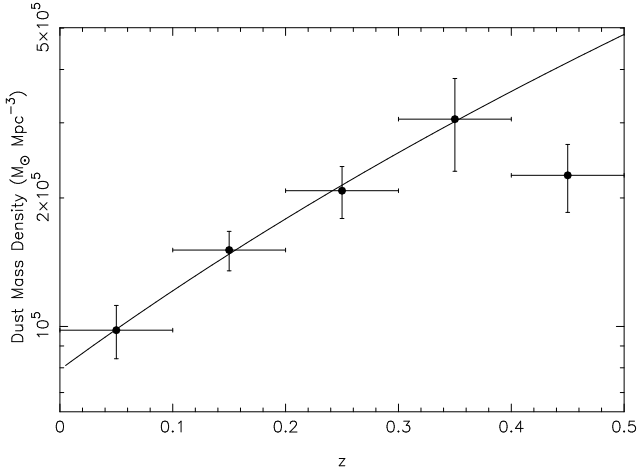}
\caption{\label{fig:dustdens} Integrated dust mass density as a
  function of redshift for H-ATLAS calculated using
  Eqn~\ref{eqn:dustdens}. The best fitting relationship excluding the
  higher redshift point is over-plotted, which is $\rho_d \propto
  (1+z)^{4.5}$. }
\end{figure}

\noindent This assumes that we can extrapolate the Schechter function
beyond the range over which it has been directly measured. Given the
low value of $\alpha$ used ($\sim -1$) the resulting integral is
convergent and so whether we extrapolate or not has negligible effect
on the resulting mass density values. The values for $\rho_d$ are
listed in Table~\ref{tab:schec} and shown as a function of redshift in
Figure~\ref{fig:dustdens}. There is clearly evolution in the cosmic
dust mass density out to $z\sim 0.4$ of a factor $\sim 3$ which can be
described by the relationship $\rho_d \propto (1+z)^{4.5}$. In the
highest redshift bin the dust mass density appears to drop (despite
the increase in $M_d^{\ast}$), but we again caution that this may be
due to incompleteness/photo-z bias in the final redshift bin. This
measure of the dust mass density at low redshift can be compared to
that made by Driver et al. (2007). They used the optical B-band disk
luminosity density from the Millennium Galaxy Catalogue scaled by a
fixed dust mass-to-light ratio from Tuffs et al. (2004). Their quoted
value for the dust density is $\rho_d = 3.8\pm1.2 \times 10^5\,\rm
Mpc^{-3}$ at $z<0.1$ but this is for a $\kappa$ value from Draine \&
Li (2001) which is lower than that used here by 70 percent. Scaling
their result to our $\kappa$, and correcting the density of our lowest
redshift bin by the factor 1.4 from Driver et al. (2011) (to allow for
the under-density of the GAMA-9hr field relative to SDSS at $z<0.1$)
we have values of $\rho_d = 2.2\pm0.7\times 10^5 \,\rm{Mpc^{-3}}$
(optical based) and $\rho_d = 1.4\pm0.2\times 10^5 \,\rm{Mpc^{-3}}$
(DMF) which are in rather good agreement given the very different ways
in which these estimates have been made.

We can also calculate the dust mass density parameter $\Omega_{\rm dust}$ from 
\[ 
\Omega_{\rm dust} = \frac{\rho_d}{\rho_{\rm crit}}
\]
\noindent where $\rho_{\rm crit}=1.399\times 10^{11}\, \rm
\msun\,Mpc^{-3}$ is the critical density for $h=0.71$.  This gives
values of $\Omega_{\rm dust}=0.7-2 \times 10^{-6}$ depending on
redshift. Fukugita \& Peebles (2004) estimated a theoretical value of
$\Omega_{\rm dust}=2.5\times 10^{-6}$ today based on the estimated
density of cold gas, the metallicity weighted luminosity function of
galaxies and a dust to metals ratio of 0.2. This is a little higher
than our (density corrected) lowest redshift estimate of $1.0\pm 0.14
\times 10^{-6}$ but not worryingly so. M\'{e}nard
et al. (2010) also estimate a dust density in the halos of galaxies
through a statistical measurement of reddening in background quasars
when cross-correlated with SDSS galaxies. They estimate a dust density
of $\Omega_{\rm dust}^{\rm halo}= 2.1\times 10^{-6}$ for a mean
redshift of $z\sim 0.35$ and suggest that this is dominated by
$0.5\,L_{\ast}$ galaxy halos. Comparing this to our measure of the
dust {\em within\/} galaxies at the same redshift ($\Omega_{\rm
  dust}^{\rm gals}=2\times 10^{-6}$) we see that at this redshift
there is about the same amount of dust outside galaxies in their halos
as there is within. We note here that dust in the halos of galaxies
will be so cold and diffuse that we will not be able to detect it in
emission with H-ATLAS and so it is not included in our DMF. The
decrease in $\rho_d$ at recent times could be due to dust being
depleted in star formation, destroyed in galaxies by shocks or also
lost from galaxies (and from our detection) to the halos. We will
return to this interesting observation in Section~\ref{models}.

\begin{figure}
\centering
\subfigure{\includegraphics[width=0.35\textwidth, angle=-90]{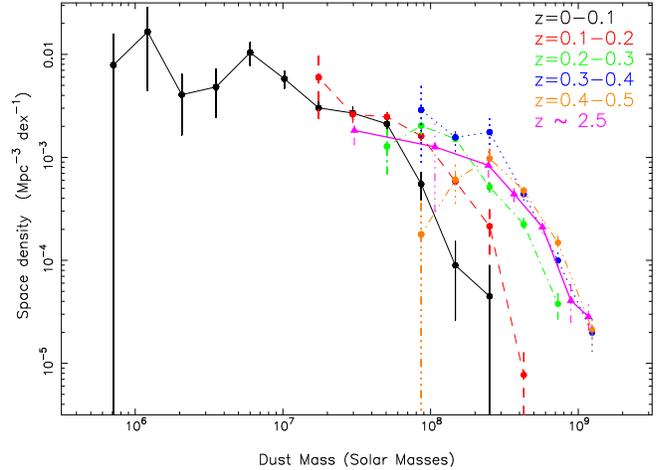}}
\caption{Comparison of the H-ATLAS dust mass function in five
    redshift slices (as in Fig~\ref{fig:dmfcomp}) and the
    high redshift, $z\sim 2.5$, DMF from D03 (magenta dashed line).}
\label{fig:cudssdmf}
\end{figure}

We can compare the DMF from H-ATLAS to that at even higher redshifts,
as traced by the 850\mic\ selected SMG population. An estimate of the
DMF for these sources at a median redshift of $z\sim 2.5$ was
presented in DEE03, using the $1/V_{\rm max}$ method. In
Fig~\ref{fig:cudssdmf} we show this higher-$z$ DMF alongside the
H-ATLAS data, where the $z\sim 2.5$ DMF is the magenta solid line
with filled triangles. The DEE03 higher-z DMF has been scaled to the
same cosmology and value of $\kappa_{250}$ as used here.  At $z\sim
2.5$, observed 850\mic corresponds to rest-frame $\sim 250$\mic\ and
so our lower-z H-ATLAS sample and the one at $z\sim 2.5$ are selected
in a broadly similar rest-frame band. DEE03 used a dust temperature of
25 K to estimate the dust mass, which allowed for some evolution over
the low-$z$ SLUGS value of 20\,K. The $z\sim 2.5$ sources from DEE03
are all ULIRGS and these higher luminosity sources do show enhanced
dust temperatures in the local Universe (Clements, Dunne \& Eales
2010; da Cunha et al. 2010b). It is also consistent with the cold,
extended dust and gas component ($T=25-30$\,K) of the highly lensed
SMG at $z=2.3$ (Swinbank et al. 2010; Danielson et al. 2011) and other
lensed sources discovered by {\em Herschel} (Negrello et al. 2010). If
we were to recompute the $z\sim 2.5$ DMF using a temperature of 20\,K
instead, this would shift the points along the dust mass axis by a
factor $\sim 1.7$.

For either temperature assumption, the $z\sim 2.5$ DMF is broadly
consistent with the H-ATLAS DMF in the two highest redshift bins
($z=0.3-0.5$). The fits to the high-z DMF are shown in
Table~\ref{tab:schec} and the dust density at $z\sim 2.5$ is also
consistent with that in the $z=0.3-0.5$ range from H-ATLAS. If true,
this implies that the rapid evolution in dust mass may be confined to the
most recent 4--6 billion years of cosmic history. Notwithstanding the
earlier statement that this trend needs to be confirmed with a larger
sample, dust masses are unlikely to continue rising at this pace
because the dust masses at very high redshifts (Micha{\l}owski et
al. 2010; Pipino et al. 2010) are not very different to those we see
here. 

This implies that the evolution in the 250\mic LF is due at least in
part to a larger interstellar dust content in galaxies in the past as
compared to today, at least out to $z\sim 0.4$ (corresponding to a
look-back time of 4 Gyr). However, an increase in star-formation rate
is also an important factor as if the dust mass increased at a
constant SFR we would expect to see a decline in dust temperature with
redshift. Our observations thus point to an increase in both dust mass
and star formation activity. If the evolution in the DMF is
interpreted as pure luminosity (or mass) evolution (as opposed to
number density evolution), then this corresponds to a factor 4--5
increase in dust mass at the high mass end over the past 4 Gyr. Since
dust is strongly correlated to the rest of the mass in the
interstellar medium (ISM) (particularly the molecular component), this
also implies a similar increase in the gas masses over this period. In
contrast, we know that the stellar masses of galaxies do not increase
with look-back time, showing very little evolution in the mass range
we are dealing with (predominantly $L_{\ast}$ or higher) (Pozzetti et
al. 2007; Wang \& Jing 2010). The evolution of the DMF is therefore
telling us something quite profound about the evolution of the dust
content of galaxies, and by inference, the gas fractions of galaxies
over this period.

\section{The dust content of H-ATLAS galaxies} 
\label{duststars}

\begin{figure*}
\centering
\subfigure{\includegraphics[width=0.99\textwidth]{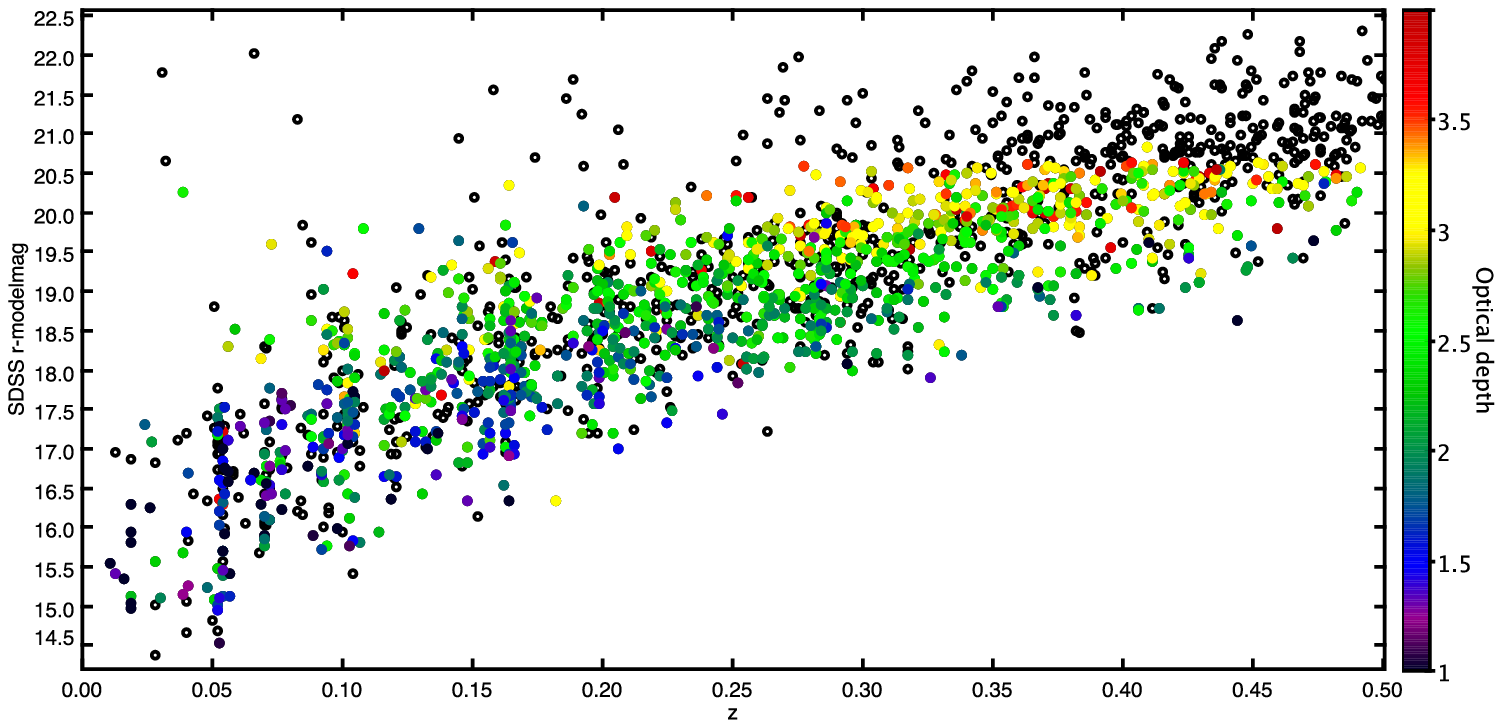}}
\subfigure{\includegraphics[width=0.99\textwidth]{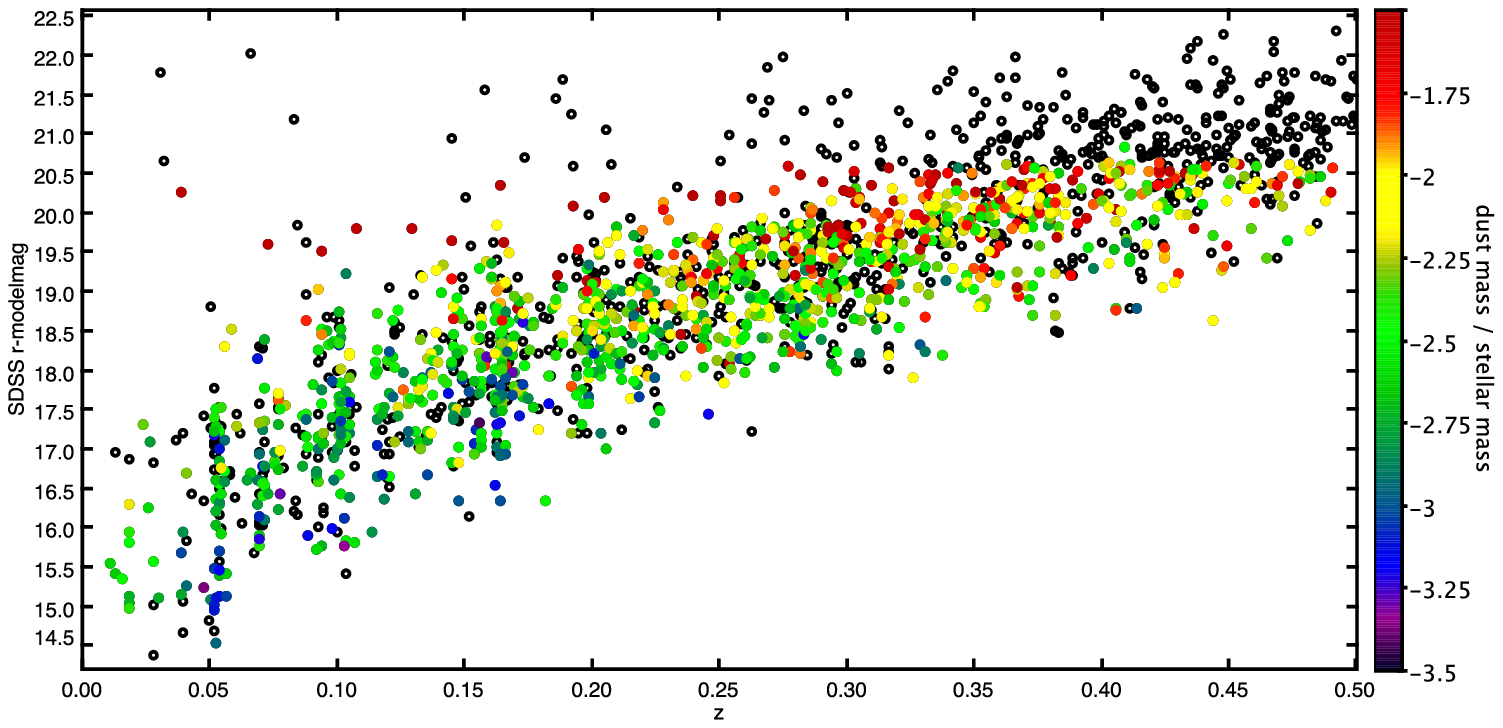}}\\
\caption{$r$-mag versus redshift for the H-ATLAS sources. Black open
  circles represent H-ATLAS sources which are too faint for an SED fit
  using the DCE08 model at the current time, or which were not in the
  region covered by GAMA photometric catalogues. Coloured points
  denote the values of either V-band optical depth ({\bf top}) or
  dust-to-stellar mass ratio ({\bf bottom}) from the DCE08 fits. The
  limit of reasonable completeness in the optical for the SED fits is
  $z\sim 0.35$. Beyond this redshift, averaged values of optical depth
  or dust-to-stellar mass ratio will be biased low because only the
  brightest optical galaxies in that redshift bin will have SED fits
  (and these tend to have less obscuration).}
\label{fig:limits}
\end{figure*}

There are two ways in which we can quantify the dust content: the
amount of light absorbed by dust (or opacity), and the dust-to-stellar
mass ratio. Both of these are derived from the DCE08 SED model fits
for galaxies which were bright enough ($r\leq 20.5$) that aperture
matched photometry was extracted by GAMA (Hill et al. 2011). Due to
this being shallower than the depth to which we can ID the H-ATLAS
sources we have to take care not to introduce selection biases when
making these comparisons. Figure~\ref{fig:limits} shows $r$-mag as a
function of redshift for the H-ATLAS sources and again highlights that
H-ATLAS does not detect low stellar mass (or low absolute $M_r$)
sources. The panels in Fig.~\ref{fig:limits} have colour coded points
  for sources where SED fits were made, and the colours represent
  either the V-band optical depth (top) or the dust-to-stellar mass
  ratio (bottom). At $z\sim 0.35$ the optical sample which has SED
  fits becomes incomplete, with only the brighter fraction of the
  galaxies having SED fits at a given redshift. This can lead to a
  lowering of the average optical depth, or dust-to-stellar mass ratio
  in bins at $z>0.35$, since the brighter galaxies (higher stellar
  masses) tend to have lower values of optical depth or
  dust-to-stellar mass. Thus in the following discussion we limit our
  model comparisons to the data with $z<0.35$. We hope to extend the SED
  fitting to the fainter sources in future work.

\begin{figure}
\centering
\subfigure{\includegraphics[width=0.35\textwidth, angle=-90]{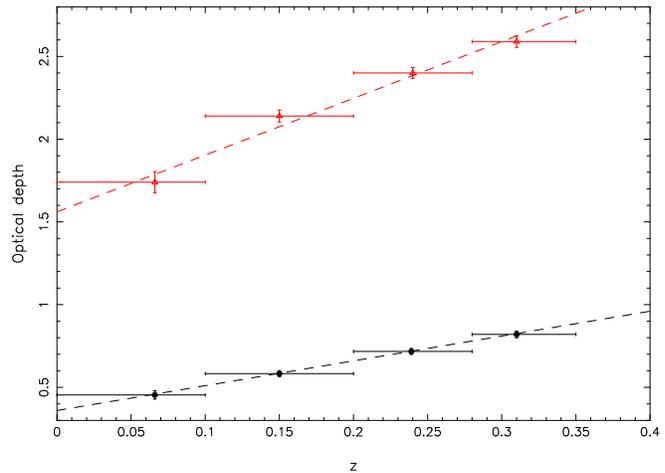}}\\
\caption{Upper red points: Mean V-band optical depth in the birth
  clouds (from the DCE08 SED fits of Smith et al. in prep) as a
  function of redshift with the best linear fit. Lower black points:
  V-band optical depth in the ISM ($\mu\tau_v$ from DCE08).}
\label{fig:tau}
\end{figure}

First we plot the amount of optical light obscured by dust: the V-band
opacity. This is derived from the DCE08 SED model fits, and is
calculated both in the birth clouds where stars are born ($\tau_V$
from DCE08) and also in the diffuse ISM ($\mu\tau_V$ from
DCE08). Figure~\ref{fig:tau} shows the evolution of both forms of
V-band optical depth from the model fits, indicating that galaxies are
becoming more obscured back to $z\sim 0.4$. Choi et al. (2006), Villar
et al. (2008) and Garn et al. (2010) also find a higher dust
attenuation in high redshift star forming galaxies. This is sometimes
attributed to an increase of SFR with look-back time (Garn et
al. 2010) and an attendant increase in dust content rather than to a
change in dust properties. It is also possible that the apparent
increase of optical depth with increasing redshift is related to the
correlation between IR luminosity and dust attenuation (Choi et
al. 2006), whereby more IR luminous galaxies tend to be more
obscured. The average IR luminosity of our sample increases strongly
with redshift (due both to the flux limit of the survey and the strong
evolution of the LF) and it is currently not possible for us to
disentangle the effects of redshift from those of luminosity since we
do not have a large enough sample to make cuts in redshift at fixed
luminosity. Regardless of which is the driver, the observational
statement remains that a sub-mm selected sample will contain more
highly attenuated galaxies at higher redshifts. This is in contrast to
some UV selected samples which show either no trend with redshift or a
decline of attenuation at higher-z, due to their selection effects
(Burgarella et al. 2007; Xu et al., 2007; Buat et al. 2009). This just
highlights the obvious -- that FIR and UV selected samples are
composed of quite different objects.

Our relationships with redshift are as follows:
\[\rm {birth\, clouds}: \, \tau_V = 3.43 z + 1.56  \]
\[\rm {diffuse\, ISM}:\,\mu \tau_V = 1.50 z + 0.36 \]

\noindent which implies that the attenuation from the birth clouds is
rising faster with increasing redshift than that in the diffuse
ISM. At higher redshifts we are therefore finding that the birth
clouds are producing a larger fraction of the attenuation in the
galaxy than at low redshift. We find this trend interesting but
further work is required to explain and confirm it, firstly ensuring
in a larger sample that it is not again related to the luminosity
(more luminous sources also have higher relative attenuation from the
birth clouds). Including Balmer line measurements in the DCE08 fits
will also better constrain the optical depth in the birth clouds.

\begin{figure}
\centering
\includegraphics[width=0.5\textwidth]{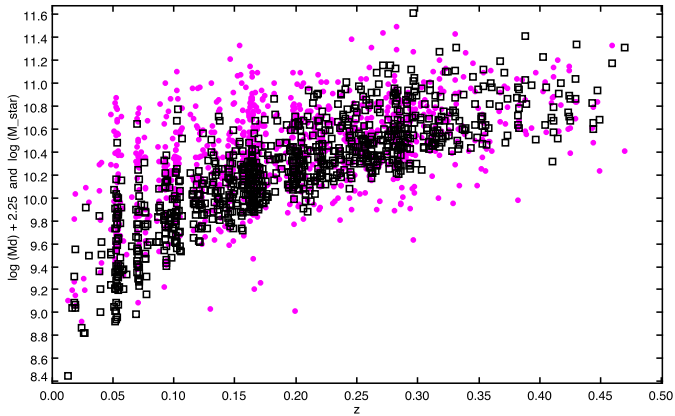}
\caption{Stellar mass (magenta) and dust mass scaled by 178 (black
  open squares) versus redshift. The dust mass is scaled to make the
  dust and stellar lower limits approximately coincide at low-z. This
  illustrates the different trends of dust and stellar mass with
  redshift, with the dust mass evolving more rapidly than the stellar
  mass (as is also evident from the DMF). At lower redshifts there are
  many galaxies with higher stellar masses than the scaled dust mass,
  while at high redshifts both stellar and dust masses are comparable
  with the same scaling.}
\label{fig:mdmstarz}
\end{figure}

Secondly, we can look at dust and stellar mass together using the
stellar masses from the DCE08 SED fits. Figure~\ref{fig:mdmstarz} shows
the variation of dust and stellar mass with redshift, where the dust
mass has been scaled up by a factor 178 in order to roughly make \Md
and $M_{\ast}$ equivalent at the lower boundary at low-z. Magenta
points show stellar mass, open black squares are the scaled dust
mass. The stellar mass remains fairly constant with redshift, while
there is a distinct lack of high dust mass objects in the local
Universe (as is shown also by the DMF). The dust-to-stellar mass ratio
as a function of redshift is shown in Figure~\ref{fig:mdstarz} and
discussed in more detail in the next section.

\section{Modelling the evolution of dust}
\label{models}
In this Section we will attempt to explain the evolution we see in the
dust content of H-ATLAS sources and in the DMF. We do this using a
chemical and dust evolution model which traces the yield of heavy
elements and dust in a galaxy as its gas is converted into stars. A
full treatment of the evolution of galaxies will be considered in
Gomez et al. in prep. Here we will consider the elementary model of
Edmunds (2001; see also Edmunds \& Eales 1998) in which one assumes
that the recycling of gas and dust in the interstellar medium is
instantaneous. Details of the model are given in Appendix A, but in
brief, a galaxy is considered to be a closed box with no loss or
addition of gas during its evolution. The evolution of the galaxy is
measured in terms of $f$, its gas fraction, which represents the
fraction of the baryonic mass in the form of gas. Gas is converted
into stars using a star formation prescription $\psi(t)=kg(t)^{1.5}$,
where $g$ is the gas mass and $k$ is the star formation efficiency
(inversely proportional to the star formation time-scale). We define
an effective yield $p=p'/\alpha \sim 0.01$ where $\alpha \sim 0.7$ is
the mass fraction of the ISM locked up in stars (Eq.\ref{eq:lock})
and $p'$ is the yield returned from stars for a given initial mass
function (IMF). We can interpret $p$ as being the true mass fraction
of heavy elements returned per stellar generation, since some fraction
of the generated heavy elements is locked up in low mass stars and
remnants. In the first instance, we use the Scalo form of the IMF (Scalo 1986)
for Milky Way evolution (e.g. Calura et al.\ 2008). The metal mass
fraction of a galaxy is tracked through $p$ and therefore follows
metals incorporated into long lived stars and remnants or cycled
through the ISM where they are available to be made into dust. The
parameters which determine how many of the available metals are in the
form of dust relate to the sources of dust in a galaxy and we consider
three of these:
\begin{enumerate}
\item{Massive stars and SNe: $\chi_1$ is the efficiency of dust
  condensation from new heavy elements made in massive star winds or
  supernovae.}
\item{Low-intermediate mass stars (LIMS): $\chi_2$ is the efficiency
  of dust condensation from the heavy elements made in the stellar
  winds of stars during their RG/AGB phases.}
\item{Mantle growth in the ISM: We can also assume that grains accrete
  at a rate proportional to the available metals and dust cores in
  dense interstellar clouds (Edmunds 2001). $\epsilon$ is the fraction
  of the ISM dense enough for mantle growth, $\eta_c$ is the
  efficiency of interstellar depletion in the dense cloud (i.e. if all
  the metals in the dense clouds are accreted onto dust grains then
  $\eta_c=1$).}
\end{enumerate}
Morgan \& Edmunds (2003) used observations of dust in low-intermediate
mass stars to show that $\chi_2\sim 0.16$ yet theoretical models
following grain growth in stellar atmospheres (e.g. Zhukovska et
al.\ 2008) suggest higher values of $\chi_2 \sim 0.5$. We adopt the
higher value here, but note that there is some considerable
uncertainty on $\chi_2$.  For core-collapse supernovae (using
theoretical models of dust formation e.g. Todini \& Ferrara et
al. 2001) Morgan \& Edmunds suggest that $\chi_1 \sim 0.2$; this
agrees with the highest range of dust masses published for Galactic
supernova remnants (Dunne et al.\ 2003; 2009, Morgan et al.\ 2003;
Gomez et al.\ 2009). If core-collapse SNe are not significant
producers of dust (e.g. Barlow et al.\ 2010) or if most of their dust
is then destroyed in the remnant (Bianchi \& Schneider 2007) then this
fraction decreases to $\chi_1 \le 0.1$, making it difficult to explain
the dust masses we see in our Galaxy or in high-redshift submillimetre
bright galaxies with stellar sources of dust (e.g. Morgan \& Edmunds
2003; Dwek et al. 2007; Micha{\l}owski et al.\ 2010).

For mantles we arbitrarily set $\epsilon=0.3$ and from interstellar
depletion levels in our Galaxy and following Edmunds (2001), we set
$\eta_c \sim 0.7$ (that is, we assume that if the clouds are dense,
then it is likely that the dust grains accrete mantles).  In this
scenario, the dust is formed during the later stages of stellar
evolution and uses up the available metals in dense clouds.  The
addition of accretion of metals onto grain cores with the parameters
described here will double the peak dust mass reached by a galaxy.
Assuming no destruction of grains, a closed box model and mantle
growth gives the highest dust mass attainable for galaxies. 

Dust destruction can be added to this elementary model by assuming
some fraction $\delta$ of interstellar grains are removed from the ISM
as a mass $ds$ is forming stars. We use two destruction scenarios: one
with a constant destruction rate $\delta=0.3$ (Edmunds 2001) and the
second where $\delta$ is proportional to the Type-II SNe rate (which
gives a similar result to Dwek's approximation for MW IMF; Dwek et
al. 2011). We also allow a mantle growth proportional to SFR since one
would expect that the efficiency will depend on the molecular fraction
of the ISM (which in turn is related to the SFR; Papadopoulos \&
Pelupessy 2010).

Finally, we relax the closed-box assumption and include outflows in
the model (Appendix A) since galactic-scale outflows are thought to be
ubiquitous in galaxies (Menard et al. 2010 made a remarkable detection
of dust reddening in the halos of galaxies which implies at least as
much dust is residing in the halos as in the disks). Here we test
outflows in which enriched gas is lost at a rate proportional to one
and four times the SFR (more powerful outflows are unlikely, since in
the latter case, the galaxy would only retain approximately 20
per\,cent of its initial gas mass).

\subsection{Evolution of Dust to Stellar Mass}
\label{sec:submdmstar}

\begin{figure}
\includegraphics[trim=0mm -0.5mm 0mm 0mm,clip=true,width=8.5cm]{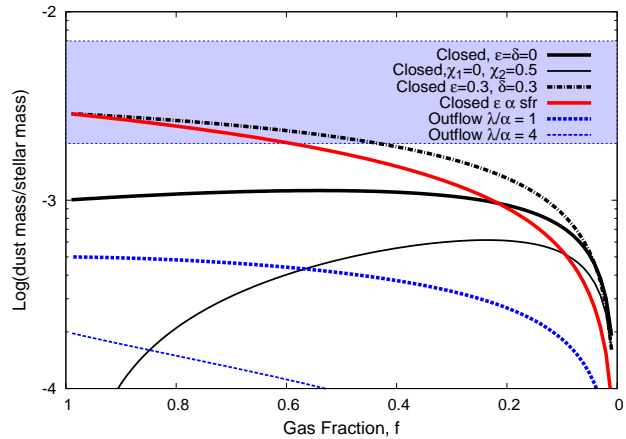}
\caption{Variation of dust-to-stellar mass ratio as a function of gas
  fraction.  The shaded box region is the range of values observed for
  the H-ATLAS galaxies.  The models are (i) a closed box with no gas
  entering/leaving the system with dust from both supernovae
  $\chi_1=0.1$ and LIMS stars $\chi_2=0.5$ (thick solid; black); (ii)
  with dust from LIMS only $\chi_1=0,\chi_2=0.5$ (thin solid; black);
  (iii) model (i) now including mantle growth (dot-dashed; black);
  (iv) A model with mantle growth, where the mantle rate is
  proportional to the SFR (solid; red); (v) and (vi) a model which has
  outflow with gas lost at a rate proportional to one or four times
  the SFR ($\lambda/\alpha$) (dashed; blue). }
\label{fig:mdmstar_mstar}
\end{figure}

The dust-to-stellar mass ratio of the models discussed here is shown
in Figure~\ref{fig:mdmstar_mstar} over the life-time of the galaxy as
measured by the gas fraction, $f$.  The shaded region shows the range
of values of $M_{\rm d}/M_*$ estimated for the H-ATLAS galaxies, which
have a peak value of $7\times 10^{-3}$ at $z=0.31$ and then decreases
as the galaxy evolves in time (to lower gas fractions) to $2\times
10^{-3}$. This global trend is reproduced by the closed box model
where dust is contributed by {\em both} massive stars and LIMS, or via
mantle growth, however the models struggle to produce values of
$M_{\rm d}/M_*$ as high as observed.  We also plot in
Figure~\ref{fig:mdmstar_mstar}, the variation of $M_{\rm d}/M_*$ if
low-intermediate mass star-dust is the only stellar contributor to the
dust budget ($\chi_1=0,\chi_2=0.5$). It is clear that the LIMS dust
source cannot reproduce the values of dust/stellar mass seen in the
H-ATLAS sources alone. Either significantly more dust is contributed
to the ISM via massive stars/SNe than currently inferred, or a
significant contribution from accretion of mantles in the ISM is
required (indeed we would need significantly more dust accretion in
the ISM than dust produced by LIMS). The simple model also suggests
that the H-ATLAS galaxies must be gas rich ($f>0.4$) in order to have
dust-to-stellar mass ratios this high. (Typical gas fractions for
spiral galaxies today are $f\sim 0.1-0.2$.)

\begin{figure*}
\centering
\includegraphics[trim=0mm -1.0mm 0mm 0mm,clip=true,width=8.85cm]{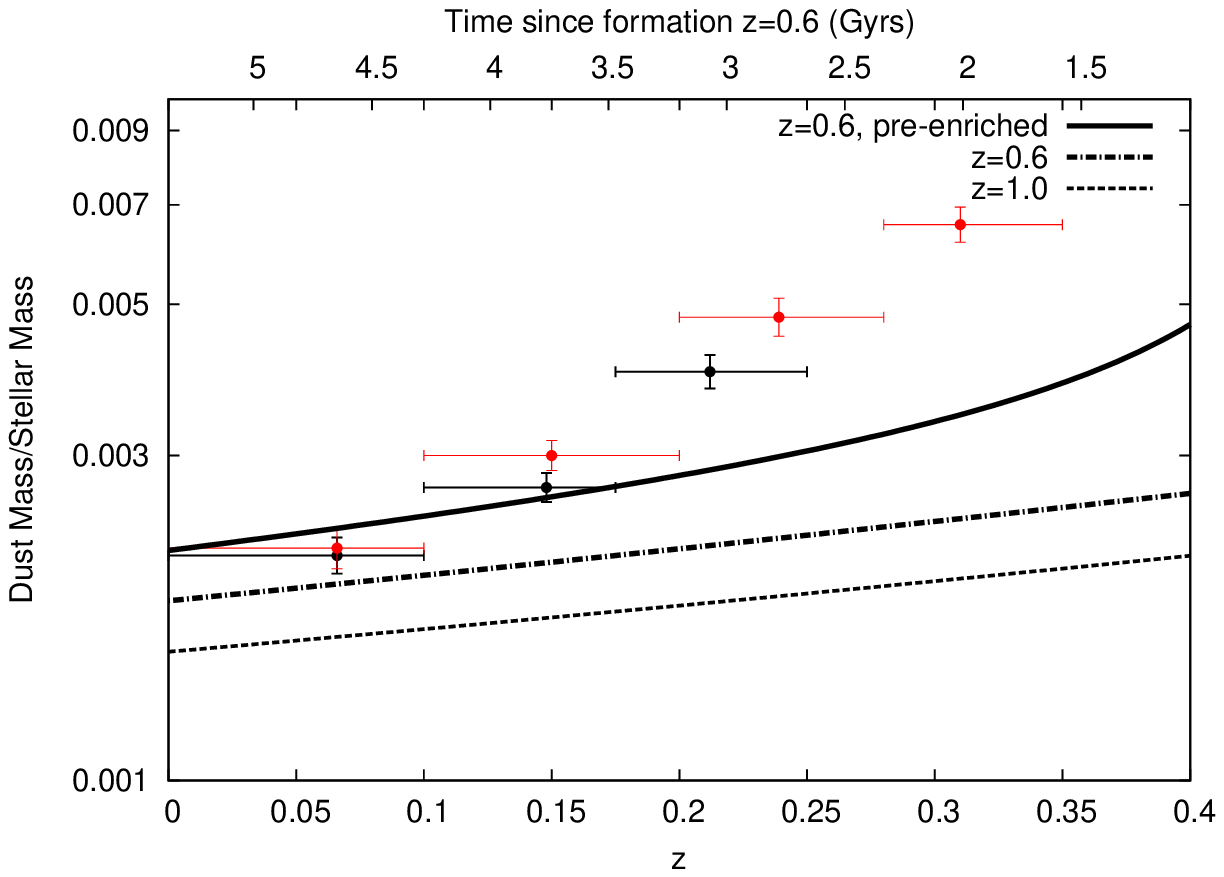}
\includegraphics[trim=5mm -1.0mm 0mm  0mm,clip=true,width=8.5cm]{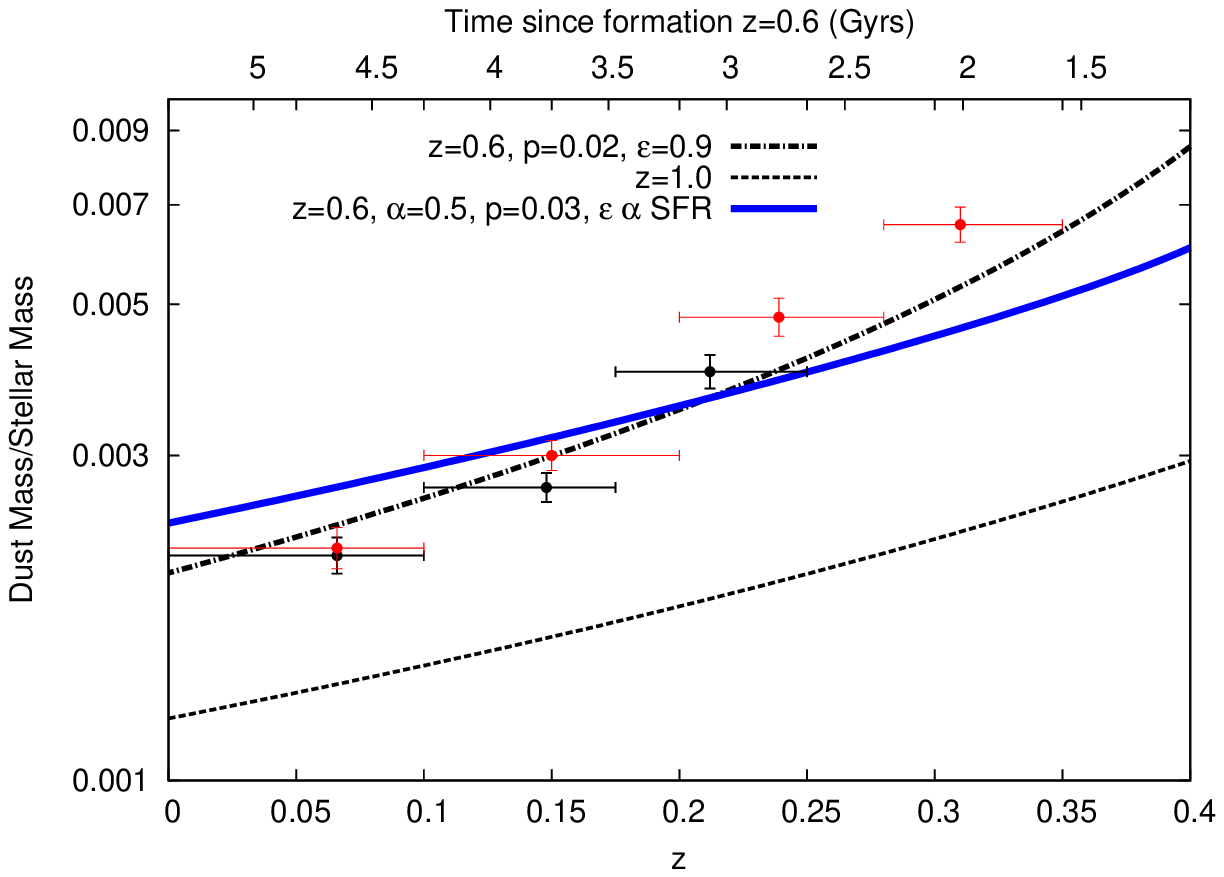}
\caption{{\bf Left:} The dust-to-stellar mass ratio as a function of
  redshift. Stellar and dust masses are derived from the SED fits
  using the models of DCE08 and are discussed in detail in
  Section~\ref{masses} and Smith et al. (2011b).  Black points show
  those sources with spectroscopic redshifts, while red points include
  photometric redshifts. Each sample is limited in redshift to the
  point where the optical flux limit is not biasing the selection to
  low dust-to-mass ratios. The model lines for the dust model
  (Section.~\ref{sec:submdmstar}) corresponding to the Milky Way
  including mantle growth and destruction are over-plotted with
  formation redshifts of $z=0.6$ (dot-dashed) and $z=1$ (dotted).  A
  model including pre-enrichment of $Z_i \sim 0.1Z_{\odot}$ with
  formation timescale at $z=0.6$ is also shown (solid; black).  {\bf
    Right:} Same as left including pre-enrichment, but models are now
  tuned to match the data for the $z=0.6$ formation time.  With
  pre-enrichment, we require $\chi_1=0.1,\,\chi_2=0.5$, $p=0.02$,
  $\epsilon=0.9$ and SF efficiency $k=1.5\,\rm Gyr^{-1}$ to `fit' the
  data points (black dot-dashed) or $\chi_1=\chi_2=\epsilon=0.5$,
  $p=0.02$ (not shown).  Also shown is a model with mantle
  growth varying with SFR and a top-heavy IMF described by $\alpha
  =0.5, \, p=0.03$ (solid; blue).  Adding outflow or destruction rates
  which vary with SFR would make the decline in $M_{\,\rm d}/M_*$ more
  pronounced at lower redshifts (later evolutionary times).  }
\label{fig:mdstarz}
\end{figure*}

We can also consider the evolution of dust-to-stellar mass as a
function of time (Eq.~\ref{app:md}). This is shown in
Fig~\ref{fig:mdstarz}a using dust production and yield parameters
appropriate for spiral galaxies like the Milky Way
($p=0.01,\,\alpha=0.7,\,\chi_1=0.1,\,\chi_2=0.5,\,\epsilon\eta_c =
0.24,\, k=0.25\,\rm Gyr^{-1}$). We compare the model for two formation
times of $z=0.6$ and $z=1$, where formation time in this model can
simply mean the time of the last major star formation event.  In this
scenario, we would expect any previous star formation to have already
pre-enriched the ISM with some metallicity $Z_i$, therefore increasing
the available metals for grain growth in the ISM.

From Fig~\ref{fig:mdstarz}a, we see that the MW model does not match
the variation of dust/stellar mass from H-ATLAS observations even if
we increase the mantle growth or the amount of dust formed by stars,
since the increase in dust-to-stellar content with gas fraction (as we
look back to larger redshifts and earlier times in the evolution of
the galaxy) is simply not rapid enough. Fig~\ref{fig:mdstarz}b shows
the same two formation times but now we have tuned the parameters to
match the data for a formation at $z=0.6$. In order to do this we have
to increase the SF efficiency parameter ($k=1.5\,\rm Gyr^{-1}$) to
produce a steeper relationship as observed. An increase in $k$
compared to the MW model is hardly surprising, since these higher
values are typical of star-forming spirals with initial
SFRs\footnote{depending on the initial gas mass of the galaxies} of
$\psi \sim 50\,\rm M_{\odot}\,yr^{-1}$ which is in agreement with the
observations of H-ATLAS sources at higher redshifts. However,
increasing $k$ then dramatically reduces the actual dust content at
any epoch due to removal of the ISM through the increase in star
formation efficiency. To explain the high $M_{\rm d}/M_*$ values for
the H-ATLAS sample, we would then need to increase the dust
condensation efficiencies (i.e. the amount of metals which end up in
dust) to a minimum of 60 percent and the effective yield $p$ of heavy
elements from stars would need increase by at least a factor of
two. This is much higher than observed condensation efficiencies for
LIMS or massive stars/SNe although the difference could come from
mantle growth. An increase in the effective yield can only be achieved
through the IMF. The stellar masses of H-ATLAS galaxies are based on
the Chabrier IMF (Chabrier 2003), which has $\alpha \sim 0.6$
(compared to $\alpha \sim 0.7$ for Scalo). However, to significantly
increase the yield from the stellar populations, we would require a
top-heavy IMF (e.g. Harayama, Eisenhauer \& Martins 2008). In
comparison to the MW-Scalo IMF, the effective yield $p$ can increase
by a factor of 4 and more material is returned to the ISM ($\alpha <
0.5$).  A model with these `top-heavy' parameters is shown in
Figure~\ref{fig:mdstarz}b (solid blue), and reproduces the H-ATLAS
observations without the need for extremely efficient mantle growth or
higher dust contribution from SNe.  A top-heavy IMF also frees up more
gas and metals in the ISM throughout the evolution of the galaxy with
time, i.e. $f\sim 0.5$ at $z=0.4$ compared to the $f\sim 0.3$ for a
Scalo IMF, providing a consistent picture with the observed high
dust-to-stellar mass ratios and the expected high gas fraction for
H-ATLAS sources.

If we assume an earlier formation time, or time since last star
formation phase, the model cannot reproduce the H-ATLAS observations
and would require even more extreme values for the dust condensation
efficiency and/or yield.  This suggests a time for the last major star
formation episode for H-ATLAS galaxies to be somewhere in the past 5-6
Gyr (which is consistent with the detailed SED modelling of Rowlands
et al. in prep).

In summary, from this simple model, it is difficult to explain the
high dust-to-stellar mass ratios in the H-ATLAS data even by assuming
we are observing these galaxies at their peak dust mass unless (i) the
fraction of metals incorporated into dust is higher (although we would
require $\chi >70$\,per cent of all metals to be incorporated into
dust) or $\chi >50$\,per cent with pre-enrichment; (ii) The yield is
significantly increased via a top heavy IMF.  An IMF of the form
$\phi(m) \propto m^{-1.7}$ would increase the yield and hence dust
mass by a factor of four, easily accounting for the highest $M_{\rm
  d}/M_*$ ratios. Such IMFs have been postulated to explain
observations of high-$z$ sub-mm galaxies, highly star-forming galaxies
in the local Universe and galaxies with high molecular gas densities
(Baugh et al. 2005; Papadolpoulos 2010; Gunawardhana et al. 2011).
(iii) H-ATLAS galaxies are rapidly consuming their gas following a
relatively recent major episode of star formation (at $z\sim 0.6$).

\subsection{Evolution of the DMF}

\begin{figure}
\includegraphics[trim=0mm -0.5mm 0mm 0mm,clip=true,width=8.5cm]{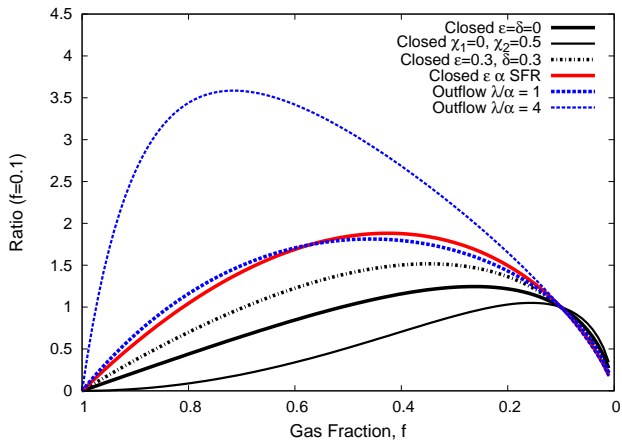}
\caption{Ratio ($R$) of dust mass at gas fraction $f$ to that at
  $f=0.1$ (today). The models are (i) a closed box with no gas
  entering/leaving the system with dust from both supernovae
  $\chi_1=0.1$ and LIMS stars $\chi_2=0.5$ (thick solid; black); (ii)
  with dust from LIMS only $\chi_1=0,\chi_2=0.5$ (thin solid; black);
  (iii) including mantle growth (dot-dashed; black); (iv) A model with
  mantle growth proportional to the SFR (solid; red). (v) and (vi) A
  model which has outflow with gas lost at a rate proportional to one
  or four times the SFR ($\lambda/\alpha$) (dotted; blue). It is worth
  noting that for higher returned fraction from stars to the ISM
  (i.e. $\alpha=0.5$), the ratio decreases for all models ($R<3$ for
  the extreme outflow).}
\label{fig:ratio}
\end{figure}

We now turn to the evolution of the dust mass itself as evidenced from
the DMF (Fig~\ref{fig:dmfcomp}) which shows an increase in the dust
mass of the most massive sources of a factor 4--5 in a relatively
small time-scale ($0<z<0.5,\, \Delta t < 5$ Gyr).  To show the maximum
change in dust mass in galaxies in the model, we plot the ratio ($R$)
of dust mass at time $t$ to that at the present day, assuming a gas
fraction of $f\sim 0.1$ today (Figure~\ref{fig:ratio}).  For a closed
box model, there is little evidence for the dust mass in a given
galaxy changing by more than a factor of 1.5 in the past compared to
its present day value.

It is clear that including outflows produces a better fit to the
variation of dust mass observed in the DMF, with the maximum change in
dust mass approaching the observed change in DMF with $R\sim 4$ for
the extreme outflow model. However, in this case, the peak $M_{\rm
  d}/M_*$ is at least an order of magnitude below the observed values
predicting only $2\times 10^{-4}$ (see
Fig~\ref{fig:mdmstar_mstar}). In this scenario, we would require
$\chi>0.8,\epsilon\eta>0.8$ and $p>0.03$. Such high dust condensation
efficiencies from stellar sources are not observed in the MW, and a
yield as high as $p=0.03$ would again, imply a top heavy IMF. For an
outflow model with $\lambda/\alpha = 1.0$, the parameters
$\chi>0.6,\epsilon\eta>0.3$ and $p>0.02$ would be required to produce
the H-ATLAS dust-to-stellar mass ratios, these are more reasonable
values yet this outflow rate is not sufficient to account for the
increase in dust mass seen in the DMF (reaching a maximum $R \sim
1.5$; Fig~\ref{fig:ratio}). We believe that outflows must be present
at some level (Alton, Davies \& Bianchi 1999) and the observation made
earlier that there is as much dust in galaxy halos as there is in
galaxies themselves is strong circumstantial evidence for some outflow
activity. Given that there are other ways (e.g. radiation pressure on
grains; Davies et al. 1998) to remove dust from disks, we can attempt
to derive a rough upper limit for the outflow required to produce as
much dust in halos at $z\sim 0.35$ as found by M\'{e}nard et
al. (2010). We integrate the dust mass lost from outflows during the
evolution of the galaxy and compare this to the dust mass in the
galaxy at $z=0.3-0.4$ for various values of outflow and star formation
efficiency $k$. The results are shown in Table~\ref{tab:outflow}. This
assumes no dust destruction in either the halo or the disk, and as
such is a very simple model. Equality in dust mass inside and outside
galaxies can be achieved by $z=0.3$ by having moderate outflow $<4
\times \rm SFR$ and $ 0.25 < k < 1.5 \rm Gyr^{-1}$. This is not to say
that all galaxies need have similar evolution; it is quite likely that
H-ATLAS sources are more active and dusty and as such may contain more dust
in their halos than the average SDSS galaxy probed by M\'{e}nard.
This simple exercise merely gives some idea of what sort of 'average'
chemical evolution history is required to reproduce the observation.

\begin{table}
\centering
\caption{\label{tab:outflow} $t$ is the age since formation of the galaxy
  at $z=0.6$. Outflow = 1 and 4 is outflow proportional to 1 and 4
  times the star formation rate. `Halo/Disk' is the ratio of the integrated
  dust mass lost in outflow from $t_{\rm form}$ to $t$ divided by the
  dust mass in the galaxy at $t$.}
\begin{tabular}{ccccc}
\hline
\multicolumn{1}{c}{z}&\multicolumn{1}{c}{t}&\multicolumn{2}{c}{Outflow = 1}&\multicolumn{1}{c}{Outflow = 4}\\
\multicolumn{1}{c}{}&\multicolumn{1}{c}{(Gyr)}&\multicolumn{1}{c}{$k=0.25\,\rm Gyr^{-1}$}&\multicolumn{1}{c}{$k=1.5\,\rm{Gyr^{-1}}$}&\multicolumn{1}{c}{$k=0.25\,\rm{Gyr^{-1}}$}\\
\multicolumn{1}{c}{}&\multicolumn{1}{c}{}&\multicolumn{1}{c}{Halo/Disk}&\multicolumn{1}{c}{Halo/Disk}&\multicolumn{1}{c}{Halo/Disk}\\
\hline
0.5  &  0.5  & 0.09  & 0.42 &  0.33\\
0.4 &  1.0  &  0.2   & 0.96 &  0.73\\
0.3  & 2.2  &  0.4  & 3.03 & 1.95 \\
0.2 & 3.2 &  0.5  & 5.47 &  3.24\\
0.1 & 3.5 & 0.6 & 12.2 & 5.13\\
\hline
\end{tabular}
\end{table}

We now have a conundrum in that the observed evolution in dust mass
requires significant outflow of material, however such outflow leads
to the lowest values of dust-to-stellar mass ratio and cannot be
reconciled to the observations without extreme alterations to the
condensation efficiencies for dust or the stellar yields. Including
dust destruction and mantle growth models which vary with the SFR
alleviates this somewhat since both decrease the dust mass more
significantly at later times. The change in dust mass over the same
period compared to the elementary model with constant $\epsilon$ and
$\delta$ is then more pronounced, but not enough to explain the
evolution in the DMF.

One solution to this is if the galaxies with the highest dust masses
at $z\sim 0.4-0.5$ are not the progenitors of the H-ATLAS sources at
$z\sim 0.1$. We speculate on a scenario where the low redshift spiral
galaxies ($z<0.15$) which do fit the MW model in
Fig~\ref{fig:mdstarz}a comprise one population and the higher redshift
(more dusty) objects are a rapidly evolving star-burst population with
much higher star formation efficiencies (higher $k$), higher dust
condensation efficiencies and/or top-heavy IMFs. The fate of the high
redshift dusty population is that they rapidly consume their gas (and
dust) in star formation and by low redshift they are no longer
detected in H-ATLAS as their gas and dust is exhausted
($f<0.05$). Today they would lie in the faint end of the DMF, mostly
below the limits to which we can currently probe. They would need to
be large stellar mass objects (since their stellar masses are already
large at $z=0.5$) but have little gas and dust today. They could
plausibly be intermediate mass ($\rm log \,M_{\ast} = 10.5-11.5$)
early type galaxies (ETG) in the local Universe, although they would
still be relatively young since they were forming stars actively at
$z=0.4-0.6$. Such depleted objects could have had much more dust in
the past with ratios of $>4$ for the closed box scenario and the model
with mantle growth proportional to SFR. In fact, the dust content of
such galaxies in the past could be even higher since the build up of a
hot X-ray ISM in ETG rapidly destroys any remaining dust (e.g. Jones
et al. 1994). This is an attractive solution as severe outflows are
then no longer required to reproduce the strong dust mass evolution
seen in the DMF. Such a scenario predicts a population of early type
galaxies with moderate dust content and moderate ages ($<5-6$ Gyr) as
the last remnants of their ISM is depleted and the dust gradually
destroyed. H-ATLAS has in fact discovered some promising candidates
for this transitional phase which are discussed in detail by Rowlands
et al. in prep.

Although a closed model does not reproduce the complexity of dust and
metal growth within galaxies, we note that this elementary model
including mantle growth predicts the {\em highest} dust masses for
galaxies with the same initial gas mass and SFRs.  Inflows and
outflows of material simply reduce the dust fraction in the ISM. A
full treatment of the build up of metals in galaxies from stars of
different initial masses further compounds this since relaxing the
instantaneous approximation would produce less dust at earlier times
(at larger values of $f$). The difficulties we have in producing the
observed dust evolution with this elementary treatment are thus only
going to be exacerbated once a more complex treatment is adopted and
therefore our conclusions about the requirements for higher yields and
condensation efficiencies are conservative. To address the issues
above, in particular, the importance of the star formation history and
the role of the IMF, a more complex model of dust and chemical
evolution is required which allows mantle growth, destruction and even
the shape of the IMF to depend on the star formation rate of
galaxies. This is beyond the scope of this paper and the reader is
referred to Gomez et al. (in prep) for a more complete investigation of
the origin and evolution of dust in galaxies.

\subsection{Final caveat}
There is one important way in which the observed dust masses could be
over-estimated; through the dust mass absorption coefficient
$\kappa$. This normalises the amount of emission from dust to the mass
of material present and is dependent on the optical properties and
shapes of the dust grains (for a more thorough review of the
literature see Alton et al. 2004). The value of $\kappa$ used here is
based on that measured in the diffuse ISM of the Milky Way (Boulanger
et al. 1996; Sodroski et al. 1997; Planck Collaboration 2011a) and
also on nearby galaxies by James et al. (2002). This value is some 70
percent higher than that predicted by some models of dust, including
the silicate-graphite-PAH model of Li \& Draine (2001), but lower than
those measured in environments where dust may be aggregated, icy
mantles or `fluffy' (Matthis \& Whiffen 1989; Ossenkopf \& Henning
1994; Krugel \& Siebenmorgen 1994). Latest results from {\em Planck\/}
(Planck Collaboration 2011a) do see a variation in the dust emissivity
with temperature which is expected if there is grain growth in the ISM. It
is thus not inconceivable that $\kappa$ could be globally higher in
galaxies with larger fractions of their ISM in states which lend
themselves to the growth of grains, or where larger fractions of
grains have a SNe origin, or are undergoing destruction by shocks. For
example Ossenkopf \& Henning (1994) show that in only $10^5$ years of
grain evolution in dense environments ($10^6-10^8\, \rm{cm^{-3}}$) the
dust emissivity can increase by a factor $\sim 5$ due to the freeze
out of molecular ice mantles and coagulation. The same authors also
show that changing the ratio of carbon to silicate dust can change the
emissivity by $\sim 40$ percent. Such a change in global dust
composition could reflect the time dependence of evolution of various
dust sources (e.g. SN-II dominating in early time) or metallicity
changes favouring O or C-rich AGB phases. The mechanism for changing
the fraction of the ISM in the densest phases conducive to mantle
growth could be triggered star formation and feedback (e.g. following
an interaction). The fraction of gas in dense clumps has been found to
increase markedly in parts of GMCs which are affected by feedback from
recently formed OB stars (Moore et al. 2007). Draine et al. (2007)
find that for local SINGS galaxies there is no need to consider
ice-mantles in the modelling of the dust emission, but similar
modelling has not been attempted for higher redshift and more sub-mm
luminous sources such as the H-ATLAS sources.

A measurement of $\kappa$ at {\em Herschel} wavelengths (but for local
normal galaxies) has been attempted by Weibe et al. (2009) and Eales
et al. (2010b). Both works, however, suggest a much lower value for
$\kappa$, which would {\em increase} the dust masses estimated here by
a factor $\sim 3$. Given the already difficult task in modelling the
dust masses, we do not believe that $\kappa_{250}$ can be
significantly lower than the values assumed here. A determination of
$\kappa$ for H-ATLAS galaxies is ideally required (as these are {\em
  sub-mm selected\/} sources which may preferentially have higher
$\kappa$). Should an enhanced $\kappa$ at higher redshifts be the
explanation for the large sub-mm luminosities of H-ATLAS galaxies then
this has important implications for the interpretation of all high-z
SMG and {\em Herschel} observations. A change in $\kappa$ will lead to
a change in the opacity of galaxies since the interaction of the
grains with optical/UV photons will be altered. A strong test is to
look at the effects of different $\kappa$ on the
attenuation-inclination relation in the optical as differing values of
$\kappa$ in the sub-mm will (for a fixed observed sub-mm flux) produce
different values for the dust opacity in the optical--UV (see Popescu
et al. 2011 for further details). For galaxies in the Millennium
Survey (Driver et al. 2007) the Li \& Draine (2001) values of $\kappa$
(which are lower than those used here by 70 percent) gave the best
consistency with the observed attenuation-inclination relation,
however it will be interesting to see the results of similar modelling
for H-ATLAS sub-mm selected sources (Andrae et al. in prep). One
result of an increasing $\kappa$ with redshift would be a flattening
of the attenuation-inclination relation with redshift.

A thorough investigation of all the implications using radiative
transfer modelling is required but a change in $\kappa$ is likely
to affect dust masses and the outputs of semi-analytic models which
try to predict the SMG populations. If the FIR luminosity of high-z
galaxies is not dominated by obscured star formation (i.e. there is a
contribution from low opacity diffuse ISM or `leaky' star forming
regions) then a change in $\kappa$ may also lead to a bias in SFR
estimated via FIR luminosities. Very high dust masses and sub-mm
fluxes for SMG in the early Universe have proved challenging for dust
formation models and semi-analytic models of galaxy formation. In
addition to exploring additional sources of dust and IMF variations to
explain the SMG populations, it is worth considering of the
possibility of {\em dust grain property\/} evolution as well.


\section{Conclusions}
\label{conclusions}
We have estimated the dust mass function for the Science
Demonstration Phase data from the {\em Herschel}-ATLAS survey, and
investigated the evolution of the dust mass in galaxies over the past
5 billion years. We find that:
\begin{itemize}
\item{There is no evidence for evolution of dust temperature out to
  $z=0.5$ in this 250\mic selected sample.}
\item{The dust mass function and dust mass density shows strong evolution out to
  $z=0.4-0.5$. In terms of pure mass evolution this corresponds
  to a factor 4--5 increase in the dust masses of the most massive
  galaxies over the past 5 billion years}
\item{Similar strong evolution is found in the ratio of
  dust-to-stellar mass and V-band optical depth - {\em Herschel}-selected
  galaxies were more dusty and more obscured at $z=0.4$ compared to today.}
\item{In order to account for the evolution of the dust content we need
  to radically alter chemical and dust evolution models. We cannot
  reproduce these trends with Milky Way metal or dust yields or star
  formation efficiencies.}
\item{H-ATLAS 250$\mu$m selected sources are highly efficient at
  converting metals into dust, either through mantle growth or through
  a bias in the IMF towards higher mass stars. They must also be
  observed following an episode of star formation (either recent
  formation or recent major burst) where the gas has been consumed at
  a much faster rate than galaxies like the Milky Way today.}
\item{As dust and gas (particularly molecular gas associated with SF)
  are tightly correlated in galaxies, this increase in dust content is
  suggestive of galaxies being more gas rich at $z=0.5$. According to
  the simple chemical model, we are possibly witnessing the period of
  growth toward peak dust mass when gas fractions are $\sim 0.5$ or
  higher. This strong decline in gas and dust content may be an
  explanation for the decrease in star-formation rate density in
  recent times as measured in many multi-wavelength surveys.}
\end{itemize}

This study uses only 3 percent of the area of the H-ATLAS data. Future
improvements will come from the wider area coverage of the full
survey, reducing uncertainties due to cosmic variance and small number
statistics. Use of deeper optical/IR data from forthcoming surveys
such as VISTA-VIKING, pan-STARRS, DES and VST-KIDS will also allow us
to push to earlier times and higher redshifts to find the epoch of
maximum dust content in the Universe.

\section*{Acknowledgments}  HLG acknowledges useful discussions
  with Mike Edmunds. The {\it {\em Herschel}}-ATLAS is a project with
  {\it {\em Herschel}}, which is an ESA space observatory with science
  instruments provided by European-led Principal Investigator
  consortia and with important participation from NASA. The H-ATLAS
  web-site is http://www.h-atlas.org. GAMA is a joint
  European-Australasian project based around a spectroscopic campaign
  using the Anglo- Australian Telescope. The GAMA input catalogue is
  based on data taken from the Sloan Digital Sky Survey and the UKIRT
  Infrared Deep Sky Survey. Complementary imaging of the GAMA regions
  is being obtained by a number of independent survey programs
  including GALEX MIS, VST KIDS, VISTA VIKING, WISE, {\em
    Herschel}-ATLAS, GMRT and ASKAP providing UV to radio
  coverage. GAMA is funded by the STFC (UK), the ARC (Australia), the
  AAO, and the participating institutions. The GAMA website is:
  http://www.gama-survey.org/. This work received support from the
  ALMA-CONICYT Fund for the Development of Chilean Astronomy (Project
  31090013) and from the Center of Excellence in Astrophysics and
  Associated Technologies (PBF06)

\section*{References}
Abazajian K.N., et al., 2009, ApJS, 182, 543\\
Adouze J, Tinsley B.M. 1976, Ann. Rev. Astron. Astro. 14, p43\\
Alton P. B., Bianchi S., Rand R. J., Xilouris E. M., Davies J. I., Trewhella M., 1998, ApJ, 507, L125\\
Alton P. B., Bianchi S. \& Davies J. I., 1999, A\&A, 343, 51\\
Alton P. B., Xilouris E. M., Misiriotis A., Dasrya K. M., Dumke M., 2004, A\&A, 425, 109\\ 
Bianchi S.,Davies J. I. \& Alton P. B., 1999, A\&A, 344, L1\\
Bianchi S., Schneider R., 2001, MNRAS, 378, 973\\
Baldry I. et al., 2010, MNRAS, 404, 86 \\
Barlow M. et al., 2010, A\&A, 518, L138\\
Baugh C. M. et al., 2005, MNRAS, 356, 1191\\
Bendo G. et al., 2010, A\&A, 518, L65\\
Blain A. W., Smail I., Ivison R. J., Kneib J.-P., 1999, MNRAS, 303, 632\\
Boselli et al., 2010, A\&A, 518, L61\\
Boulanger F. et al., 1996, A\&A, 312, 256\\
Braine J., Guelin M.,Dumke M., Brouillet N., Herpin F., Wielebinski R., 1997, A\&A, 326, 963\\
Bruzual G. \& Charlot S., 2003, MNRAS, 344, 1000\\
Buat V., Takeuhci T., Burgarella D., Giovannoli E., Murata K. L., 2009, A\&A, 507, 693\\
Burgarella D., Le Floc\'h E., Takeuchi Y., Huang J. S., Buat V., Rieke G. H., Tyler K. D., 2007, MNRAS, 380, 968\\
Calura F., Pipino A. \& Matteucci F., 2008, A\&A, 479, 669\\
Calzetti D. et al., 2007, ApJ, 666, 870\\
Cannon R. et al., 2006, MNRAS, 372, 425\\
Chabrier G., 2003, PASP, 115, 763\\
Charlot S. \& Fall M., 2000, ApJ, 539, 718\\
Choi P. I. et al., 2006, ApJ, 637, 227\\
Clements D. L., Dunne L. \& Eales S. A, 2010, MNRAS, 403, 274\\
Collister A.A. \& Lahav O., 2004, PASP, 116, 345\\
Coppin K. et al., 2008, MNRAS, 384, 1597\\
Croom S.M. et al., 2009, MNRAS, 392, 19\\
da Cunha E., Charlot S. \&  Elbaz D., 2008, MNRAS, 388, 1595\\
da Cunha E., Eminian C., Charlot S., Blaizot J., 2010a, MNRAS, 403, 1894 \\
da Cunha E., Charmandaris V., Diaz-Santos T., Armus L., Marshall J. A., Elbaz D., 2010b, A\&A, in press (arXiv:1008.2000)\\
Dale D. A., Helou G., Contursi A., Silbermann N. A., Kolhatkar S., 2001, ApJ, 549, 215\\
Danielson A. L. R. et al., 2011, MNRAS, 410, 1687\\ 
Davies J. I., Alton P. B., Bianchi S., Trewhella M., 1998, MNRAS, 300, 1006\\
Driver S., Popescu C. C., Tuffs R. J., Liske J., Graham A. W., Allen P. D., de Propris R., 2007, MNRAS, 379, 1022\\
Driver S. et al., 2011, MNRAS, 413, 971\\
Driver S. \& Robotham A., 2010., MNRAS, 407, 2131\\
Dunne L. et al., 2000, MNRAS, 315, 115\\
Dunne L. \& Eales S. A., 2001, MNRAS, 327, 697\\
Dunne L., Eales S. A. \& Edmunds M. G., 2003, MNRAS, 341, 589 \\
Dunne L., Eales S. A., Ivison R. J., Morgan H., Edmunds M. G., 2003, Nature, 424, 285\\
Dunne L. et al., 2009, MNRAS, 394, 1307\\
Dwek E. 1998, ApJ, 501, 643\\
Dwek E., Galliano F. \& Jones A. P., 2007, ApJ, 662, 927\\
Dwek E., Cherchneff I., 2011, ApJ, 727, 63\\
Dye S. et al., 2010, A\&A, 518, L10 \\
Eales S. A. \& Edmunds M. G., 1996, MNRAS, 299, L29\\
Eales S. A. et al., 2009, ApJ, 707, 1779\\
Eales S., et al., 2010a, PASP, 122, 499\\
Eales S. A. et al., 2010b, A\&A, 518, L62\\
Edmunds M. G., 2001, MNRAS, 328, 223\\
Ferrarotti A. S. \& Gail H.-P., 2006, A\&A, 447, 553\\
Fukugita M. \& Peebles P. J. E., 2004, ApJ, 616, 643\\
Gall C., Andersen A. C. \& Hjorth J., 2011, A\&A, 528, 13\\
Garn T. et al., 2010, MNRAS, 402, 2017\\
Gehrz R., 1989, IAUS, 135, 445\\
Gomez H. et al., 2009, MNRAS, 397, 1621\\
Griffin M. et al., 2010, A\&A, 518, L3\\
Gruppioni C. et al., 2010, A\&A, 518, L27\\
Gunawardhana M. et al., 2011, MNRAS, in press (arXiV:1104.2379)\\
Harayama Y., Eisenhauer F. \& Martins F., 2008, ApJ, 675 1319\\
Helou G. 1986, ApJ, 311, L33\\
Hill D. et al., 2011, MNRAS, 412, 765\\
Hippelein H., Haas M., Tuffs R. J., Lemke D., Stickel M., Klaas U., Volk H. J., 2003, A\&A, 407, 137\\
Hopkins A. M., 2004, ApJ, 615, 209\\
Huchra J., Davis M., Latham D., Tonry J., 1983, ApJS, 52, 89\\
Ibar E. et al., 2010, MNRAS, 409, 38\\
Inoue A. K., 2003, PASJ, 55, 901\\
James A. et al., 2002, MNRAS, 335, 753\\
Jones A. P., Tielens A., Hollenbacj D. J., McKee C. F., 1994, ApJ, 433, 797\\
Jones D.H. et al., 2009, MNRAS, 399, 683\\
Kennicutt R. C., 1998, ApJ, 498, 541\\
Kennicutt R. C. et al., 2009, ApJ, 703, 1672\\
Krause O. et al., 2004, Nature, 432, 596\\
Krugel, E., \& Siebenmorgen, R. 1994, A\&A, 288, 929 \\
Lawrence A. et al., 2007, MNRAS, 379, 1599\\
Le FLoc'h E. et al., 2005, ApJ, 632, 169\\
Li A. \& Draine B. T., 2001, ApJ, 554, 778\\
Madau P., Ferguson H. C., Dickinson M. E., Giavalisco M., Steidel C. C., Fruchter A., 1995, MNRAS, 283, 1388\\
Mannucci, F., Cresci, G., Maiolino, R., Marconi, A., Gnerucci, A., 2010, MNRAS, 418, 2115\\
Mathis, J.S., \& Whiffen, G., 1989, ApJ, 341, 808\\
Meijerink R., Tilanus R. P. J., Dullemond C. P., Israel F. P., van der Werf P. P., 2005, A\&A, 430, 427\\ 
M\'{e}nard B., Scranton R., Fukugita M., Richards G., 2010, MNRAS, 405, 1025\\
Micha{\l}owski M.J., Murphy E.J., Hjorth J., Watson D., Hjorth
  J., Gall C., Dunlop J.S., 2010, A\&A, 522, 15\\
Moore T. et al., 2007, MNRAS, 379, 663\\  
Morgan H. L. \& Edmunds M. G., 2003, MNRAS, 343, 427\\
Morgan H., Dunne L., Eales S. A., Ivison R. J., Edmunds M. G., 2003, ApJ, 597, L33\\
Negrello M. et al., 2010, Science, 330, 800 \\
Ossenkopf, V., \& Henning, Th., 1994, A\&A, 291, 943\\
Page M. J. \& Carrera F. J., 2000, MNRAS, 311, 433\\
Papadopoulos P. P., 2010, ApJ, 720, 226\\
Papadopoulos P. P. \& Pelupessy F. I., 2010, ApJ, 717, 1037\\
Pascale E. et al., 2011, MNRAS, in press (arXiV:1010.5782)\\
Pilbratt G. et al., 2010, A\&A, 518, L1\\
Planck Collaboration, 2011a, A\&A, accepted (arXiV:1011.2036)\\
Planck Collaboration, 2011b, A\&A accepted (arXiV:1011.2045)\\ 
Poglitsch A. et al., 2010, A\&A, L2 \\
Popescu C. C., Tuffs R. J., Volk H. J., Pierini D., Madore B. F., 2002, ApJ, 567, 221\\
Popescu C. C., Tuffs R. J., Dopita M. A., Fischer J., Kylafis N. D., Madore B. F., 2011, A\&A, 527, 109\\
Pozzetti L. et al., 2007, A\&A, 474, 443\\
Rigby E.E. et al., 2011, MNRAS, in press (arXiV:1010.5787)\\
Rho J. et al., 2008, ApJ, 673, 271\\
Romano D., Chiappini C., Matteucci F., Tosi M., 2005, A\&A., 430, 491, \\
Sargent B. A. et al., 2010, ApJ, 716, 878\\
Saunders W., Rowan-Robinson M., Lawrence A., Efstathiou G., Kaiser N., Ellis R. S., Frenk C. S., 1990, MNRAS, 242, 318\\
Saunders W. et al., 2000, MNRAS, 317, 55\\
Scalo, J. M. 1986, Fund. Cosmic Phys., 11, 1\\
Schechter P., 1976, ApJ, 203, 297\\
Schmidt M., 1968, ApJ, 151, 393\\
Serjeant S. \& Harrison D., 2005, MNRAS, 356, 192\\
Seymour N., Symeonidis M., Page M. J., Huynh M., Dwelly T., McHardy I., Rieke G., 2010, MNRAS, 402, 2666\\
Sodroski T. J., Odegard N., Arendt R. G., Dwek E., Weiland J. L., Hauser M. G., Kelsall T., 1997, ApJ, 480, 173\\
Soifer B. T., Boehmer L., Neugebauer G., Sanders D. B., 1989, AJ, 98, 766\\
Smith D.J.B. et al., 2011, MNRAS, in press (arXiv:1007.5260)\\
Stevens J. A., Amure M. \& Gear W. K., 2005, MNRAS, 357, 361\\
Stickel M., Klaas, U. \& Lemke D., 2007, A\&A, 466, 831\\
Sutherland R. \& Saunders R., 1992, MNRAS, 259, 413\\
Swinbank M. et al., 2010, Nature, 464, 733\\
Symeonidis M., Page M. J., Seymour N., Dwelly T., Coppin K., McHardy I., Rieke G. H., Huynh M., 2009, ApJL, 660, L73\\
Symeonidis M., Page M. J. \& Seymour N., 2011, MNRAS, 411, 983\\
Tielens A. G. G. M., 1998, ApJ, 499, 267\\
Todini P \& Ferrara A., 2001, MNRAS, 325, 726\\
Tuffs R.J, Popescu C.C., Volk H.J., Kylafis N.D., Dopita M.A., 2004, A\&A, 419, 835\\
VanDalfsen M.L., \& Harris W.E., 2004, ApJ, 127, 368\\
Villar V. et al., 2008, ApJ, 677, 169\\
Vlahakis C., Dunne L. \& Eales S. A., 2005, MNRAS, 364, 1253 \\
Wang L., \& Rowan-Robinson M., 2009, MNRAS, 398, 109\\
Wang L. \& Jing Y. P., 2010, MNRAS, 402, 1796\\
Weibe D. V. et al., 2009, ApJ, 707, 1809\\
Willmer C. N. A. et al., 2009, ApJ, 138, 146\\ 
Xu K. C. et a., 2007, ApJS, 173, 432\\
Zhukovska S., Gail H.-P. \& Trieloff M., 2008, A\&A, 479, 453\\


\section*{Appendix A: Chemical Evolution Modelling}

This simple chemical evolution model describes the star, gas, metal
and dust content of a galaxy making the instantaneous recycling
approximation. The mass fraction of metals, $Z$ in this model changes
as a mass $ds$ of the ISM is formed into stars  assuming no inflows or outflows via the following
equation (Edmunds 2001):

\begin{equation}
d(Zg)=\alpha p ds+(1-\alpha)Zds - Zds
\end{equation}

where $g$ is the gas mass and $\alpha$ (Eq.~\ref{eq:lock}) is the
fraction of mass from a generation of star formation which is locked
up in long-lived stars or remnants $m_R$ as determined by the
initial mass function ($\phi(m)$):  

\begin{equation}
\alpha = 1-\int^{m_2}_{m_1} \left[m-m_R(m)\right]\phi(m)dm 
\label{eq:lock}
\end{equation}

$p$ is the effective yield of heavy elements from stars $p =p'/\alpha
\sim 0.01$ where $\alpha \sim 0.7$ in agreement with Milky Way values
for a Scalo IMF.

In a closed box model (i.e. no inflow or outflow of material), the
total mass of the system ($M_{\rm tot} =\rm gas+stars$) is unity so
that the fraction of gas in a galaxy (the ratio of gas mass to total
baryonic mass) is $f=g$.  In this scenario, the initial conditions
are: $Z=0$ at $g=f=1$ and the gas mass of the galaxy is given by
$g=1-\alpha s$. The analytic solution for the metal mass fraction $Z$
is (Eq.~\ref{eq:metal}):

\begin{equation}
Z=-p{\rm ln}f.
\label{eq:metal}
\end{equation}

An early episode of star-formation prior to the evolution of the
closed box would pre-enrich the gas and increase the interstellar
metallicity (pre-enrichment is often invoked to explain the
metallicities of globular clusters in the Milky Way).  We can include
pre-enrichment of the ISM with metals $Z_i$ using

\begin{equation}
Z=Z_i-p{\rm ln}f
\end{equation}

where $Z_i \sim 0.1-0.2\,Z_{\odot}$ (VanDalfsen \& Harris 2004).
Correspondingly, the dust mass fraction $y$ varies with $ds$ via:

\begin{equation}
d(yg)=\alpha p \chi_1 ds+(1-\alpha)\chi_2Zds - yds
\label{eq:dusty}
\end{equation}

where $\chi$ is a parameter to describe the fraction of the mass of
interstellar metals in dust grains from supernovae remnants or their
massive star progenitors ($\chi_1$), and/or from the stellar
atmospheres of low-intermediate mass stars (LIMS: $\chi_2$). The
analytic solution is given in Eq.~\ref{eq:complex} for $y=0$ at $g=1$
and for $\alpha=0.7$ (typical locked up fraction for a Scalo IMF):

\begin{equation}
y = 2.3\left[{\left(\chi_1-\chi_2\right)\left(1-f^{0.43}\right)\over{{\rm ln}(1/f)}} +0.43\chi_2\right] p {\rm ln(1/f)}
\label{eq:complex}
\end{equation}

For the special case where $\chi_1=\chi_2=\chi$, Eq.~\ref{eq:complex}
reduces to:

\begin{equation}
y = \chi  p {\rm ln(1/f)}.
\label{eq:simpler}
\end{equation}

We can add an additional term to the dust mass from stars by assuming
that grains accrete at a rate proportional to the available metals and
dust cores in dense interstellar clouds (following Edmunds 2001):

\begin{equation}
y = \chi p {\rm ln}(1/g) + \epsilon \eta_c(z-y)
\end{equation}

where $\epsilon$ is the fraction of the ISM dense enough for mantle
growth (here we set this arbitrarily to $0.3$), $\eta_c$ is the
efficiency of interstellar depletion in the dense cloud (i.e. if all
the metals in the dense clouds are accreted onto dust grains then
$\eta_c=1$). 

Dust destruction via supernova shocks can be added to this elementary
model by assuming some fraction $\delta$ of interstellar grains are
removed from the ISM as a mass $ds$ is forming stars (therefore adding
a term $-\delta ds$ to Eq.~\ref{eq:dusty}). In this work, we test
both a constant fraction with $\delta=0.3$ (appropriate for MW-type
galaxies and therefore provides a testcase with a minimum destruction
level expected for the H-ATLAS spirals) and a function that varies
proportionally to the SFR (since a higher SFR equates to a higher Type
II SN rate).
 
\subsection*{Outflow}

We include a simple model for outflow of gas, in which gas is added or
lost from the system at rates proportional to the star formation
rate. For large galaxies this outflow rate is assumed to be less than
four times the SFR ($\lambda/\alpha \leq 4$; see Eales \& Edmunds 1996
for discussion; this corresponds to a galaxy which retains only $\sim$
20\,per\,cent of its original mass).  We do not consider inflow of
unenriched material since this only slightly reduces the dust mass
w.r.t. the closed box model and doesn't significantly change the
evolution of a galaxy (Edmunds 2001). One can imagine a scenario with
inflow of pre-enriched material (e.g. merger), providing new material
for star formation, even at later times when the original gas mass of
the galaxy has been consumed through the star formation efficiency
parameter $k$.  Modelling the effects of this on the dust mass is
beyond the scope of the simple model presented here.

Outflows remove dust from the
interstellar medium via $-{\lambda} y ds$.  The solution is given by
Eq.~\ref{eq:outflow} if destruction $\delta=0$:

\begin{equation}
y ({\rm outflow}) = {y\over{1+\lambda/\alpha}}
\label{eq:outflow}
\end{equation}

The gas mass $g$ is related to the gas fraction $f$ in this model by:

\begin{equation}
\label{outflowg}
g ({\rm outflow}) =   {f \over{1+(\lambda/\alpha)(1-f)}},
\end{equation}

the metallicity mass fraction, $Z$:

\begin{equation}
\label{outflowz}
Z ({\rm outflow}) =   -{p{\rm ln}(g) \over{1+\lambda/\alpha}},
\end{equation}

and the total mass of the system is:

\begin{equation}
M_{\rm tot} ({\rm outflow}) =   {1+(\lambda/\alpha)g \over{1+\lambda/\alpha}}.
\end{equation}

\subsection*{Dust and Stellar Mass}

The dust mass per unit stellar mass for the elementary model for equal
$\chi$ with no mantles, destruction or outflow, is given by
Eq.~\ref{app:md}:

\begin{equation}
{M_{\rm d}\over{\alpha s}}= {-\chi p g {\rm ln} (g)\over{1-g}}
\label{app:md}
\end {equation} 


We can rewrite Eq.~\ref{app:md} as a function of time, since SFR $\psi(t)$ is related to the gas mass via is related to the gas mass via
\begin{equation}
\psi(t)=kg(t)^{1.5}
\end{equation}

where $k$ is the star formation efficiency measured in $\rm Gyr^{-1}$
and the variation of $g$ with time is
\begin{equation}
g= \left( {1.5\over{\alpha k t +1.5}}\right)^2 
\label{eq:simple}
\end{equation}

High values of
$k$ will result in a higher SFR and a more rapid build up of the final
stellar mass for the
same initial gas mass.  

For outflow models, the dust mass fraction and the gas mass is reduced
as described in Eqs.~\ref{eqn:outflow} - \ref{outflowz}.
\end{document}